%% file: ms.tex
\documentclass[12pt]{article}
\pdfoutput=1

\usepackage[letterpaper,margin=1.0in]{geometry}
\usepackage[toc,page]{appendix}

\usepackage[round]{natbib}
\usepackage[pdftex]{graphicx}
\graphicspath{{./img/}}
\DeclareGraphicsExtensions{.pdf,.jpeg,.png}

\usepackage{newtxtext}

\usepackage{amsmath}
\usepackage{amsfonts}
\usepackage{mathtools}
\usepackage{dsfont}
\usepackage{bm}
\usepackage{accents}
\usepackage{ifthen}
\usepackage{xcomment}
\usepackage{booktabs}
\usepackage[linesnumbered,ruled,vlined]{algorithm2e}
\usepackage{tikz}
\usetikzlibrary{bayesnet}

\usepackage{array}

\usepackage{threeparttable}

\usepackage[caption=false,font=footnotesize,labelfont=sf,textfont=sf]{subfig}

\usepackage{url}

\DeclareMathOperator*{\argmax}{arg\,max}

\DeclareMathOperator{\Dir}{Dir}
\DeclareMathOperator{\Mult}{Mult}

\DeclareMathOperator{\sign}{sign}
\DeclareMathOperator{\diag}{diag}
\newcommand{\eat}[1]{}

\newcommand{\bbeta}{\bm \beta}

\newcommand{\bsigma}{\bm \Sigma}
\newcommand{\bphi}{\bm \phi}
\newcommand{\bpsi}{\bm \psi}
\newcommand{\btheta}{\bm \theta}

\newcommand{\bpi}{\bm \pi}

\newcommand{\bw}{{\bm w}}
\newcommand{\bz}{{\bm z}}

\newcommand{\bG}{{\bm G}}
\newcommand{\bb}{{\bm b}}
\newcommand{\bmu}{{\bm \mu}}
\newcommand{\bvarphi}{{\bm \varphi}}
\newcommand{\bxi}{{\bm \xi}}
\newcommand{\blambda}{{\bm \lambda}}
\newcommand{\brho}{{\bm \rho}}
\newcommand{\btau}{{\bm \tau}}
\newcommand{\bomega}{{\bm \omega}}
\newcommand{\Real}{\mathbb{R}}
\newcommand{\simplex}{\mathbb{S}}
\newcommand{\lgamma}{\log\Gamma}
\newcommand{\Exp}{\mathbb{E}}
\newcommand{\given}{\,|\,}

\newcommand{\iid}{\stackrel{\text{iid}}{\sim}}
\newcommand{\ind}{\stackrel{\text{ind}}{\sim}}

\newboolean{include-notes}
\setboolean{include-notes}{true}
\ifthenelse{\boolean{include-notes}}%
{}%
{\newxcomment[]{notes}}

\newboolean{include-notes-to-stay}
\setboolean{include-notes-to-stay}{false}
\ifthenelse{\boolean{include-notes-to-stay}}%
{\newenvironment{notes-to-stay}
	{\begin{color}{magenta}}{\end{color}}}%
{\newxcomment[]{notes-to-stay}}

\providecommand{\keywords}[1]
{
	\small	
	\textbf{\textit{Keywords---}} #1
}

\title{Analyses of Multi-collection Corpora via 
	Compound Topic Modeling}

\author{Clint~P.~George\thanks{Assistant Professor, Indian 
		Institute of Technology Goa. The work performed at the 
		Informatics Institute, University of Florida. 
		E-mail: \texttt{clint@iitgoa.ac.in}
	}, 
	Wei~Xia\thanks{Department of Statistics, University of 
	Florida}, 
	and~George~Michailidis\thanks{Founding 
		Director of the Informatics Institute and Professor of 
		Statistics, University of Florida
	}
}

\begin{document}

\maketitle

\begin{abstract}
As electronically stored data grow in daily life, obtaining novel 
and relevant information becomes challenging in text mining. 
Thus people have sought statistical methods based on term 
frequency, matrix algebra, or topic modeling for text mining. 
Popular topic models have centered on one single text collection, 
which is deficient for comparative text analyses. We consider a 
setting where one can partition the corpus into subcollections. 
Each subcollection shares a common set of topics, but there exists 
relative variation in topic proportions among collections. 
Including any prior knowledge about the corpus (e.g. organization 
structure), we propose the compound latent Dirichlet allocation 
(cLDA) model, improving on previous work, encouraging 
generalizability, and depending less on user-input parameters. To 
identify the parameters of interest in cLDA, we study Markov chain 
Monte Carlo (MCMC) and variational inference approaches 
extensively, and suggest an efficient MCMC method. We evaluate cLDA 
qualitatively and quantitatively using both synthetic and 
real-world corpora. The usability study on some real-world corpora 
illustrates the superiority of cLDA to explore the underlying 
topics automatically but also model their connections and 
variations across multiple collections. 
\end{abstract}

%

\input{introduction}

\input{hierarchical-model}

\input{posterior-inference}
\input{experiments}

\input{discussion}


%
\begin{appendices}
	\input{app-posterior}
	\input{app-mmala}

	\input{app-est-eta-gamma}
	\input{app-vem}
	\input{supp-compare-ags-mgs-vem}

	\input{app-perplexity}

	\input{app-nips-results}
\end{appendices}

\section*{Acknowledgments}

This work is supported by grants NIH \#$7$ 
R$21$ GM$101719$-$03$ to George Michailidis and  
Institute of Education Sciences 
\#R$305160004$, and the University of Florida 
Informatics Institute.

%
\bibliographystyle{plainnat}
\bibliography{ms}

\end{document}

%% file: introduction.tex
\section{Introduction}
\label{sec:introduction}
Newspapers, magazines, scientific journals, and 
social media messages being composed in daily living produce 
routinely an enormous volume of text data. The corresponding 
content comes from diverse backgrounds and represent distinct 
themes or ideas; modeling and analyzing such heterogeneity in 
large-scale is crucial in any text mining frameworks. Typically, 
text mining aims to extract relevant and interesting information 
from the text by the process of structuring the written text 
(e.g. via semantic parsing, stemming, lemmatization), inferring 
hidden patterns within the structured data, and finally, 
deciphering the results. To address these tasks in an unsupervised 
manner, numerous statistical methods such as TF-IDF 
\citep{SaltonWongYang:1975}, latent semantic indexing  
\citep[LSI]{Deerwester:1990}, and probabilistic topic models, e.g. 
probabilistic LSI~\citep[pLSI]{Hofmann:1999}, latent Dirichlet 
allocation~\citep[LDA]{BleiNgJordan:2003}, have emerged in the 
literature. With minimal human effort, they can be generalized to 
arbitrary text in any natural language.  

Topic models such as LDA are developed to capture 
the underlying semantic structure of a corpus (i.e. a document 
collection) based on co-occurrences of words. They consider 
documents as bags of words, considering the order of words in a 
document uninformative. LDA (\figurename~\ref{fig:gm}) assumes that 
(a) a topic is a latent (i.e., unobserved) distribution on the 
corpus vocabulary, (b) each document in the corpus is described by 
a latent mixture of topics, and (c) each observed word in a 
document, there is a latent variable representing a topic from 
which the word is drawn. This model is suitable for documents 
coming from a single collection.  This assumption is insufficient 
for comparative analyses of text, especially, for partition-able 
text corpora, as we describe next. 

Suppose, we deal with corpora of (i) articles accepted in various 
workshops or consecutive proceedings of a conference or (ii) 
blogs/forums from people different countries. It may be of interest 
to explore research topics across multiple/consecutive proceedings 
or workshops of a conference or cultural differences in blogs and 
forums from different countries \citep{Zhai:2004}. Suppose, we have 
\textsl{political} news articles from different sources such as New 
York Times, Washington Post, The Wall Street Journal, and Reuters. 
Although, all talk about topic politics, modeling articles from 
each news source as a collection may help to explore each sources' 
article style, policies, region influences, culture, etc. Moreover, 
the category labels, workshop names, article timestamps, or geotags 
can provide significant prior knowledge regarding the unique 
structure of these corpora. We aim to include these prior 
structures and characteristics of the corpus into a probabilistic 
model adding less computational burden, for expert analyses of the 
corpus, collection, and document-level characteristics.

\begin{figure}[t!] 
	\centering 
	\includegraphics[width=.65\linewidth]
	{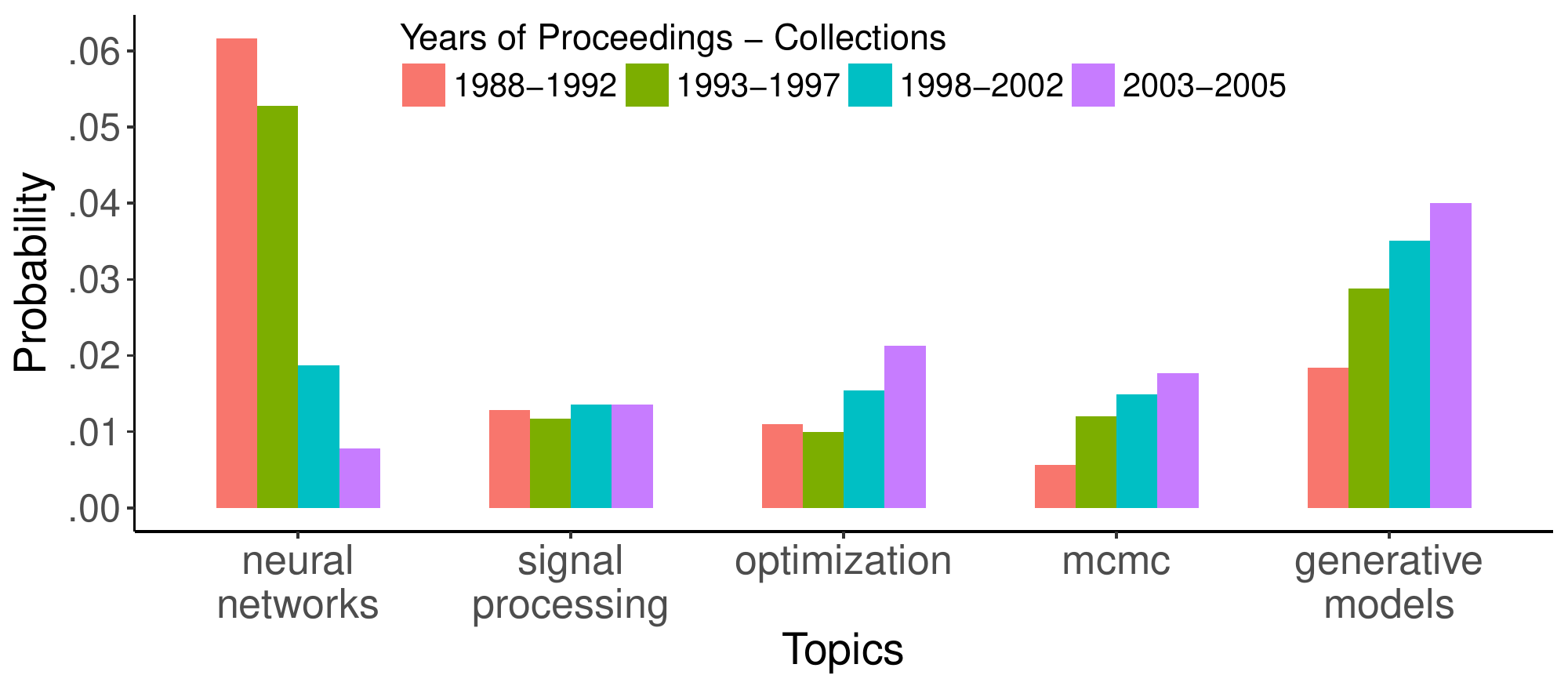}
	\caption{
		Estimated topic proportions of five topics for the four 
		time spans 
		of the NIPS conference proceedings. 
	}
	\label{fig:nips-clda-sel-pi}
\end{figure}

For example, \figurename~\ref{fig:nips-clda-sel-pi} gives the 
results of an experiment on a real document corpus employing the 
proposed model in this paper. The corpus consists of articles 
accepted in the proceedings of the NIPS conference for the years 
from $1988$ to $2005$. We wish to analyze topics that evolve over 
this timespan. We thus partitioned the corpus into four collections 
based on time (details in Section~\ref{sec:experiments}). 
The plot shows estimates of topic proportions for all four 
collections. Evidently, some topics got increased attention  (e.g.  
Markov chain Monte Carlo (MCMC), generative models) and some topics 
got decreased popularity (e.g. neural networks) over the years of 
the conference. For some topics, the popularity is relatively 
constant (e.g. signal processing) from the beginning of NIPS. The 
algorithm used is unsupervised, and only takes documents, words, 
and their collection labels as input. 

Assuming a flat structure to all documents in a corpus, 
LDA or its variants is not well suited to the multi-collection 
corpora setting: (1) 
Ignoring predefined structure or organization of collections 
may end up having topics that describe only some, not all of the 
collections. 
(2) There is no direct way to define which set of topics describe 
the common information across collections and which topic describes 
information specific to a particular collection. A crude solution 
is to consider each collection as a separate corpus and fit an LDA 
model for each corpus. There are reasons why one may want to avoid 
it: (i) one needs to solve the nontrivial alignment of topics in 
each model, for any useful comparison of topics among collections 
(Moreover, the topics inferred from individual corpus partitions 
and the whole corpora can themselves be different) and (ii) 
information loss due to modeling collections separately, especially 
for small datasets (details, Section \ref{sec:eval-performance}). 

In this paper, we introduce the compound latent Dirichlet 
allocation (cLDA, Section~\ref{sec:hierarchical-model}) model that 
incorporates any prior knowledge on the organization of documents 
in a corpus into a unified modeling framework. cLDA assumes a 
shared, collection-level, latent mixture of topics for documents in 
each collection. This collection-level mixture is used as the base 
measure to derive the (latent) topic mixture for each document. All 
collection-level and document-level variables share a common set of 
topics at the corpus level, which enables us to perform exciting 
inferences, as shown in \figurename~\ref{fig:nips-clda-sel-pi}. 
cLDA exhibits a certain degree of supervision incorporating the 
collection membership for each word (and document) in a corpus 
implicitly in the modeling framework.  cLDA can thus aid visual 
thematic analyses of large sets of documents, and include corpus, 
collection, and document specific views.

The parameters of interest in cLDA are hidden and are inferred via 
posterior inference. However, exact inference is intractable in 
cLDA; and, non-conjugate relationships in the model further make 
approximate inference challenging. Popular approximate posterior 
inference methods in topic models are MCMC and variational 
methods.  MCMC methods enable us to sample hidden variables of 
interest from the intractable posterior with convergence 
guarantees. We consider two MCMC methods for cLDA: (a) 
one uses the traditional auxiliary variable updates within Gibbs 
sampling, and (b) the other uses Langevin dynamics within 
Gibbs sampling, a method that received recent attention. Our 
experimental evidence suggests the former method, which gives 
superior performance with only a little computational overhead 
compared to the collapsed Gibbs sampling algorithm for 
LDA~\citep[CGS]{GriffithsSteyvers:2004} (details in 
Sections~\ref{sec:pi-sampling-illustration}, 
\ref{sec:experiments}, \ref{sec:algorithm-illustration}, and 
\ref{app:nips-results}), and is the main focus in this paper. 
Variational methods are often used in topic modeling as they give 
fast, parallel implementations by construction. Although they 
converge rather quickly, our studies show that (Section 
\ref{sec:pi-sampling-illustration} and Section 
\ref{sec:algorithm-illustration}) their solutions are suboptimal 
compared to the results of the other two MCMC schemes.

The contributions of this paper are three-fold: (i) we propose a 
probabilistic model cLDA that can capture the topic structure of a 
corpus including organization hierarchy of documents, (ii) we study 
efficient methods for posterior inference in cLDA, and (iii) we 
perform an empirical study of the real-world applicability of the 
cLDA model---for example, (a) analyzing topics that evolves 
overtime, (b) analyzing patterns of topics on customer reviews, and 
(a) summarizing topic structure of document collections in a 
corpus---via three text corpora used in the research community. 
Also, note that the inference about collection-level topic mixtures 
may be of interest to the general perspective of posterior sampling 
on the probability simplex in statistics.

The remainder of the paper is organized as follows.
Section~\ref{sec:hierarchical-model} formally defines the cLDA
hierarchical model. Section~\ref{sec:inference} describes 
algorithms for posterior inference and evaluates correctness of the 
algorithms using a synthetically corpus. In 
Section~\ref{sec:experiments}, we assess the performance of the 
cLDA model and conclude that it exhibits superior performance, both 
quantitatively (e.g. via perplexity---a popular scheme for 
evaluating the predictive performance of topic models 
\citep{WallachEtal:2009}, and external 
measures such as topic coherence \citep{MimnoEtAl:2011}) and 
qualitatively. We also 
compare cLDA with other popular models in the literature, and 
provide a usability study for cLDA in this section. 
Section~\ref{sec:discussion} concludes this paper with a  
summary of our work.

%% file: hierarchical-model.tex
\section{A Compound Hierarchical Model} 
\label{sec:hierarchical-model}


We first set up some terminology and notation.  Vectors are denoted 
by bold, lower case alphabets (e.g. $\bpi$) and scalar values are 
denoted 
by normal, lowercase letters (e.g. $\pi_{jk}$). Matrices or 
tensors are denoted by bold, upper case Latin alphabets (e.g. 
$\bG$) or bold Greek alphabets without subscripts (e.g. $\bbeta$).
There is a vocabulary $\mathcal{V}$ of $V$ terms in the corpus;
in general, $\mathcal{V}$ is considered as the union of all the 
word tokens in all the documents of the corpus, after removing 
stop-words and normalizing tokens (e.g. stemming). 
The number of topics $K$ is assumed to be known. (Discussion of how 
to handle this issue in practice is Section~\ref{sec:experiments}.) 
By definition, a topic is a distribution over ${\cal V}$, i.e., a 
point in the $V$-$1$ dimensional simplex $\simplex_V$.  We will 
form a $K \times V$ matrix $\bbeta$, whose $k^{\text{th}}$ row is 
the $k^{\text{th}}$ topic (how $\bbeta$ is formed will be described 
shortly).  Thus, the rows of $\bbeta$ are vectors $\bbeta_1, 
\ldots, \bbeta_K$, all lying in $\simplex_V$. There are $J$ 
collections in the corpus and for $j = 1, 2, \ldots, J$, collection 
$j$ has $D_j$ documents. For $d = 1, \ldots, D_j$, document $d$ in 
collection $j$ (i.e. document $jd$) has $n_{jd}$ words, $w_{jd1}, 
\ldots, w_{jdn_{jd}}$. 
Each word is represented by the index or id of the corresponding 
term from the vocabulary. We represent document $jd$ by the vector 
$\bw_{jd} = (w_{jd1}, \ldots, w_{jdn_{jd}})$, collection $j$ by the 
concatenated vector $\bw_j = (\bw_{j1}, \ldots, \bw_{jD_j})$, and 
the corpus by the concatenated vector $\bw = (\bw_{1}, \ldots, 
\bw_{J})$. 

\begin{figure}[ht!] 
\centering 
\begin{tikzpicture}[x=4.5cm,y=2cm]

\node[obs]                      (w)       {$w_{jdi}$} ; %
\node[latent, left=.4 of w]     (z)       {$z_{jdi}$} ; %
\node[latent, left=.4 of z]     (theta)   {$\btheta_{jd}$}; %
\node[const, above=.7 of theta](gamma)   {$\gamma$};
\node[latent, left=.4 of theta, style=dashed] (pi)      {$\bpi_j$}; 
\node[const, above=.7 of pi]   (alpha)   {$\alpha$};

\node[latent, above=.6 of w]        (beta)     {$\bbeta_k$}; %
\node[const, left=.55 of beta]    (eta)     {$\eta$}; %

\edge[semithick,style=dashed]    {alpha}   {pi} ; 
\edge[semithick,style=dashed]    {pi}      {theta} ; 
\edge[semithick]                 {gamma}   {theta} ; 
\edge[semithick]                 {theta}   {z} ; 
\edge[semithick]                 {eta}     {beta} ; 
\edge[semithick]                 {z, beta}  {w} ; 

\plate {word-plate} { %
	(w) %
	(z) %
} {$i = 1, \ldots, n_{jd}$}; %

\plate {doc-plate} { %
	(word-plate) %
	(theta) %
} {$d = 1, \ldots, D_j$} ; %

\plate [style=dashed]{collection-plate} { %
	(doc-plate) %
	(pi) %
} {$j = 1, \ldots, J$} ; %

\plate {topic-plate} { %
	(beta) %
} {$k = 1, \ldots, K$} ; %

\end{tikzpicture}
\caption{Graphical model of the latent Dirichlet allocation (LDA) 
model (the inner structure) and the compound latent Dirichlet 
allocation (cLDA) model (the outer dashed-structure is the extension 
to LDA): Nodes denote random variables, shaded nodes denote observed 
variables, edges denote conditional dependencies, and plates denote  
replicated processes.\label{fig:gm}} 
\end{figure}
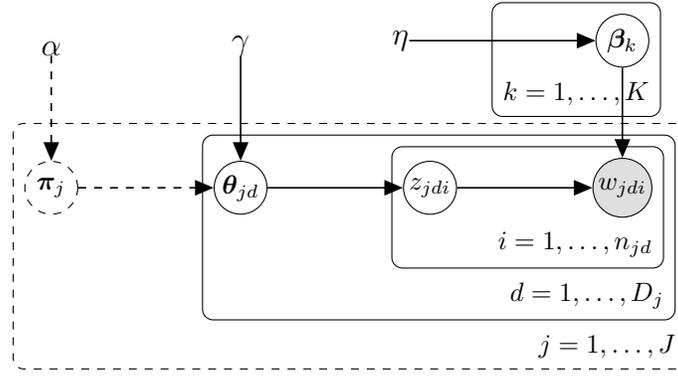

We use $\Dir_L(a \omega_1, \ldots, a \omega_L)$ to 
denote the finite-dimensional Dirichlet distribution on the 
($L$-$1$)-dimensional simplex. This has two parameters, a scale 
(concentration) parameter $a$ and a base measure $(\omega_1, \ldots, 
\omega_L)$ on the ($L$-$1$)-dimensional simplex. Thus 
$\Dir_L(a, \ldots, a)$ denotes an $L$-dimensional Dirichlet 
distribution with the constant base measure $(1, \ldots, 1)$. 
$\Mult_L(b_1, \ldots, b_L)$ to represents the multinomial 
distribution with number of trials equal to $1$ and probability 
vector $(b_1, \ldots, b_L)$. Let $h = (\eta, \alpha, \gamma) \in 
(0, \infty)^3$ be the hyperparameters in the model. We formally 
define the cLDA model as (see \figurename~\ref{fig:gm}).
\begin{eqnarray}
\bbeta_k &\iid& \Dir_V(\eta, \ldots, \eta), 
\text{ for topic } k = 1, \ldots, K \label{eq:cLDA-e} \\ 
\bpi_j &\iid& \Dir_K(\alpha, \ldots, \alpha), 
\text{ for collection } j = 1, \ldots, J \label{eq:cLDA-d} \\
\btheta_{jd} &\iid& \Dir_K(\gamma \pi_{j1}, \ldots, \gamma 
\pi_{jK}), \text{ for document } jd 
\label{eq:cLDA-c}\\ 
z_{jdi} &\iid& \Mult_K(\btheta_{jd}), 
\text{ for each word } w_{jdi} \label{eq:cLDA-b}\\ 
w_{jdi} &\ind& \Mult_V(\bbeta_{z_{jdi}}) \label{eq:cLDA-a}
\end{eqnarray}

%
%
%
%

The distribution of $z_{jd1}, \ldots, z_{jdn_{jd}}$ 
will depend on the document-level variable $\btheta_{jd}$ that  
represents a distribution on the topics for document $jd$. A single 
$\btheta_{jd}$ for document $jd$ 
encourages different words in document $jd$ to share the same 
document level characteristics. The distribution of $\btheta_{j1}, 
\ldots, \btheta_{jD_j}$ will depend on the collection-level 
variable $\bpi_{j}$ which indicates a distribution on the topics 
for collection $j$. A single $\bpi_{j}$ for collection $j$ 
encourages different documents in collection $j$ to have the same 
collection level properties. 
A single $\bbeta$ is shared among all documents, which  
encourages documents in various collections in the corpus 
to share the same set of topics.  
Note that the standard LDA model~\citep{BleiNgJordan:2003} is a 
special case of the proposed cLDA model, where there exist single  
parameters $\bpi_{j}$ and $\bbeta$ that fail to capture 
potential heterogeneity amongst the predefined collections, 
an objective that cLDA is designed for.

\subsection*{Related Work}

\cite{Zhai:2004} refer to the problem of  
multi-collection corpora modeling as comparative text mining (CTM), 
which uses pLSI as a building block. Comparing with CTM model, cLDA 
employs an efficient and generalizable LDA-based framework that has 
several advantages over pLSI---for example, LDA incorporates 
Dirichlet priors for document topic structures in a natural way to 
deal with newly encountered documents. Furthermore, cLDA combines 
collection-specific characteristics in a natural probabilistic 
framework enabling efficient posterior sampling, depending less on 
user-defined parameters as in CTM. 

The cLDA model shares some similarities, but also exhibits 
differences from: Hierarchical Dirichlet 
Process~\citep{TehEtal:2006} and Nested Chinese Restaurant 
Process~\citep{Blei:2004}, which are introduced to learn document 
and topic hierarchies from the data non-parametrically. In 
nonparametric models, as we observe more and more data, the data 
representations grow structurally. Instead of manually specifying 
the number of topics K, these nonparametric topic models infer K 
from the data by assuming a Dirichlet process prior for document 
topic distributions and topics. A major hurdle in these frameworks 
is that inference can be computationally challenging for large 
datasets. Our experimental evidence also shows that HDP produces  
too many fragmented topics, which may lead the practitioner to bear 
the additional burden of post processing (Section 
\ref{sec:eval-performance}). Here, 
similar to LDA, cLDA is a parametric model, i.e., 
the data representational structure is fixed and does not grow as 
more data are observed. cLDA assumes that K is fixed and can be 
inferred from the data directly, e.g., empirical Bayes 
methods~\citep{George:2015}, or by cross-validation. We thus 
have simple Dirichlet priors in cLDA without adding much burden to 
the model and inference (details appear in 
Section~\ref{sec:inference}).

In light of adding supervision, several modifications to LDA model 
have been proposed in the literature. Supervised LDA 
\citep[sLDA]{McauliffeBlei:2008} is an example; for each document 
$d$, sLDA introduces a response variable $y_d$ that is assumed to 
be generated from document $d$'s empirical topic mixture 
distribution.\eat{For regression, variable $y_d$ is normally 
distributed with the mean as the empirical distribution. For 
classification, one can assume a discrete likelihood for the class 
labels $y_d$.} In practice, the posterior inference in such a 
setting can be inefficient due to the high non-linearity of the 
discrete distribution on the empirical 
parameters~\citep{ZhuXing:2014}.  cLDA, on the other hand, proposes 
a generative framework incorporating the collection-level 
characteristics, without much computational burden (details,  
Section~\ref{sec:experiments}). Also, the objectives of these 
related models are different from the focus of this paper.

\begin{notes-to-stay}
One can view the hierarchical model of cLDA as a special instance 
of the model proposed by \citet{Kim:2013}. \citet{Kim:2013} 
extended the LDA model to have multiple layers in the hierarchical 
model, for corpora with many categories and subcategories. 
In cLDA, we are interested in corpora with just one layer of top 
level categories. This model is adequate for many real-world 
corpora such as books with various chapters and conference 
proceedings from different years. Additionally, we will show that 
\citet{Kim:2013}'s variational methods for the multilayer 
hierarchical model is suboptimal compared to the proposed MCMC 
methods for two layered data.  

A popular scheme for evaluating the performance of topic models is 
perplexity (or the predictive probability of held-out documents), 
which we use here to assess cLDA's performance  quantitatively. 
In the text domain, one can use trained cLDA models for tasks such 
as classifying text documents and summarizing document collections 
and corpora. cLDA also gives excellent options to visualize and 
browse documents. The model and algorithm we describe can be used 
to form a hierarchy of documents in a collection including relevant 
topic information. One can then use the document hierarchy to 
enlighten one's perception of the contents of the collection. 
\end{notes-to-stay}

%% file: posterior-inference.tex
\section{Posterior Sampling}
\label{sec:inference}

The parameters of interest in the cLDA model, i.e., (a) corpus-level 
topics, (b) collection-level mixture of topics, (c) document-level 
mixture of topics, and (d) topic indices of words are hidden. 
We identify these hidden variables given the observed word 
statistics and document organization hierarchy in the corpus via 
posterior inference.

Let $\btheta = (\btheta_{11}, \ldots, \btheta_{1D_1}, \ldots, 
\btheta_{J1}, \ldots, \btheta_{JD_J})$, $\bz_{jd} = (z_{jd1},\ldots, 
z_{jdn_{jd}})$ for $d = 1, \ldots, D_j$, $j = 1, \ldots, J$, $\bz = 
(\bz_{11}, \ldots, \bz_{1D_1}, \ldots, \bz_{J1}, \ldots, 
\bz_{JD_J})$, and $\bpi = (\bpi_1, \ldots, 
\bpi_J)$. We will then use $\bpsi$ to denote 
the latent 
variables $(\bbeta, \bpi, \btheta, \bz)$ in the cLDA model. 
For any given $h$, \eqref{eq:cLDA-e}--\eqref{eq:cLDA-a} 
in the hierarchical model induce a 
prior distribution $p_h(\bpsi)$ on $\bpsi$. 
Equation~\eqref{eq:cLDA-a} gives the 
likelihood $\ell_{\bw}(\bpsi)$. The words 
$\bw$ and their document and collection labels 
are observed. We are interested in 
$p_{h,\bw}(\bpsi)$, the posterior distribution of $\bpsi$ given $\bw$ 
corresponding to the prior $p_h(\bpsi)$. Applying Bayes rule, we can 
write the posterior distribution 
$p_{h,\bw}$ of $\bpsi$ as
\begin{equation}
p_{h,\bw}(\bpsi) \propto \ell_{\bw}(\bpsi) p_h(\bpsi) \\
\label{eq:posterior-psi}
\end{equation}
Using \eqref{eq:cLDA-e}--\eqref{eq:cLDA-a} of the 
hierarchical model, we can write 
\eqref{eq:posterior-psi} as (Section \ref{app:posterior} gives 
additional details)
\begin{equation}
\begin{split}
  p_{h,\bw}(\bpsi) & \propto 
	\Biggl[ \prod_{j=1}^J \prod_{d=1}^{D_j} 
	\frac{\prod_{k=1}^K \theta_{jdk}^{n_{jdk} + \gamma \pi_{jk} - 
	1}}{\prod_{k=1}^K {\Gamma(\gamma \pi_{jk})}} 
	  \Biggr]  
	\Biggl[ \prod_{j=1}^J \prod_{k=1}^K \pi_{jk}^{\alpha - 1} \Biggr] \\ 
	& \hspace{.5cm} \Biggl[ \prod_{k=1}^K \prod_{v=1}^V \beta_{kv}^{\sum_{j=1}^{J} 
	\sum_{d=1}^{D_j} m_{jdkv} + \eta - 1} \Biggr] 
\end{split}
\label{eq:posterior-psi-exp}
\end{equation}
where $n_{jdk}$ is the number of words in 
document $d$ in collection $j$ that are 
assigned to topic $k$, and $m_{jdkv}$ is  
the number of words in document $d$ in 
collection $j$ for which the latent topic is 
$k$ and the index of the word in the 
vocabulary is $v$. These count statistics 
depend on both $\bz$ and $\bw$. 
Note that the constants in the Dirichlet 
normalizing constants are absorbed into the 
overall constant of proportionality.  
Unfortunately, the normalizing constant of the 
posterior $p_{h,\bw}(\bpsi)$, is the 
likelihood of the data with all latent 
variables integrated out, is a non-trivial 
integral. This makes exact inference difficult 
in cLDA. 

Popular methods for approximate posterior inference in topic models 
are Markov chain Monte Carlo (e.g. see  
\cite{GriffithsSteyvers:2004}) and variational methods (e.g. 
see~\cite{BleiNgJordan:2003}). Although variational methods may 
give a fast and scalable approximation for the posterior, due to 
optimizing the proxy lower-bound, it may not produce optimal 
solutions as the MCMC methods in practice (e.g. see 
\cite{TehNewmanWelling:2007}). That is the case in this model 
setting, as our experimental evidence suggests. Hence, we leave the 
details of our development of variational methods (VEM) for cLDA in 
Section~\ref{app:vem}, to stay our discussion 
focused. 

\subsection{Inference via Markov chain Monte Carlo Methods}
\label{sec:mcmc}


According to the hierarchical model 
\eqref{eq:cLDA-e}--\eqref{eq:cLDA-a}, $\btheta_{jd}$'s and 
$\bbeta_{k}$'s are independent, and by 
inspecting the posterior~\eqref{eq:posterior-psi}, given 
$(\bpi, \bz)$, we get: 
\begin{equation}
\begin{split}
\btheta_{jd} & \sim \Dir_K \biggl( n_{jd1} + \gamma \pi_{j1}, 
\ldots, n_{jdK} + \gamma \pi_{jK} \biggr),  \\ 
\bbeta_{k} & \sim \Dir_V \biggl( m_{k1} + \eta, \ldots, m_{kV} 
+ \eta \biggr), 
\end{split}
\label{eq:theta-beta-posterior}
\end{equation}
where $m_{kv} = \sum_{j=1}^{J} \sum_{d=1}^{D_j} m_{jdkv}$ and  
$d = 1, \ldots, D_j, j = 1, \ldots, J, k = 1, \ldots, K$. 
Note that \eqref{eq:theta-beta-posterior}
implicitly dependent on the observed data $\bw$. 


We can integrate 
out $\btheta$'s and $\bbeta$'s to get the 
marginal posterior 
distribution of $(\bpi, \bz)$ (up to a normalizing constant) as  
\begin{eqnarray}
  \begin{split}
	p_{h,\bw}(\bpi, \bz) 
	&\propto  
	\Biggl[ 
		\prod_{j=1}^J \prod_{d=1}^{D_j} \prod_{k=1}^K
		\frac{ \Gamma(\gamma \pi_{jk} + n_{jdk})}
				 {\prod_{k=1}^K {\Gamma(\gamma \pi_{jk})}} 		
  \Biggr]  
  \\ & \hspace{4.6mm}
	\Biggl[ \prod_{j=1}^J \prod_{k=1}^K 
	\pi_{jk}^{\alpha - 1} 
	\Biggr]  
	\Biggl[ \prod_{k=1}^K 
						\frac{\prod_{v=1}^V \Gamma(m_{..kv} + \eta)}
						     {\Gamma(m_{..k.} + V\eta)} 
  \Biggr]
	\end{split}
	\label{eq:pi-z-posterior}
\end{eqnarray}

Let the vector $\bz^{(-jdi)}$ be the topic assignments of all 
words in the corpus except for word $w_{jdi}$. And, we define 
$n_{jd} := \sum_{k = 1}^{K} n_{jdk}$ and $m_{k} := \sum_{v = 
1}^{V} m_{kv}$.  By inspecting~\eqref{eq:pi-z-posterior}, we 
obtain a closed form expression for the conditional posterior 
distribution for $z_{jdi}$, given $\bz^{(-jdi)}$ and $\pi_{jk}$, 
as 
\begin{equation}
p_{h,\bw} \left(z_{jdik} = 1 \given . \right) 
\propto  
\frac{\gamma \pi_{jk} + n^{(-jdi)}_{jdk}}
{\gamma + n^{(-jdi)}_{jd}} 
\frac{\eta  + m^{(-jdi)}_{kv}}
{V\eta + m^{(-jdi)}_{k}}	
\label{eq:posterior-z}
\end{equation}
where the superscript $(-jdi)$ for the count statistics $n_{jdk}$, 
$n_{jd}$, $m_{kv}$, and $m_{k}$ means that we discard the 
contribution of word $w_{jdi}$ for counting. (see  
Section \ref{app:posterior-z}) This enables us 
to build a Gibbs sampling chain on $\bz$, by sampling $z_{jdi}$, 
given $\pi_{jk}$ and $\bz^{(-jdi)}$.

Given $\bz_j$ and the observed data $\bw_j$, 
we have the unnormalized posterior probability 
density function for $\bpi_j$ as  
\begin{equation}
\tilde{p}_{\bw_j}(\bpi_j \given \bz_j) \propto  
\prod_{d=1}^{D_j} 
\prod_{k=1}^K 
\frac{ \Gamma(\gamma \pi_{jk} + n_{jdk})}
		{\Gamma(\gamma \pi_{jk})}
\prod_{k=1}^K \pi_{jk}^{\alpha - 1}
\label{eq:posterior-pi}
\end{equation} 
Here we use the fact that $\bpi_j$'s are independent of 
$\bbeta_k$'s. We wish to sample $\bpi_j$'s from this distribution, 
however the normalized density function $p_{\bw_j}(\bpi_j \given \bz_j)$ 
is computationally intractable: the density $p_{\bw_j}(\bpi_j \given 
\bz_j)$ has a non-conjugate 
relationship between $\bpi_j$'s and 
$\btheta_{dj}$'s, which makes its normalizer intractable. Next, we 
describe two Markov chain Monte Carlo schemes that enable us to 
sample from this posterior density.  We shall only focus on the 
first scheme due to its superior performance in our numerical 
experiments. 

\subsubsection{Auxiliary Variable Sampling} 

The first scheme is based on the traditional auxiliary variable 
sampling scheme. The idea of auxiliary variable sampling is that 
one can sample from a distribution $f(x)$ for variable $x$ by 
sampling from some augmented distribution $f(x, s)$ for variable 
$x$ and auxiliary variable $s$, such that the marginal distribution 
of $x$ is $f(x)$ under $f(x, s)$. One can build a Markov chain 
using this idea in which, auxiliary variable $s$ is introduced 
temporarily and discarded, only leaving the value of $x$.  
Since $f(x)$ is the marginal distribution of $x$ under $f(x, s)$, 
this update for $x$ will leave $f(x)$ 
invariant~\citep{Neal:2000}. Suppose  
$\text{S}(n_{jdk}, s)$ denotes the unsigned 
Stirling number of the first kind. We can then 
get the expression for augmented sampling by 
plugging in the 
factorial expansion~\citep{Abramowitz:1974}
\begin{equation}
\frac{ \Gamma(\gamma \pi_{jk} + n_{jdk})} {\Gamma(\gamma \pi_{jk})} 
= \sum_{s = 0}^{n_{jdk}} \text{S}(n_{jdk}, s) {(\gamma \pi_{jk})}^s,
\label{eq:gammafn-sterling-exp}
\end{equation}
into the marginal posterior density 
\eqref{eq:posterior-pi} as  
\begin{equation}
\tilde{p}_{\bw_j}(\bpi_j \given .) \propto  
\prod_{d=1}^{D_j} 
\prod_{k=1}^K 
\text{S}(n_{jdk}, s_{jdk}) {(\gamma \pi_{jk})}^{s_{jdk}}
\prod_{k=1}^K \pi_{jk}^{\alpha - 1}
\label{eq:posterior-pi-augmented}
\end{equation}
(\cite{Newman:2009} and \cite{TehEtal:2006}  used a 
similar idea in a different hierarchical 
model.) The 
expression~\eqref{eq:posterior-pi-augmented} 
introduces auxiliary variable $s_{jdk}$. 
By 
inspecting~\eqref{eq:posterior-pi-augmented}, 
we get a closed form 
expression for sampling $\bpi_j$, for $j = 1, 
\ldots, J$:   
\begin{equation}
\bpi_j \sim \Dir_{K} \left (\sum_{d = 1}^{D_j} s_{jd1} + \alpha, 
\ldots, \sum_{d = 1}^{D_j} s_{jdK} + \alpha 
\right),
\label{eq:aux-pi}
\end{equation}
and we update the auxiliary variable $s_{jdk}$ 
by the Antoniak sampling scheme 
\citep[Appendix A]{Newman:2009}, i.e., the 
Chinese restaurant process~\citep[CRP]{Aldous:1985} with 
concentration parameter $\gamma \pi_{jk}$  and the number of 
customers $n_{jdk}$. 
The auxiliary variable $s_{jdk}$ is typically 
updated by drawing $n_{jdk}$ Bernoulli 
variables as
\begin{equation}
s_{jdk} = \sum_{l = 1}^{n_{jdk}}{s^{(l)}}, \,\,
s^{(l)} \iid \text{Bernoulli} 
\left(\frac{\gamma \pi_{jk}}{\gamma 
	\pi_{jk} + l - 1} \right)
\end{equation}
In our experience, this augmented update for collection 
level parameters will add a low computational 
overhead to the collapsed Gibbs sampling chain on $\bz$, and is easy 
to implement in practice.   
The Markov chain on $(\bpi, \bz)$ based on the auxiliary 
variable update within Gibbs sampling is given by Algorithm 
\ref{alg:gibbs-aux-pi-z}. We use AGS to denote this chain. 

\begin{algorithm}[t]
 \KwData{Observed words $\bw$ and document metadata}
 \KwResult{A Markov chain on $(\bpi, \bz)$} \vspace{.3cm}
 initialize $(\bpi^{(0)}, \bz^{(0)})$\;
 \For{Gibbs iteration $t$}{ 
 
	\tcp{Sampling word topic indices} 
	
	\For{word $w_{jdi}$, $i = 1, \ldots, n_{jd}$, $d = 1, \ldots, D_j$, $j = 1, \ldots, J$}{
		given $\bpi^{(t)}_j$, sample $z^{(t + 1)}_{jdi}$ via $p_{\bw}(z_{jdi} 
		\given \bz^{(-jdi)}, \bpi^{(t)}_j)$ given by
		\eqref{eq:posterior-z}\;
		update count statistics $n_{jdk}$, $n_{jd}$, $m_{kv}$, 
		and $m_{k}$, according to $z^{(t + 1)}_{jdi}$\;
	}

	\tcp{Auxiliary variable sampling for collection-level topic mixtures}
	\For{collection $j = 1, \ldots, J$}{
		given $(\bpi^{(t)}_j, \bz^{(t + 1)})$, update $s_{jdk}$ via 
		the Antoniak sampling scheme\;
		given $s_{jdk}$, sample $\bpi_j$ via~\eqref{eq:aux-pi}\;
		discard $s_{jdk}$ and update $\bpi^{(t + 1)}_j$ as $\bpi_j$\;
	}

 }
 \caption{Augmented Gibbs sampler (AGS)}
\label{alg:gibbs-aux-pi-z}
\end{algorithm}

\subsubsection{Metropolis Adjusted Langevin Monte Carlo} 
\label{sec:mgs}
Another option to define a Markov chain with invariant density 
$p_{\bw_j}(\bpi_j \given \bz_j)$  is to employ the 
Metropolis-Hastings (MH) algorithm \citep{MetropolisEtal:1953, 
Hastings:1970} with Langevin dynamics~\citep{Girolami:2011}. 
This scheme has become popular for sampling on the simplex 
recently~\citep{Patterson:2013}. 

Typically, MH algorithm proposes a transition $\bpi^{(t)} 
\rightarrow \bpi^{(*)}$ with density $q(\bpi^{(*)} \leftarrow 
\bpi^{(t)})$---i.e. the proposal---for the current step $t$, and 
then accept it with probability 
\begin{equation}
a (\bpi^{(t)}, \bpi^{(*)}) = \text{min} \left(1, 
\frac{\tilde{p}(\bpi^{(*)}) q(\bpi^{(t)} \leftarrow \bpi^{(*)})} 
{\tilde{p}(\bpi^{(t)}) q(\bpi^{(*)} \leftarrow \bpi^{(t)}) } 
\right)
\label{eq:mh-accept-ration}
\nonumber 
\end{equation}
where $\tilde{p}(\bpi)$ denotes the unnormalized density 
of $\bpi$. Here, we ignore the subscript $j$ and other dependencies 
for brevity. The accept-reject step ensures that the proposed 
Markov chain is reversible with respect to the stationary target 
density and satisfies detailed balance. The proposal 
distribution simulates random-walks---e.g. $q(\bpi^{(*)} 
\leftarrow \bpi^{(t)}) =  \mathcal{N}_K(\bpi^{(*)} \given 
\bpi^{(t)}, \bsigma)$, a $K$-dimensional normal distribution 
with mean $\bpi^{(t)}$ and covariance matrix $\bsigma$. A key  
challenge of MH in practice is to find a proposal with reasonable 
acceptance rate, especially when $K$ is large. 

Recent developments show that Langevin dynamics~\citep{Kennedy:1990} 
is an ideal option to define a proposal distribution. Langevin 
dynamics proposes random walks by a combination of gradient updates 
and Gaussian noise as follows. 
We denote the log-density at state $t$ by ${\cal L}(\bpi^{(t)}) := 
\log p(\bpi^{(t)})$. The Langevin diffusion with stationary 
distribution $p(\bpi)$ is defined by the stochastic 
differential equation (SDE) 
\begin{equation}
\text{d}\bpi(t) = \frac{1}{2} \nabla_{\bpi} {\cal L}(\bpi^{(t)}) 
\text{d}t + \text{d}\bb(t)   
\label{eq:langevin-sde}
\end{equation} 
where $\bb$ denotes a $K$-dimensional Brownian motion. Given the 
current state $t$, we then define a proposal based on the 
first-order Euler discretization of~\eqref{eq:langevin-sde} as 
\begin{eqnarray}
\bpi^{(*)} &=& \bmu(\bpi^{(t)}, \varepsilon) + \varepsilon 
\bxi^{(t)}, \label{eq:langevin-discrete-eq} \\
\bmu(\bpi^{(t)}, \varepsilon) &=& \bpi^{(t)} + 
\frac{\varepsilon^2}{2} \nabla_{\bpi} {\cal L}(\bpi^{(t)}) 
\nonumber\\
\bxi^{(t)} &\iid& \mathcal{N}_K(0, \mathds{1}_K) \nonumber
\end{eqnarray}
where $\varepsilon$ is a user defined step-size for the 
discretization and $\bxi^{(t)}$ is distributed according to a zero 
mean $K$-dimensional multivariate normal distribution with the 
identity covariance matrix $\mathds{1}_K$. This induces a proposal 
density $q(\bpi^{(*)} \leftarrow \bpi^{(t)}) = 
\mathcal{N}_K(\bpi^{(*)} \given \bmu(\bpi^{(t)}, \varepsilon), 
\varepsilon^2 \mathds{1}_K)$. 
Note that the discretized process~\eqref{eq:langevin-discrete-eq} 
may be 
transient and is no longer reversible with respect to the 
stationary density $p(\bpi)$~\citep{RobertsTweedie:1996}. 
However, one can ensure convergence to 
the target density $p(\bpi)$ via a MH accept-reject scheme 
\citep[p. 591]{Besag:1994}, which we denote by MALA (Section 
\ref{app:mmala}). One may notice this MALA update as a 
special case of Hamiltonian Monte Carlo~\citep[Section 
5.5.2]{Neal:2010}. 

However, the proposed MALA update for $\bpi^{(t+1)}$ has some 
shortcomings. First, the drift term $\bxi^{(t)}$ in the MALA 
proposal is based on an isotropic diffusion and may be inefficient 
for strongly correlated variables with widely differing variances, 
forcing the step size $\varepsilon$ to accommodate variates with 
smallest variance~\citep{Girolami:2011}. Second, $\bpi^{(t+1)} \in 
\simplex_K$, the probability simplex $\simplex_K$ is compact, and 
it needs to handle the cases when MALA proposes a path that's 
outside the simplex. Third, typical Dirichlet priors over the 
probability simplex put most of their probability mass on the edges 
and corners of the simplex---e.g. in models such LDA 
\citep{Patterson:2013}. Computing gradients for these models become 
unstable when the probabilities are close to zero and causes issues 
for MALA updates. Our approaches to handle these issues are 
discussed next.

To handle the first issue, \cite[Section 
5]{Girolami:2011} suggested using a preconditioning matrix 
${\bG}(\bpi)$. They defined ${\bG}(\bpi)$ as 
an arbitrary metric tensor on a Riemannian manifold induced by the 
parameter space of a statistical model. This requires us to update 
the natural gradient $\nabla_{\bpi} {\cal L}(\bpi^{(t)})$ and  
Brownian motion $\text{d}\bb(t)$ in \eqref{eq:langevin-sde} 
(details in Section \ref{app:mmala}). We thus use 
the corresponding Riemannian Manifold Metropolis Adjusted Langevin 
Algorithm \citep[MMALA]{Girolami:2011} in this paper.

To perform valid moves of $\bpi$ that lies on the probability 
simplex (the second and third issues), one needs to consider 
boundary conditions. A natural solution to handle boundaries is to 
re-parameterize $\bpi$~\citep{Patterson:2013}. 
Let $\bvarphi = (\varphi_{1}, \ldots, \varphi_{K}) \in 
\Real^K$. We take the prior on $\bvarphi$ as a product of i.i.d. 
Gamma random variables as 
\begin{equation}
\begin{split}
|\varphi_{k}| &\iid \text{Gamma} \Bigl (\alpha, 1 \Bigr ), \\
p(\bvarphi) &\propto \prod_{k = 1}^{K} {|\varphi_{k}|}^{\alpha - 
1} e^{-|\varphi_{k}|}. 
\end{split}
\end{equation}
We define $|\varphi_{.}| := \sum_{k=1}^K |\varphi_{k}|$. Let  
$\pi_{k}$ be $|\varphi_{k}| / |\varphi_{.}|$, for each $k = 1, 
2, \ldots, K$. This choice keeps the prior on $\bpi$ a    
Dirichlet density. We can then re-write the unnormalized 
conditional density~\eqref{eq:posterior-pi} on $\bpi$ in terms of 
$\bvarphi$, and derive the MMALA updates, accordingly 
(Section \ref{app:mmala}). The Markov 
chain on $(\bpi, \bz)$ induced by this scheme is denoted by MGS 
(Algorithm \ref{alg:gibbs-mmala-pi-z}).  

\begin{algorithm}[t!]
	\KwData{Observed words $\bw$ and document metadata}
	\KwResult{A Markov chain on $(\bpi, \bz)$} \vspace{.3cm}
	initialize $(\bpi^{(0)}, \bz^{(0)})$\;
	initialize $\bvarphi^{(0)}$, i.e., the re-parametrization for 
	$\bpi^{(0)}$\;
	\For{Gibbs iteration $t$}{ 
		
		\tcp{Sampling word topic indices} 
		
		\For{word $w_{jdi}$, $i = 1, \ldots, n_{jd}$, $d = 1, 
		\ldots, 
			D_j$, $j = 1, \ldots, J$ }{
			given $\bpi^{(t)}_j$, sample $z^{(t + 1)}_{jdi}$ via 
			$p(z_{jdi} 
			\given \bz^{(-jdi)}, \bw, \bpi^{(t)}_j)$ given by
			\eqref{eq:posterior-z}\;
			update count statistics $n_{jdk}$, $n_{jd.}$, 
			$m_{..kv}$, 
			and $m_{..k.}$, according to $z^{(t + 1)}_{jdi}$\;
		}
		
		\tcp{Metropolis Hastings updates for collection-level 
		topic mixtures}
		
		\For{collection $j = 1, \ldots, J$}{
			propose the MMALA update $\bvarphi_j^{(*)}$ 
			via~\eqref{eq:riemann-langevin-euler-reparam}\;
			calculate the acceptance ratio 
			$\text{a}(\bvarphi_j^{(t)}, 
			\bvarphi_j^{(*)})$, based on the unnormalized density 
			\eqref{eq:posterior-pi-re-parametrization} and 
			the transition density 
			\eqref{eq:pi-transition-density-re-parameterization} \;
			
			\eIf{$\text{Uniform}(0, 1) < \text{min}(1, 
			a(\bvarphi_j^{(t)}, \bvarphi_j^{(*)}))$ } { 
				set $\bvarphi_j^{(t+1)} = \bvarphi_j^{(*)}$ \tcp{MH 
				accept}
			}{
				set $\bvarphi_j^{(t+1)} = \bvarphi_j^{(t)}$ \tcp{MH 
				reject}
			}
			
			set $\pi_{jk} = 
			\frac{\left|\varphi^{(t+1)}_{jk}\right|}{\sum_{k=1}^{K}{\left|\varphi^{(t+1)}_{jk}
			 \right|}}, \, k = 1, 2, \ldots, K$\;
		}
		
	}
	\caption{MMALA updates within Gibbs sampler (MGS)}
	\label{alg:gibbs-mmala-pi-z}
\end{algorithm}


Note that we can easily augment any of these two chains on 
$(\bpi, \bz)$: AGS and MGS to a Markov chain on $\bpsi$ with 
invariant distribution $p_{h,\bw}(\bpsi)$ by using the conditional 
distribution of $(\bbeta, \btheta)$, given 
by~\eqref{eq:theta-beta-posterior}. 
The augmented chain will hold the  
convergence properties of the chain on $(\bpi, \bz)$.

\subsection{Empirical Evaluation of Samples $\bpi$}
\label{sec:pi-sampling-illustration}

We consider a synthetic corpus by simulating  
\eqref{eq:cLDA-e}--\eqref{eq:cLDA-a} of the 
cLDA hierarchical model with the number of collections $J = 2$ and 
the number of topics $K = 3$. 
We did this solely so that we can visualize the results of 
the algorithms. We also took the vocabulary size $V = 40$, the 
number of documents in each collection $D_j = 100$, and the 
hyperparameters $h_{\text{true}} = (\alpha, \gamma, \eta) = (.1, 1, 
.25)$. Collection-level Dirichlet sampling via 
\eqref{eq:cLDA-d} with $\alpha$  produced 
two topic distributions $\bpi^{\text{true}}_{1} = (.002, \epsilon, 
.997)$ and $\bpi^{\text{true}}_{2} = (.584, .386, .030)$, where 
$\epsilon$ denotes a small number. 

We study the ability of the proposed algorithms AGS, MGS, and VEM 
(details, Algorithms~\ref{alg:gibbs-mmala-pi-z} and 
~\ref{alg:vem}) for cLDA to recover parameters 
$\bpi^{\text{true}}_{1}$ and $\bpi^{\text{true}}_{2}$. 
We do this by comparing samples of $\bpi_j$ from the AGS and MGS 
chains on $(\bpi, \bz)$ with variational estimates of $\bpi_j$ from 
VEM iterations\footnote{An implementation of all 
algorithms 
and datasets discussed in this paper is available as an R package 
at \url{https://github.com/clintpgeorge/clda}}.  We initialized 
$\bpi^{(0)}_{1} = \bpi^{(0)}_{2} = (.33, .33, .33)$ and use 
hyperparameters $h_{\text{true}}$ for all three algorithms. Using 
the data $\bw$, we ran both chains AGS and MGS for $2000$ 
iterations, and algorithm VEM converged after $45$ EM  
iterations.

\begin{figure*}[t!] 
	\centering 
	\subfloat[AGS: $\bpi^{(2000)}_{2} = (.347, .015, .636)$]{
		\includegraphics[trim={0 1.2cm 0 
		1.2cm},clip,width=.45\linewidth]
		{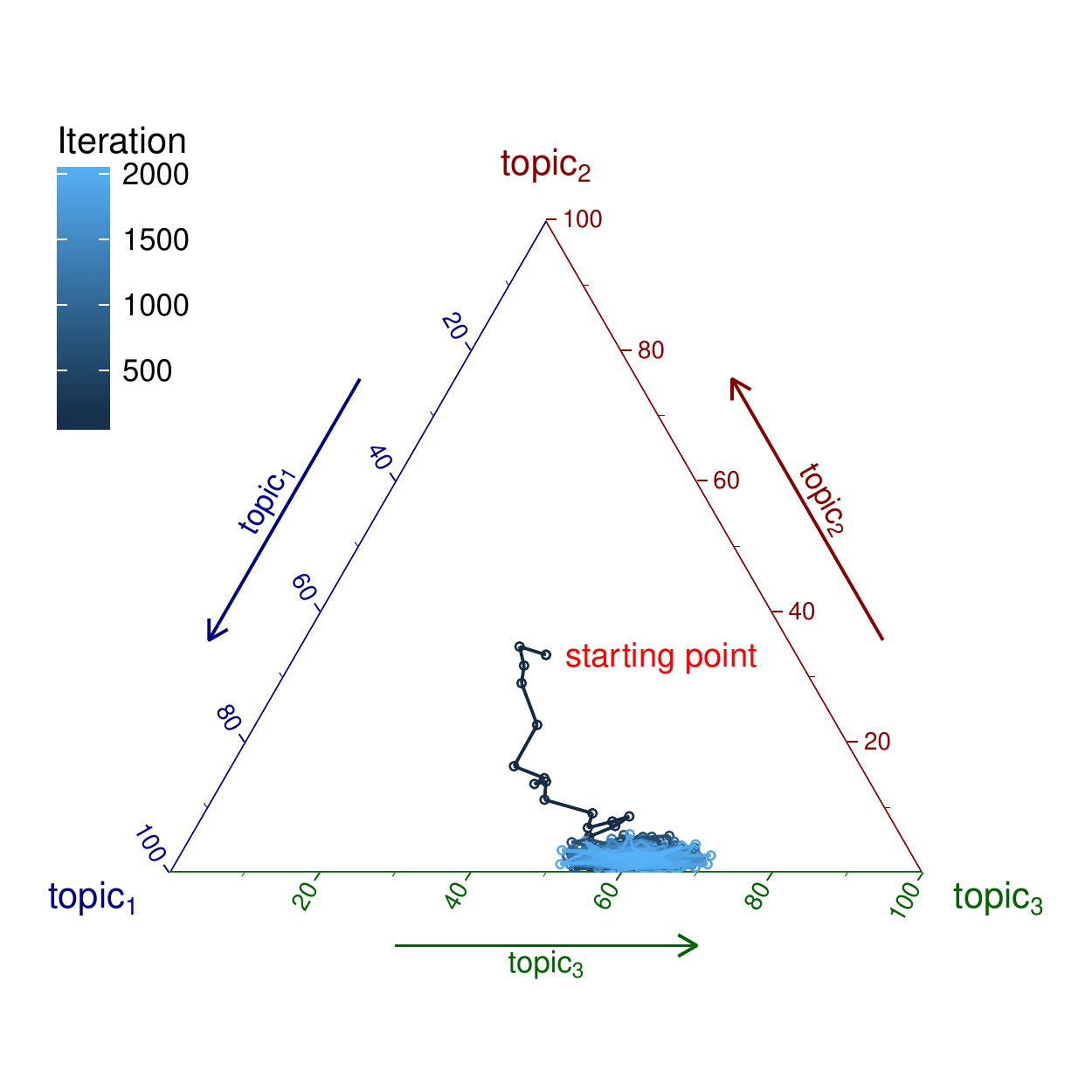}}
	\subfloat[MGS: $\bpi^{(2000)}_{2} = (.379, .005, .615)$]{
		\includegraphics[trim={0 1.2cm 0 
		1.2cm},clip,width=.45\linewidth]
		{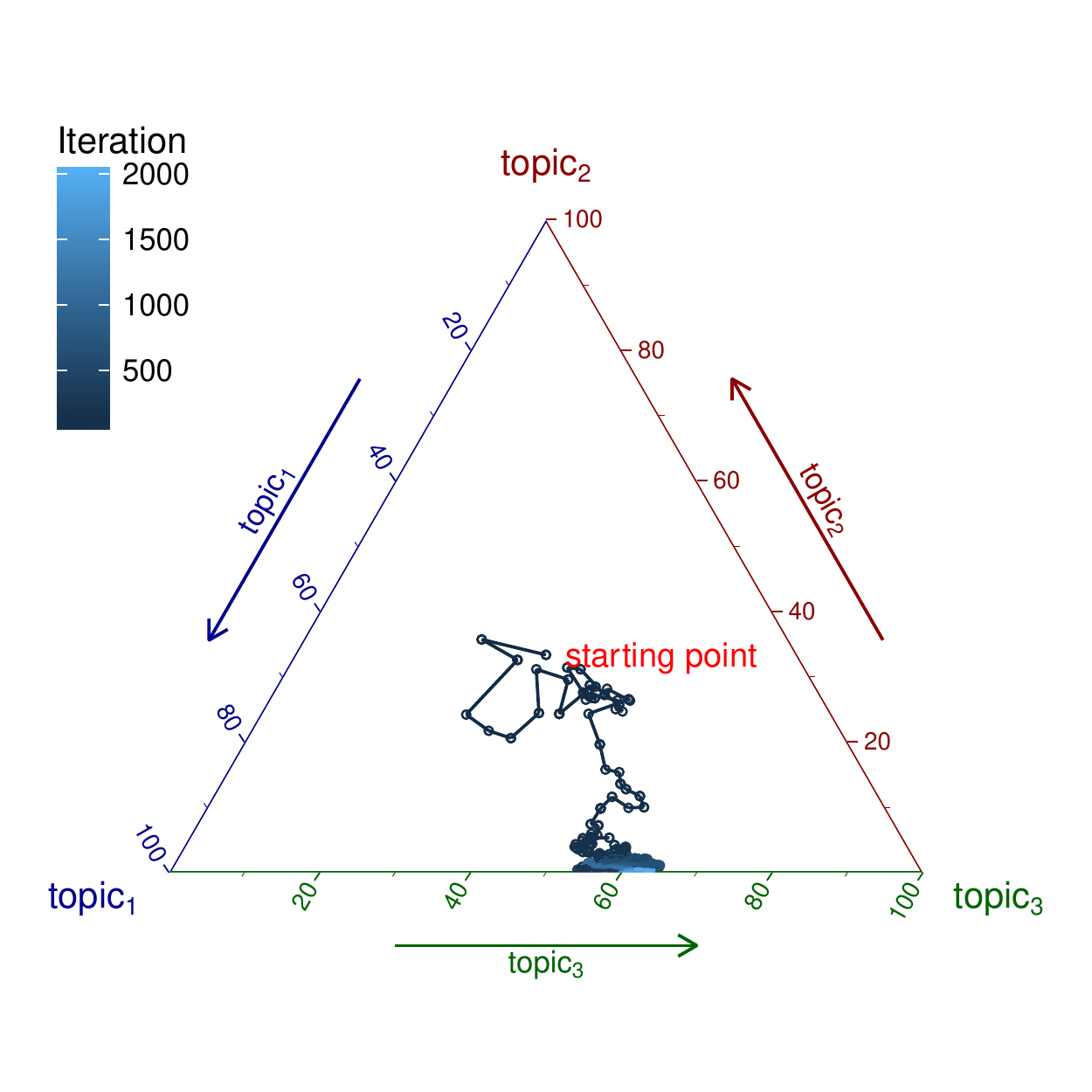}}\\
	\subfloat[VEM: $\bpi^{(45)}_{2} = (.258, .155, .585)$]{
		\includegraphics[trim={0 1.2cm 0 
			1.2cm},clip,width=.45\linewidth]
		{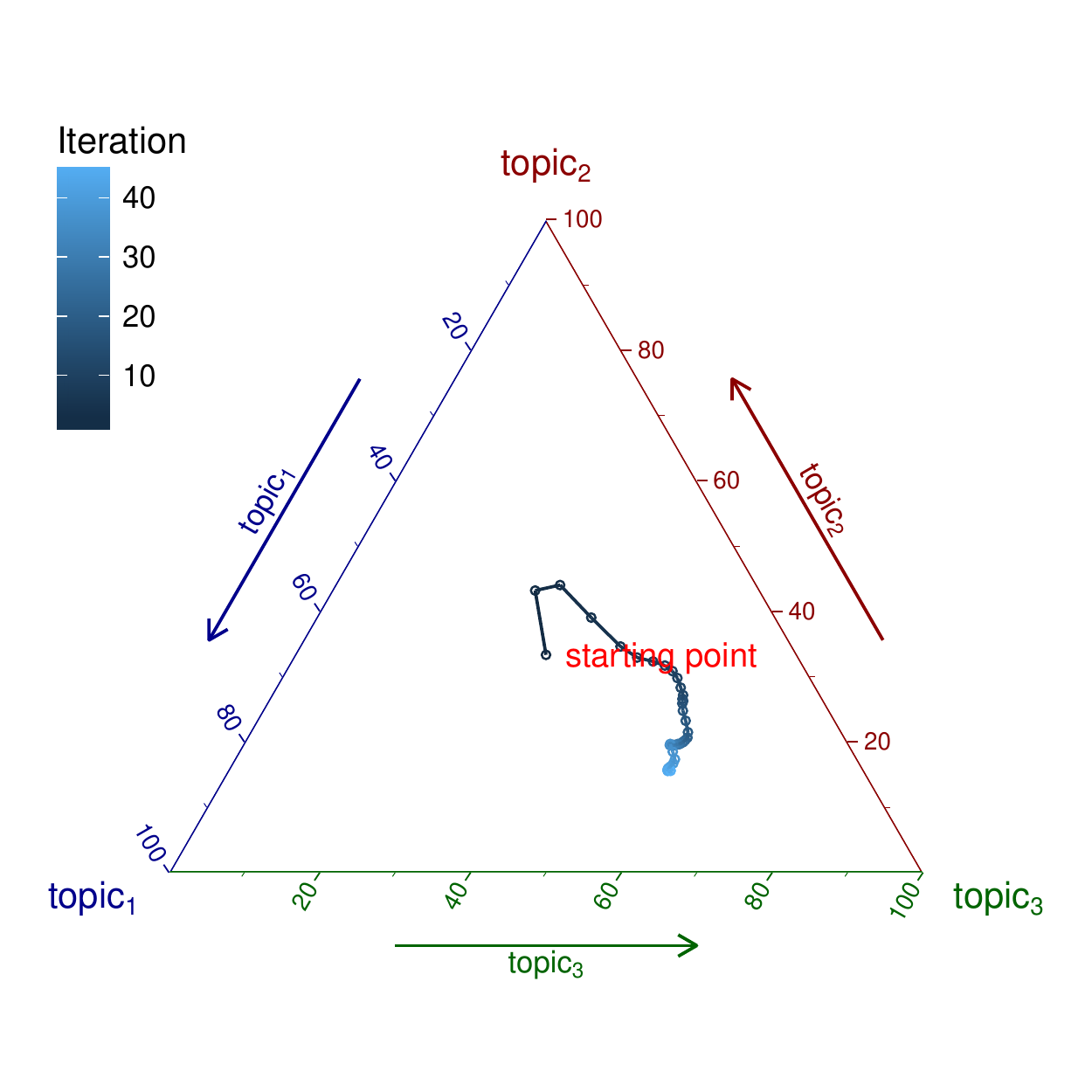}}
	\caption{Plots of values of $\bpi_{2}$ via algorithms AGS, MGS, 
	and VEM. With approximately $23$ iterations the AGS chain 
	reached the optimal region, i.e., $.07$ from the true value 
	$\bpi^{\text{true}}_{2} = (.584, .386, .030)$, but the MGS 
	chain took $139$ iterations to reach there. Algorithm VEM never 
	reached the optimal regions: at convergence, VEM hit points 
	that are $0.18$ far from $\bpi^{\text{true}}_{2}$.}
	\label{fig:pi-2-random-walk}
\end{figure*}

Table~\ref{tab:ags-mgs-vem} gives the values of $\bpi_{1}$ 
and $\bpi_{2}$ on the $2$-simplex from the last iteration of all 
three algorithms. We can see that both AGS and MGS chains 
were able to recover the values $\bpi^{\text{true}}_{1}$ and 
$\bpi^{\text{true}}_{2}$ reasonably well, although AGS chain 
has an edge.  
To converge to optimal regions $\| \bpi^{\text{true}}_1 
- \bpi^{(s)}_1 \| \leq .003$ and $\| \bpi^{\text{true}}_2 - 
\bpi^{(s)}_2 \| \leq .07$, AGS chain  took cycles $42$ and 
$23$ only; but, MGS chain required cycles $248$ and $139$, 
respectively. ($\|.\|$ denotes L-$1$ norm on the simplex 
$\simplex_K$.) 
Lastly, algorithm VEM never reached the optimal 
regions: at convergence, VEM hit points that are $0.08$ far 
from $\bpi^{\text{true}}_{1}$ and $0.18$ far from 
$\bpi^{\text{true}}_{2}$. 
Here, the reported values are after the 
necessary alignment of topics with the true set of topics, for all 
the three methods. 
\figurename~\ref{fig:pi-2-random-walk} gives  
trace plots of values of $\bpi_{2}$ on the 
$2$-simplex for all three algorithms. Note that due to the 
superior performance of AGS chain over 
MGS chain, we use it in our experimental analyses.
Additional details of this experiment is 
provided in Section~\ref{sec:algorithm-illustration}.

\begin{table}%
	\centering
	\begin{threeparttable}[b] 
		\caption{Estimated values of $\bpi$ via   	
			algorithms AGS, MGS, and VEM} 
		\label{tab:ags-mgs-vem}
		\begin{tabular}{c c c c}
			\toprule 
			Method & 
			$\hat{\bpi}_{1}$\tnote{$\dagger$} & %
			$\hat{\bpi}_{2}$\tnote{$\dagger$} & %
			Iterations \\  
			\midrule
			AGS & 
			$(\epsilon, .996, .003)$ & 
			$(.347, .015, .636)$ & 	
			$2{,}000$ \\ 
			MGS & 
			$(.001, .997, \epsilon)$ & 
			$(.379, .005, .615)$ & 	
			$2{,}000$ \\ 
			VEM & 
			$(.057, .935, .006)$ & 
			$(.258, .155, .585)$ & 	
			$45$ \\ 
			\bottomrule 
		\end{tabular}
		\begin{tablenotes} 
			\item [$\dagger$] $\bpi^{\text{true}}_{1} = (.002, 
			\epsilon, .997)$, $\bpi^{\text{true}}_{2} = (.584, 
			.386, .030)$
		\end{tablenotes} 
	\end{threeparttable} 
\end{table}

\subsection{Selecting Hyperparameters in cLDA} 
\label{sec:clda-estimate-eta-gamma}
The hyperparameters $h = (\alpha, \gamma, \eta)$ in cLDA can affect 
the prior and posterior distributions of parameters $\psi$ and 
should be selected carefully. A natural solution to select them is 
via maximum likelihood: for example, we let $m_{\bw}(h)$ denote the 
marginal likelihood of the data as a function of $h$, and use 
$\hat{h} = \argmax_h m_{\bw}(h)$. However, for models such as LDA 
and cLDA, the function $m_{\bw}(h)$ is analytically intractable 
(Section \ref{sec:inference}). Works in the literature  
\citep{BleiNgJordan:2003,Wallach:2006} suggest an approximate EM 
algorithm to estimate $\argmax_h m(h)$ for LDA, which can be 
described as follows. We consider $\bw$ ``observed data,'' and 
$\bpsi$ ``missing data.'' We have the ``complete data likelihood'' 
$p_h(\bpsi, \bw)$ available, then the EM algorithm 
\citep{DempsterLairdRubin:1977} is a natural candidate to estimate 
$\argmax_h m(h)$, since $m(h)$ is the ``incomplete data 
likelihood.'' However, the E-step in EM involves calculating an 
expectation with respect to the intractable posterior $p_{h,\bw}$ 
of the model. One solution is to approximate this expectation via 
variational inference (e.g. Variational-EM 
\citep{BleiNgJordan:2003}) or Markov chain Monte Carlo (e.g. 
Gibbs-EM \citep{Wallach:2006}). We follow the latter approach for 
cLDA, which includes: 
\begin{enumerate}
	\item Initialize $h_0$ and $\bpsi_0$
	\item Until convergence  
	\begin{enumerate}
		\item[\textbf{E-step}] Sample $\bpsi_t^{(1)}, 
		\ldots, \bpsi_t^{(S)}$ from $p_{h_t, \bw}(\bpsi)$ via 
		chain AGS or MGS 
		\item[\textbf{M-step}] $h_{(t+1)} = \argmax_h \sum_{s = 
		1}^S \log p_{h_t}(\bpsi_t^{(s)}, \bw)$ \label{e-step}
	\end{enumerate}
\end{enumerate}    

\begin{figure}[htbp!]
	\centering
	\includegraphics[width=.48\textwidth]
	{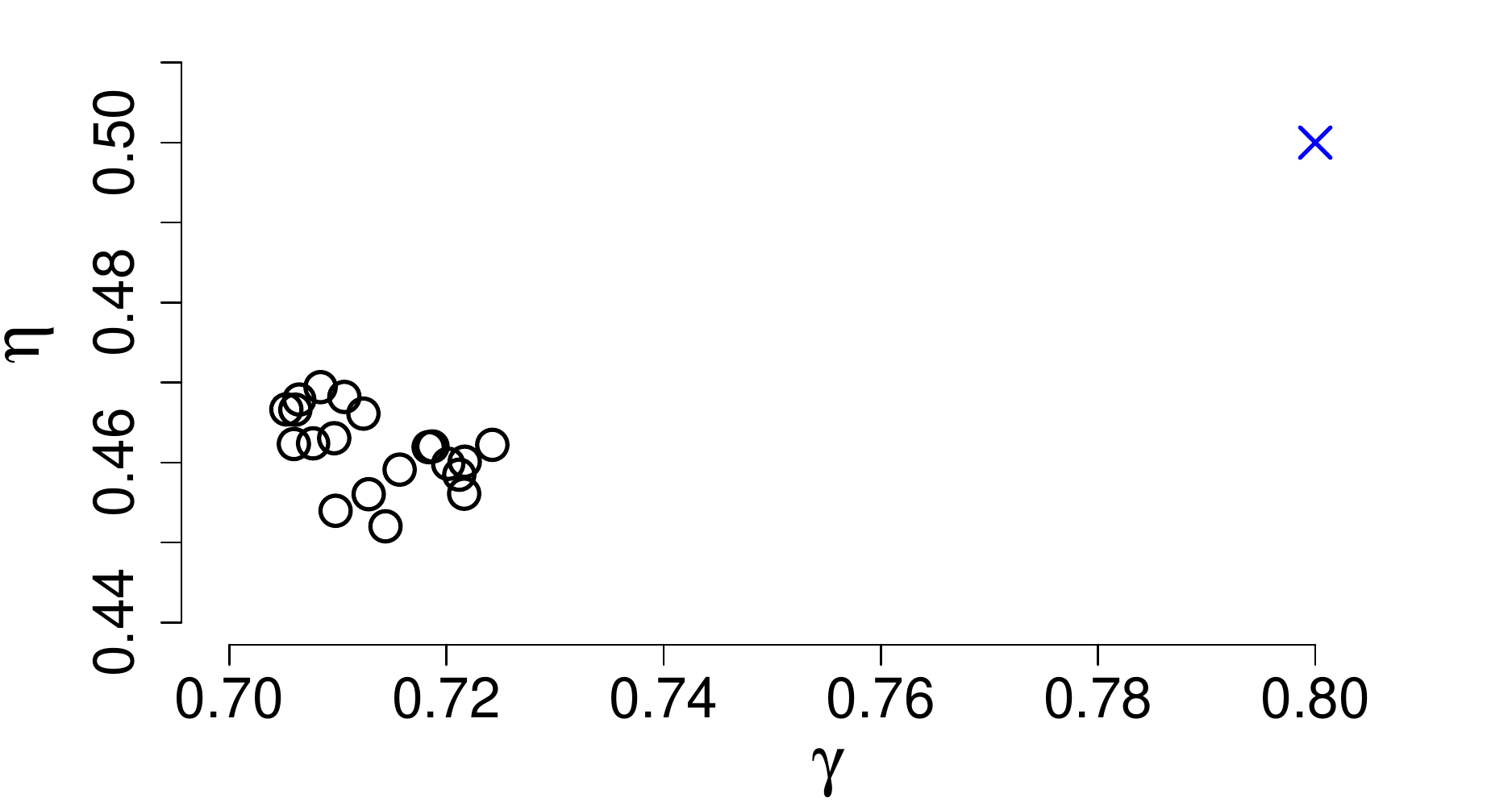}
	\caption{Independent Gibbs-EM estimates ($\circ$) of 
	hyperparameters $(\eta, \gamma)$ in cLDA with constant 
	hyperparmeter $\alpha = 1$, for a synthetic corpora with 
	hyperparameters $\alpha_\text{true} = 1, 
	\gamma_\text{true} = .8, \eta_\text{true} 
	= .5$ (\textcolor{blue}{$\times$}). }
	\label{fig:clda-gem-eta-gamma-estimates-synth}
\end{figure}

One solution to M-step is via the fixed point 
iteration \citep{minka2000estimating}. We derive the expressions for 
maximizing hyperparameter  $\gamma$ and $\eta$ as follows (see 
Section \ref{sec:est-eta-gamma}): 
\begin{equation}
\eta_{(t+1)} = \frac{\eta_t}{V} 
\frac{\sum_{s, k, v} \Big[ \Psi(m^{(s)}_{kv} + \eta_t) - 
\Psi(\eta_t)\Big]}{\sum_{s, k} \Big[ \Psi(m^{(s)}_k + V\eta_t) - 
\Psi(V\eta_t) 
\Big]}
\label{eq:gem-est-eta} 
\end{equation} 
\begin{equation}
\gamma_{(t+1)} = \gamma_t \frac{\sum_{s, j, d, k} 
\pi^{(s)}_{jk} \Big[ 
\Psi (n^{(s)}_{jdk} + \gamma_t \pi^{(s)}_{jk}) - \Psi (\gamma_t 
\pi^{(s)}_{jk}) \Big]}
{\sum_{s, j, d}  \Big[ \Psi (n^{(s)}_{jd} + \gamma_t) - \Psi 
(\gamma_t) \Big]}
\label{eq:gem-est-gamma} 
\end{equation}
where $\sum_{x, y, \ldots}$ represents $\sum_{x = 1}^X \sum_{y = 
1}^Y \ldots$. To illustrate the performance of the proposed method, 
\figurename~\ref{fig:clda-gem-eta-gamma-estimates-synth} shows 
$20$ independent Gibbs-EM estimates of 
cLDA hyperparameters $(\eta, \gamma)$ with default $\alpha = 1$. 
We used a synthetic corpus with number of collections $J = 2$, 
number of topics $K = 3$, vocabulary size $V = 
40$, collection size = $100$, and document 
size = $200$ and the ``true'' hyperparameters 
$\alpha_\text{true} = 1, \gamma_\text{true} = .8, \eta_\text{true} 
= .5$. We initialized the 
algorithm with $(\eta_0, \gamma_0) = (1, 1)$. We notice that 
Gibb-EM recovered the true hyperparameters $\eta$ and $\gamma$ with 
a margin of error in all cases. 

One approach to deal with hyperparameter $\alpha$ is to 
put a Gamma prior on $\alpha$ and include a posterior sampling 
scheme for $\alpha$. But, our sensitivity analyses on 
$\alpha$ shows no significant impact for $\alpha$ on the predictive 
power of cLDA, once we select $\eta$ and $\gamma$. We 
thus use a default value for $\alpha$ in our experiments.

%% file: experiments.tex
\section{Experimental Analysis}
\label{sec:experiments}

This section compares the performance of the proposed cLDA model 
with alternatives, LDA and HDP, on real-world corpora (Section 
\ref{sec:datasets}). We explore two performance metrics: (a) the 
quality of the inferred model, which we measure by the predictive 
power of the model (e.g. via perplexity) and external topic 
evaluation scores (e.g. topic coherence) 
(Section~\ref{sec:eval-performance}), and (b) the applicability of 
cLDA model to real-world corpora 
(Section~\ref{sec:usability-study}). We also briefly discuss 
guidelines for selecting the number of topics $K$ in cLDA 
(Section~\ref{sec:eval-performance}).

\subsection{Datasets}
\label{sec:datasets}

We use three document corpora based on: (i) the NIPS $00$-$18$ 
dataset~\citep{Chechik:2007}, (ii) customer reviews from 
Yelp\footnote{\url{http://uilab.kaist.ac.kr/research/WSDM11/}}, and 
(iii) the $20$Newsgroups  
dataset\footnote{\url{http://qwone.com/~jason/20Newsgroups}}. 
Standard corpus preprocessing involve tokenizing text and 
discarding standard stop-words, numbers, dates, and infrequent 
words from the corpus. 
The NIPS $00$-$18$ dataset is a popular dataset used in the topic 
modeling community. It consists of papers published in proceedings 
$0$ to $18$ of the  Neural Information Processing Systems (NIPS) 
conference (i.e. years from $1988$ to $2005$). Along with standard 
preprocessing, we discarded words with length less than two or with 
length two and not in (``ml", ``ai", ``kl", ``bp", ``em", ``ir", 
``eb") from the corpus. Finally, this corpus consists of $2{,}741$ 
articles and $9{,}156$ unique words. A question of interest is to 
see how various research topics evolve over the years $1988$-$2005$ 
of NIPS proceedings. Typically, new topics do not emerge in 
consecutive years.  So, we partition the NIPS $00$-$18$ corpus into 
four collections based on time periods $1988$-$1992$, 
$1993$-$1997$, $1998$-$2002$, and $2003$-$2005$ in our analyses. 

Yelp is a crowd-sourced platform for user reviews and 
recommendations of restaurants, shopping, nightlife, entertainment, 
etc. In this paper, we used a collection of restaurant reviews from 
Yelp. Each text review (i.e. a document) in the collection is 
associated with a customer rating on the scale $1$ to $5$, where 
$5$ being the best and $1$ being the worst.  We discarded reviews 
with less than $50$ words. After standard preprocessing, this 
corpus consists of $24{,}310$ restaurant reviews and $9{,}517$ 
unique words. 

The $20$Newsgroups dataset consists of approximately $20{,}000$ 
news articles that are distributed across $20$ different  
newsgroups. 
According to the subject matter of articles, these $20$ newsgroups 
are further partitioned into six groups. We took a subset of this 
dataset with the subject groups computers, recreation, science, and 
politics. After standard preprocessing, this corpus consists of 
$10{,}764$ articles and $9{,}208$ unique words. We consider the 
subject group of a document as its collection label to fit  
cLDA models. We denote this corpus by $16$Newsgroups.

\subsection{Model Evaluation and Model Selection}
\label{sec:eval-performance}

\subsubsection{Comparisons with Models in the Prior Work} 
\cite[Section 3]{WallachMimnoMcCallum:2009} proposed 
\emph{non-uniform} base measures instead of the typical 
\emph{uniform} base measures used in the Dirichlet priors over 
document-level topic distributions $\btheta_d$'s in LDA. The 
purpose was to get a better asymmetric Dirichlet prior for LDA, 
which is generally used for corpora with a flat organization 
hierarchy for documents. We can consider one such model as a 
particular case of the cLDA model proposed here: their model is 
closely related to a cLDA model with a single collection. To 
illustrate, we perform a comparative study of perplexity scores of 
cLDA with single and multi-collection in our experiments. 

Very briefly, HDP can be described as follows. Suppose we have $D$ 
populations, and that for population $d, \, d = 1, \ldots, D$, 
there are observations $w_{di} \ind F_{z_{di}, \sigma_{di}}, \, 
i = 1, \ldots, n_d$.  Here, $F_{z_{di}, \sigma_{di}}$ is a 
distribution depending on some latent variable 
$z_{di}$ and possibly also on some other known parameter 
$\sigma_{di}$ particular to individual $di$. We assume that 
$z_{di} \iid G_d, \, i = 1, \ldots, n_d$, and that for $d = 1, 
\ldots, D$, $G_d \iid {\cal D}_{G_0,\alpha}$, the Dirichlet process 
(DP) with base probability measure $G_0$ and precision parameter 
$\alpha > 0$ \citep{Ferguson:1973, Ferguson:1974}.  
In applications such as topic modeling, it is desirable to 
model the distributions of the $w_{di}$'s as mixtures, and to 
have mixture components shared among the distributions of the 
$w_{di}$'s in different populations. We can obtain this by taking  
$G_0$ itself to have a Dirichlet process prior, $G_0 \sim 
{\cal D}_{{\cal K},\gamma}$, where ${\cal K}$ is a probability 
distribution and $\gamma > 0$. 
In the topic modeling setting, for word $w_{di}$ of document $d$,  
we imagine that there exists a \emph{topic} $z_{di}$---a 
distribution on ${\cal V}$, from which the word is drawn. 
Typically, the distribution ${\cal K}$ is a member of a known 
parametric family such as Dirichlet with parameter $\omega$. The 
hierarchical model described here is a 
two-level DP~\citep{TehEtal:2006}, which is (conceptually) similar to a one-collection 
cLDA model. We thus compare the results of one-collection cLDA and 
LDA with the results of two-level DP\footnote{Code by Wang and Blei 
(2010): \url{https://github.com/blei-lab/hdp}}. We use the popular Chinese Restaurant Franchise~\citep[CRF]{TehEtal:2006} sampling scheme for inference.

\subsubsection{Criteria for Evaluation} 
We use the perplexity 
computation scheme that is specified as in \cite{Patterson:2013, 
WallachMurrayMimmo:2009}. We first divide 
documents in the corpus into  a training set 
and a held-out set. Second, we partition words 
in every document $\bw_{jd}$ in the held-out 
set to two sets of words, 
$\bw_{jd}^{\text{train}}$ and  
$\bw_{jd}^{\text{test}}$. We also use 
$\bw_{jd}^{\text{train}}$ to denote words in a 
training document $jd$. We 
define the vector $\bw^{\text{train}}$ for the 
training corpus combining 
$\bw_{jd}^{\text{train}}$, $j = 1, \ldots, J$, 
$d = 1, \ldots, D_j$. Similarly, we define the 
vector $\bw^{\text{test}}$. We compute the 
per-word perplexity for the held-out words 
$\bw^{\text{test}}$ (uses the fact that 
$z_{jdi}, \, i = 1, \ldots, n_{jd}$ are 
conditionally independent given 
$\btheta_{jd}$) as  
\begin{equation}
{\cal S}(\bw^{\text{test}} 
\given 
\bw^{\text{train}}) = \exp \frac{-\sum_{ 
w_{jdi} 
		\in  \bw^{\text{test}} } \log p (w_{jdi} \given 
		\bw^{\text{train}}
	)}{| \bw^{\text{test}} |}   
\label{eq:c-lda-perplexity}
\end{equation} 
where $| \bw^{\text{test}} |$ is the length of vector 
$\bw^{\text{test}}$. Exact computation of this score is 
intractable. However, we can estimate this score via MCMC or 
variational methods for both cLDA and LDA models (see 
Section~\ref{app:perplexity}).

\cite{MimnoEtAl:2011} suggested some generic 
evaluation scores based on human coherence judgments of estimated 
topics via topic models such as LDA. One such score is the  
\textsl{topic size}, i.e., the number of words assigned to a topic 
in the corpus. One can estimate it by samples from the posterior of 
the topic latent variable $\bz$, given observed words $\bw$ (e.g. 
via Markov chains: collapsed Gibbs sampling (CGS) for LDA 
\citep{GriffithsSteyvers:2004}, AGS, and  CRF). 

Another option is the topic \textsl{coherence score} 
\citep{MimnoEtAl:2011}, which is computed based on the most probable 
words in an estimated topic for a corpus. We find the $m$ most 
probable words $v_1^{(k)}, v_2^{(k)}, \ldots, v_m^{(k)}$ for topic 
$k$ by sorting the vocabulary words in topic $k$ in the descending 
order of topic specific probabilities (i.e. $\beta_{kt}, t = 1, 
\ldots, V$, estimated via Markov chains such as CGS, AGS, and CRF) 
and picking the top $m$ words. Let $\text{df}(v_t)$ be the document 
frequency of term $v_t$, i.e., the number of documents in the 
corpus which have the term $v_t$. Let $\text{df}(v_i, v_j)$ be the 
co-document frequency of the terms $v_i$ and $v_j$, i.e., the 
number of documents in the corpus which have both of the terms 
$v_i$ and $v_j$. We then define \textsl{coherence score}
\citep{MimnoEtAl:2011}, for topic $k = 1, 2, \ldots, K$, as 
\begin{equation}
\text{topic-coherence}_k = \sum_{i=2}^{m} \sum_{j=1}^{i} \log 
\frac{\text{df}(v^{(k)}_i, v^{(k)}_j) + 
	1}{\text{df}(v^{(k)}_j)} 
\label{eq:mimno-topic-coherence}
\end{equation}       
The intuition behind this score is that group of words 
belonging to a topic possibly co-occur with in a 
document.

\subsubsection{Comparing Performance of cLDA and LDA Models}
\begin{figure}[htbp!]
	\centering
	\subfloat[$16$Newsgroups]
	{\includegraphics[width=.66\textwidth]
		{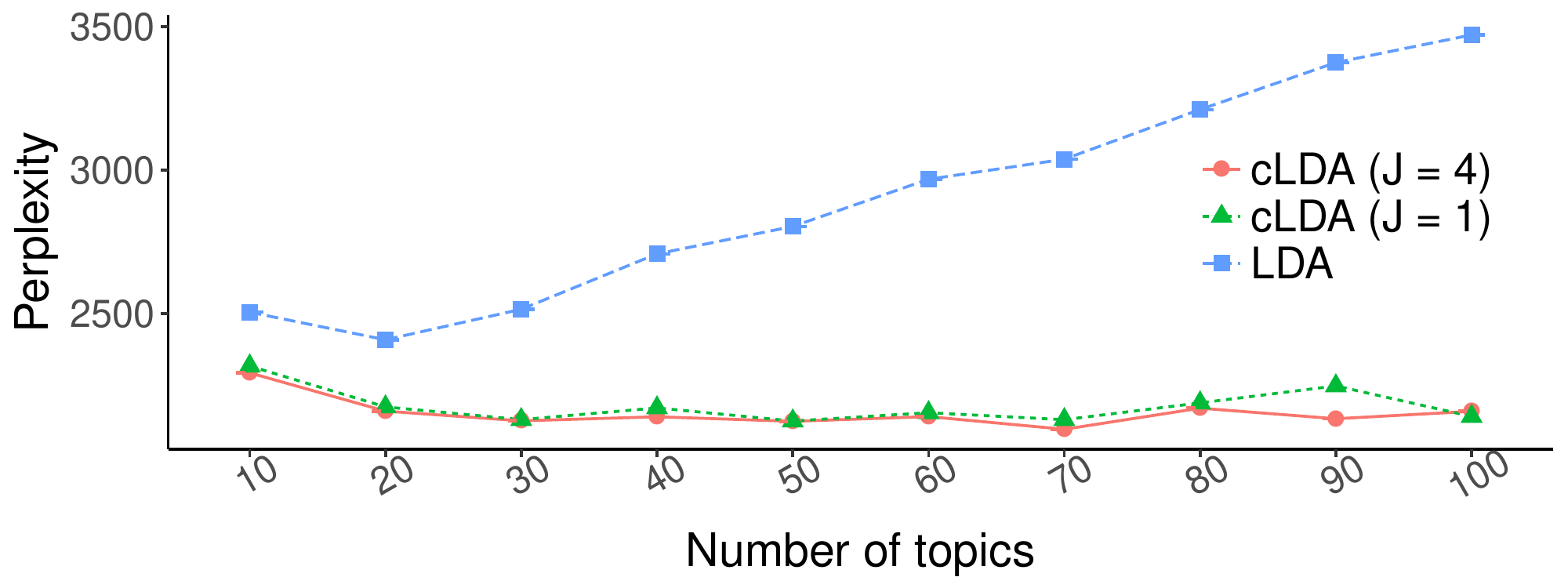}} \\ 
	\subfloat[Yelp]
	{\includegraphics[width=.65\textwidth]
		{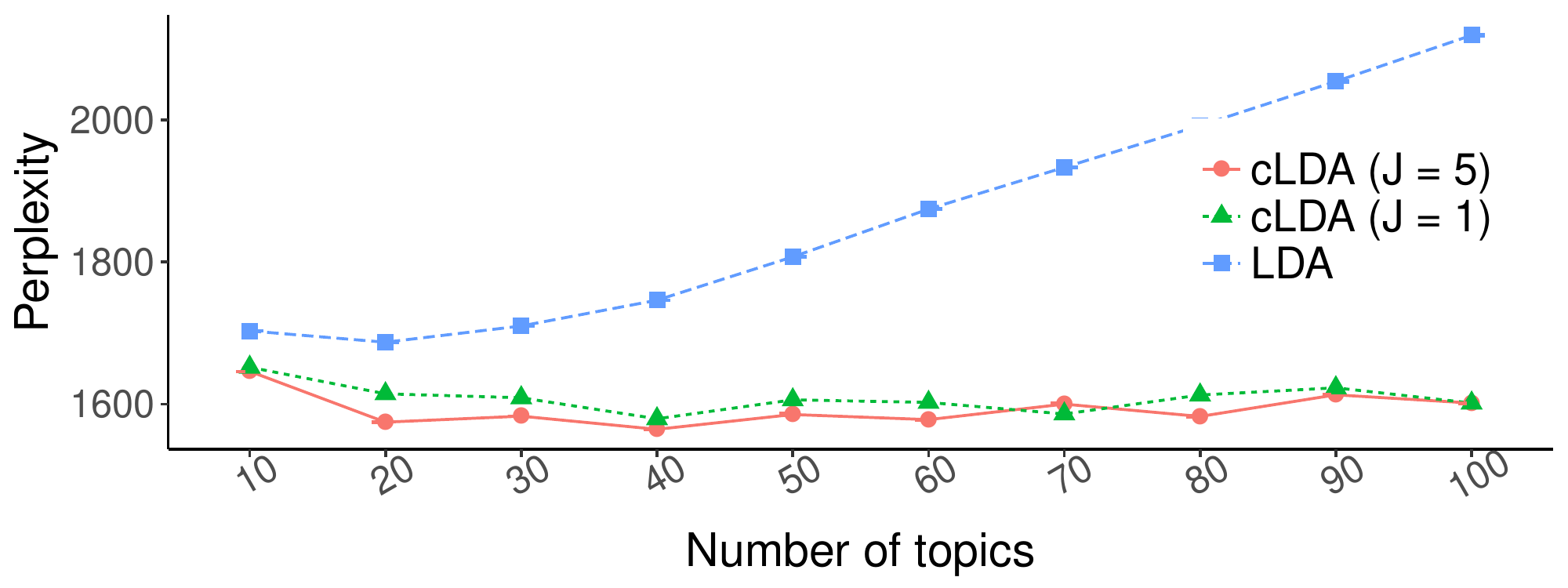}}\\
	\subfloat[NIPS $00$-$18$]
	{\includegraphics[width=.65\textwidth]
		{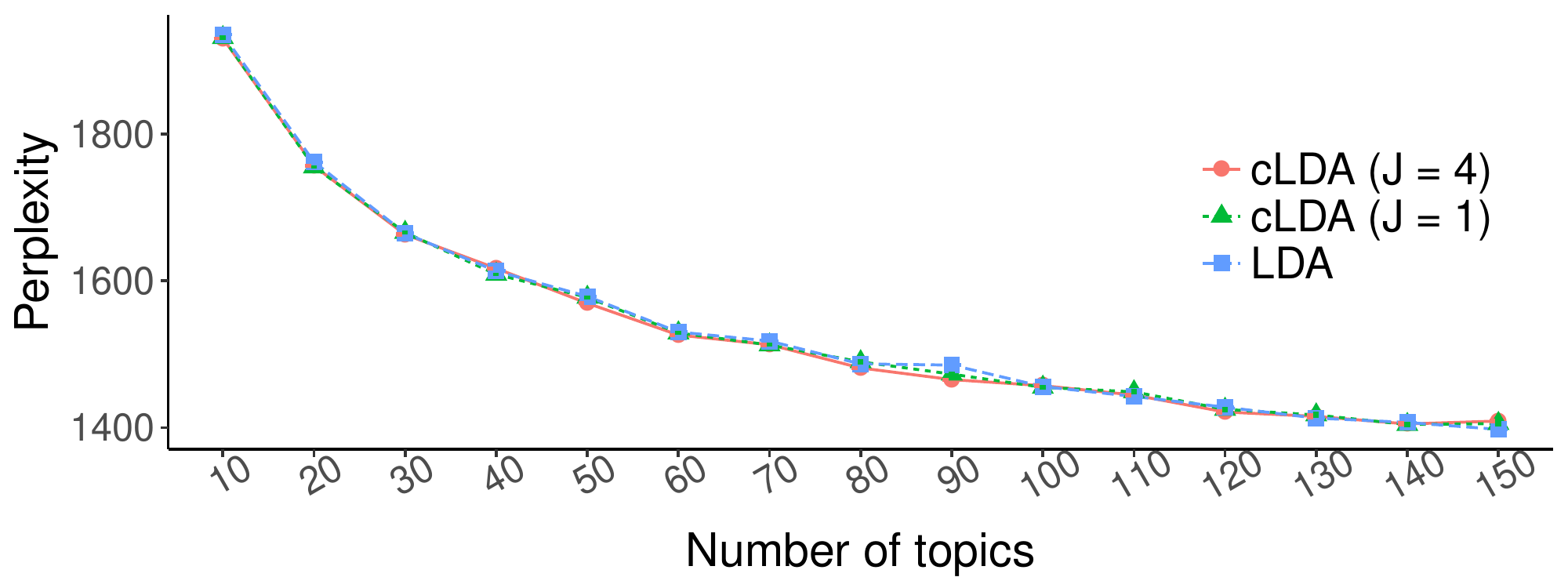}} 
	\caption{Estimated (average) per-word perplexities for models 
		LDA and cLDA with single collection ($J = 1$),  
		multi-collection  ($J > 1$) and various 
		configurations of the number of topics $K$ using corpora 
		$16$Newsgroups, Yelp, and NIPS $00$-$18$. }
	\label{fig:lda-clda-K-search-perplexity-lineplots}
\end{figure}

We first look at the criterion perplexity scores 
for both LDA and cLDA models with various values of the number of 
topics keeping each model hyperparameter fixed (and default).  
We ran the Markov chains AGS and CGS for $1000$ iterations. 
\figurename~\ref{fig:lda-clda-K-search-perplexity-lineplots} 
gives (average) per-word perplexities for the held-out (test) words using 
algorithms CGS and AGS (with single collection, i.e., $J = 1$ and  
multi-collection, $J > 1$) for corpora NIPS $00$-$18$,  
$16$Newsgroups, and yelp. From the plots, 
we see that cLDA outperforms LDA well in terms of 
predictive power, except for corpus NIPS $00$-$18$, for which 
cLDA has a slight edge over LDA. We believe the marginal performance 
for corpus NIPS $00$-$18$ is partly due to the nature of 
partitions defined in the corpus; collections 
in this corpus share many common topics, compared to corpus 
$16$Newsgroups. It also suggests that the gain in using 
cLDA is larger when we model corpora with separable collections. 

In terms of perplexity, we see a small improvement for cLDA models 
with multi-collection over single collection. But, the advantage of 
cLDA over LDA is clear, even with a single collection. Also, note 
that cLDA has a better selection of priors as well as the ability 
to incorporate document hierarchy in a corpus into the modeling 
framework.  

\textsl{Selecting $K$ and Hyperparameters $\alpha, \gamma, \eta$:} 
\figurename~\ref{fig:lda-clda-K-search-perplexity-lineplots} also 
gives some insights on the number of topics $K$ should be used for 
each corpus. Perplexities of cLDA models go down quickly with 
increase in the number of topics in the beginning of the curves, 
and then the rate of decrease go steady after reasonable number of 
topics (e.g. $K = 90$ corpus NIPS $00$-$18$). LDA, on the other 
hand, has a ``U'' curve with increase in perplexity after $K = 30$ 
or $K = 40$, for corpora $16$Newsgroups and Yelp. However, for 
corpus NIPS $00$-$18$, LDA shows a behavior similar to cLDA. We 
thus use $K = 90$ for corpus NIPS $00$-$18$, $K = 30$ for corpus 
$16$Newsgroups, and $K = 40$ for corpus Yelp, in our comparative 
analyses in this section and case studies. 

To estimate hyperparameters $\eta$ and $\gamma$ in cLDA, we use  
the Gibbs-EM algorithm (Section \ref{sec:clda-estimate-eta-gamma}) 
for all three corpora with the selected $K$ and constant 
$\alpha = 1$. Estimated $(\hat{\eta}, \hat{\gamma})$'s for cLDA 
models at convergence for corpora $16$Newsgroups, NIPS $00$-$18$, 
and Yelp are $(.05, 2.07)$, $(.026, 6.72)$, and $(.034, 3.44)$, 
respectively. Similarly, one can estimate hyperparameters $\alpha$ 
and $\eta$ in LDA based on a similar Gibbs-EM algorithm (see 
\eqref{eq:gem-est-eta} and \cite{Wallach:2006}). Estimated 
$(\hat{\eta}, \hat{\alpha})$'s for LDA models at convergence for 
corpora $16$Newsgroups, NIPS $00$-$18$, and Yelp are $(.053, 
.048)$, $(.027, .06)$, and $(.029, .12)$, respectively. Note that 
estimates $\hat{\eta}$'s for LDA and cLDA are quite similar for 
all three corpora.

\begin{figure}[htbp!]
	\centering
	\subfloat[Topic size]
	{\includegraphics[width=.65\textwidth]
		{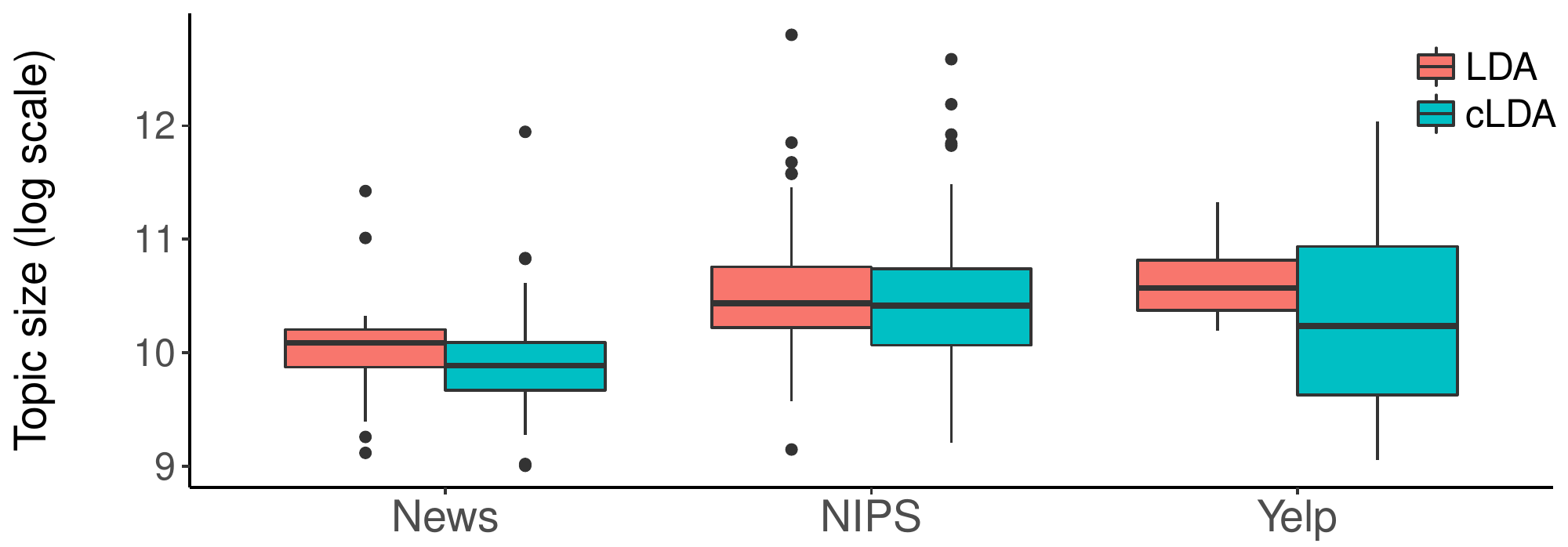}} \\ 
	\subfloat[Topic coherence]
	{\includegraphics[width=.65\textwidth]
		{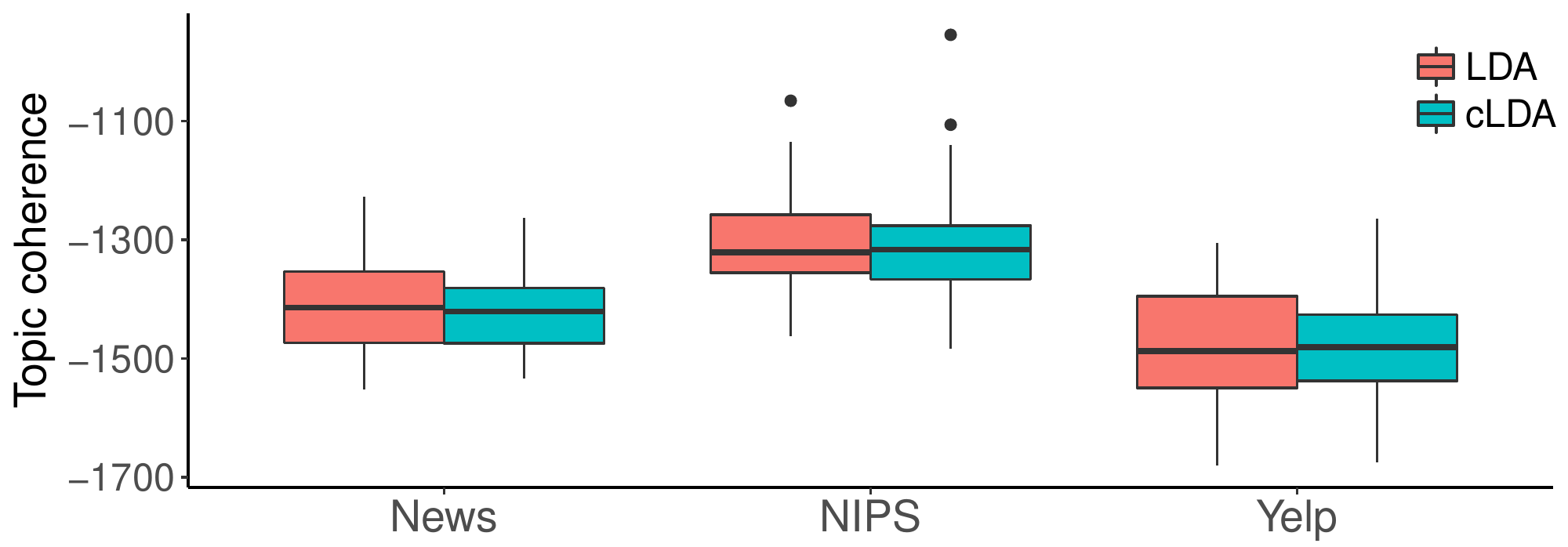}}
	\caption{Estimated topic sizes and coherences for models 
		LDA and cLDA  for corpus $16$Newgroups, NIPS $00$-$18$, and Yelp. }
	\label{fig:lda-clda-topic-coherences-sizes}
\end{figure}

We now evaluate distributions of topics learned by cLDA and LDA 
models for corpora $16$Newsgroups, NIPS $00$-$18$, and Yelp with 
chosen $K$ and hyperparameters $h$, quantitatively. 
\figurename~\ref{fig:lda-clda-topic-coherences-sizes} shows 
boxplots of estimated topic sizes and coherences for models LDA and 
cLDA for all three corpora. Note that LDA models have uniform topic 
sizes compared to cLDA models. Coherence scores of cLDA topics are 
better than LDA topics except for corpus $16$Newsgroups, which we 
think, is partially due to having relatively easily separable 
topics compared to the other two complex corpora.       

\textsl{Execution time:} In our experience, both AGS (with a 
reasonable number of collections) and CGS chains have comparable 
computational cost for all three corpora (details in 
Table~\ref{tab:computation-costs-extended}). 
On average, the combined 
execution-time of multiple CGS runs for a corpus that is 
partitioned on collection labels was fairly equivalent to the 
execution-time of a single CGS run on the whole corpus. We 
implemented all these algorithms single-threaded on the  
\texttt{R}-\texttt{C++} (using the efficient \texttt{Rcpp} and 
\texttt{RcppArmadillo} libraries) programming environment. 

\begin{table*}%
	\centering
	\begin{threeparttable}[b] 
		\caption{Average per iteration execution time (in seconds) for  
			various topic modeling algorithms. } 
		\label{tab:computation-costs-extended}
		\begin{tabular}{c c r r r r}
			\toprule 
			Corpus & Number of Topics & CGS & AGS & 
			MGS\tnote{a} & VEM\tnote{b} \\  
			\midrule
			$16$Newsgroups & $64$ & $0.6101$ & $0.6879$ & $1.8616$ & 
			$52.3518$\\ 
			NIPS $00$-$18$ & $90$ & $3.9568$ & $4.2588$ & $12.5580$ & 
			$320.9931$ 
			\\ 	
			Yelp & $60$ & $1.2997$ & $1.4743$ & $3.9524$ & $164.0285$ \\
			\bottomrule 
		\end{tabular}
		\begin{tablenotes} 
			\item [a] MGS chains took approximately twice or more the 
			execution time of CGS chains for all three corpora.
			\item [b] We can see that per iteration running cost for VEM is 
			relatively high for these corpora. We believe that this due to 
			the single-threaded vanilla implementation for VEM; it may be 
			improved by an efficient parallel implementation, which we 
			leave to future research. 	
		\end{tablenotes} 
	\end{threeparttable} 
\end{table*}

\subsubsection{Comparing Performance of cLDA with LDA and HDP}
\begin{figure}[htbp!]
	\centering
	\includegraphics[width=.65\textwidth]
	{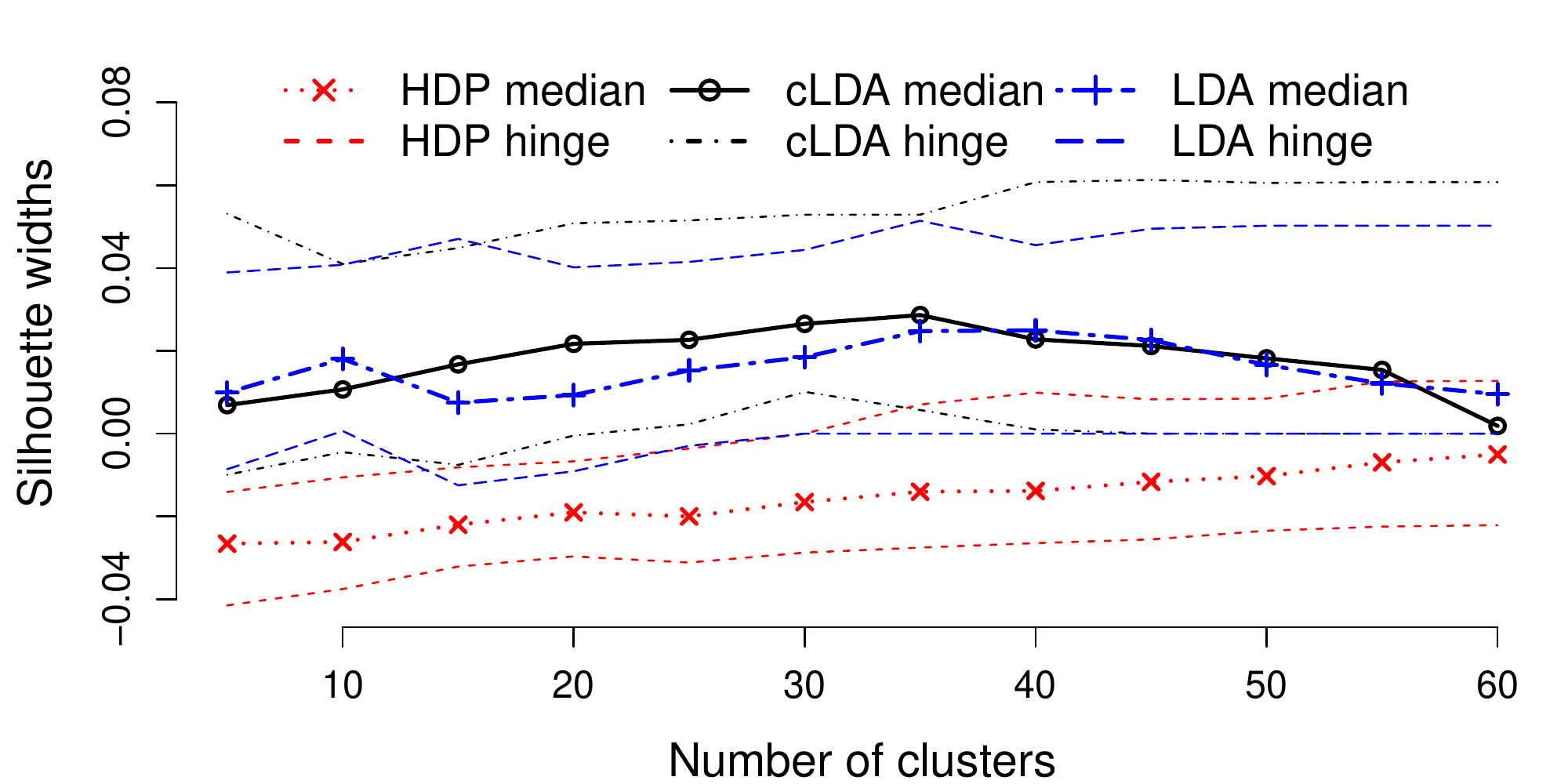}
	\caption{Boxplot statistics (median, lower hinge, and upper 
		hinge) of silhouette widths computed on \emph{topics}' 
		\texttt{hclust} clusters (i.e. hierarchical clustering of 
		topic distributions) with various values of the number 
		of clusters,  for corpus NIPS $00$-$18$ 
	}
	\label{fig:clda-lda-hdp-topic-hclust-silhouette}
\end{figure}

\begin{figure}[htbp!]
	\centering
	\includegraphics[width=.65\textwidth]
		{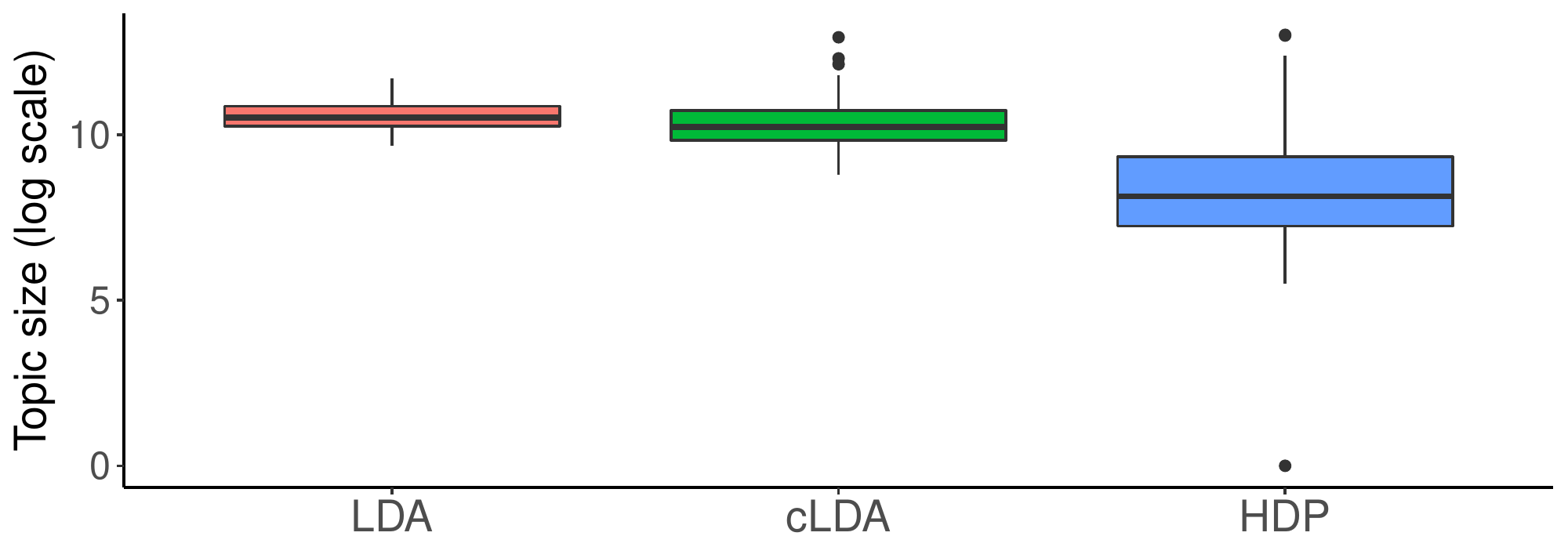}
	\caption{Estimated topic sizes for models 
		LDA, cLDA, and HDP  for corpus NIPS $00$-$18$. }
	\label{fig:lda-clda-hdp-topic-sizes}
\end{figure}

We now compare the performance of one-collection cLDA and LDA with 
two-level DP  using corpus NIPS $00$-$18$. We ran all the three 
chains AGS, CGS, and CRF for $1{,}000$ iterations and used the last 
sample from each chain for our analysis. We used fixed $K = 90$  
for both algorithms AGS and CGS, but algorithm CRF inferred 
$204$ topics from the corpus. We first evaluate the quality of 
topics learned for each method, by applying the \texttt{hclust} 
algorithm on topic distributions. \texttt{hclust} first computes a 
topic-to-topic similarity matrix based on the \emph{manhattan} 
distance. Then, it builds a hierarchical tree in a bottom-up 
approach. 
We noticed that the hierarchies formed by cLDA and LDA are 
comparable. By looking closely at topic word distributions, we 
notice that cLDA produces better topics, exploiting the asymmetric 
hierarchical Dirichlet prior on document $\btheta_d$'s that which 
is absent in LDA. (details, Section 
\ref{app:nips-results}). HDP, on the other hand, found too many 
redundant topics, as shown in 
\figurename~\ref{fig:lda-clda-hdp-topic-sizes}. To evaluate 
clusters induced by \texttt{hclust} quantitatively, we look at 
silhouette widths computed on topics' \texttt{hclust} clusters. We 
favor methods with high silhouette widths. 
\figurename~\ref{fig:clda-lda-hdp-topic-hclust-silhouette} gives 
boxplot statistics (i.e. median, lower hinge, and upper hinge) of 
silhouette widths computed on topics' \texttt{hclust} clusters with 
various values of the number of clusters (a user specified value in 
\texttt{hclust}). Overall, cLDA topics outperform HDP topics, and 
cLDA topics are comparable or better than LDA topics. Our analysis 
on clustering on learned document topic proportions, i.e., 
$\btheta_d$s, also show similar results 
(\figurename~\ref{fig:clda-lda-hdp-doc-hclust-silhouette})

\subsection{Usability Study}
\label{sec:usability-study}

\begin{figure*}[t!] 
	\centering
	\subfloat[cLDA: t$8$ (ANN)\label{fig:nips-clda-lda-topics:a}]{
		\includegraphics[width=.3\linewidth]
		{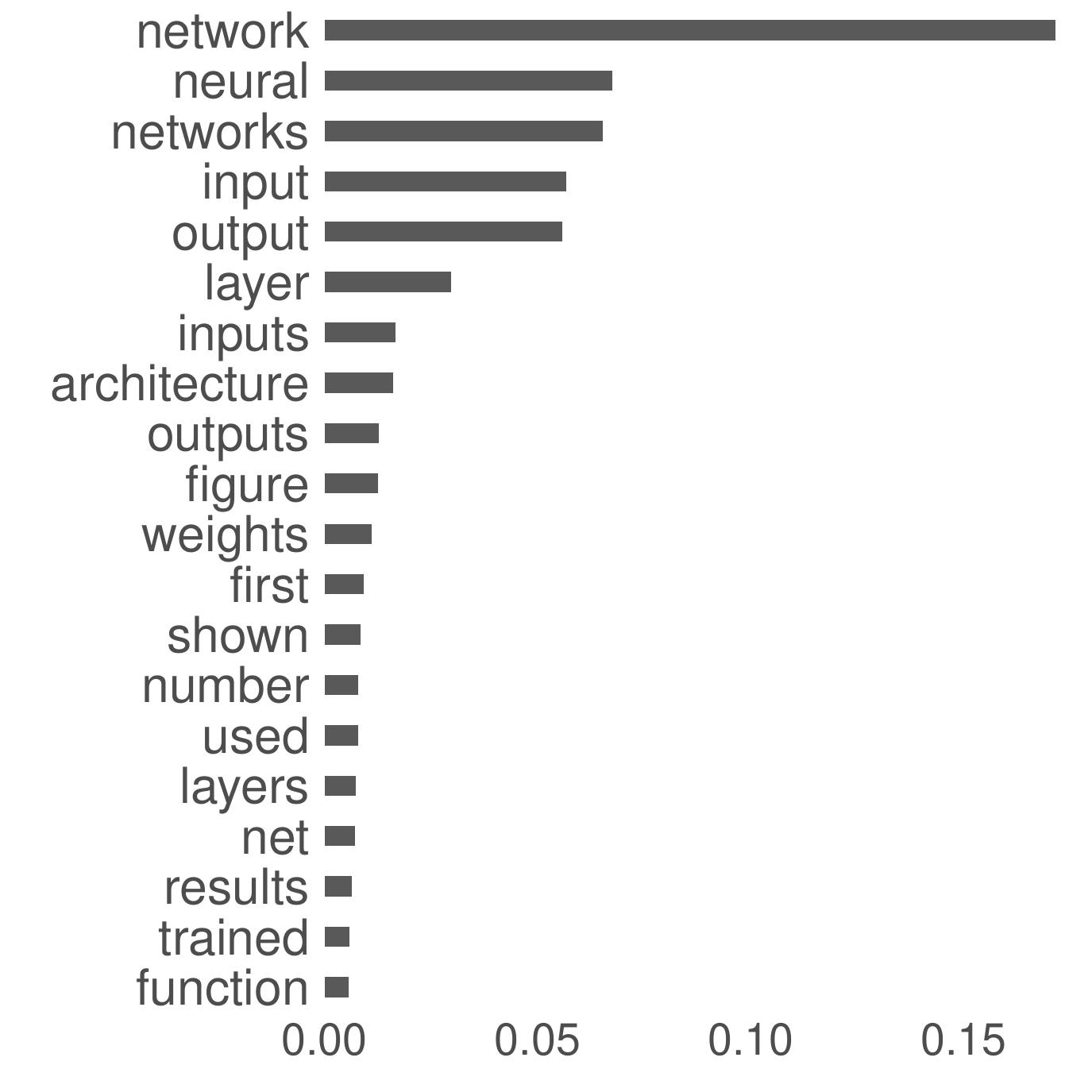}}\hfill%
	\subfloat[cLDA: t$57$ (MCMC)\label{fig:nips-clda-lda-topics:b}]{
		\includegraphics[width=.3\linewidth]
		{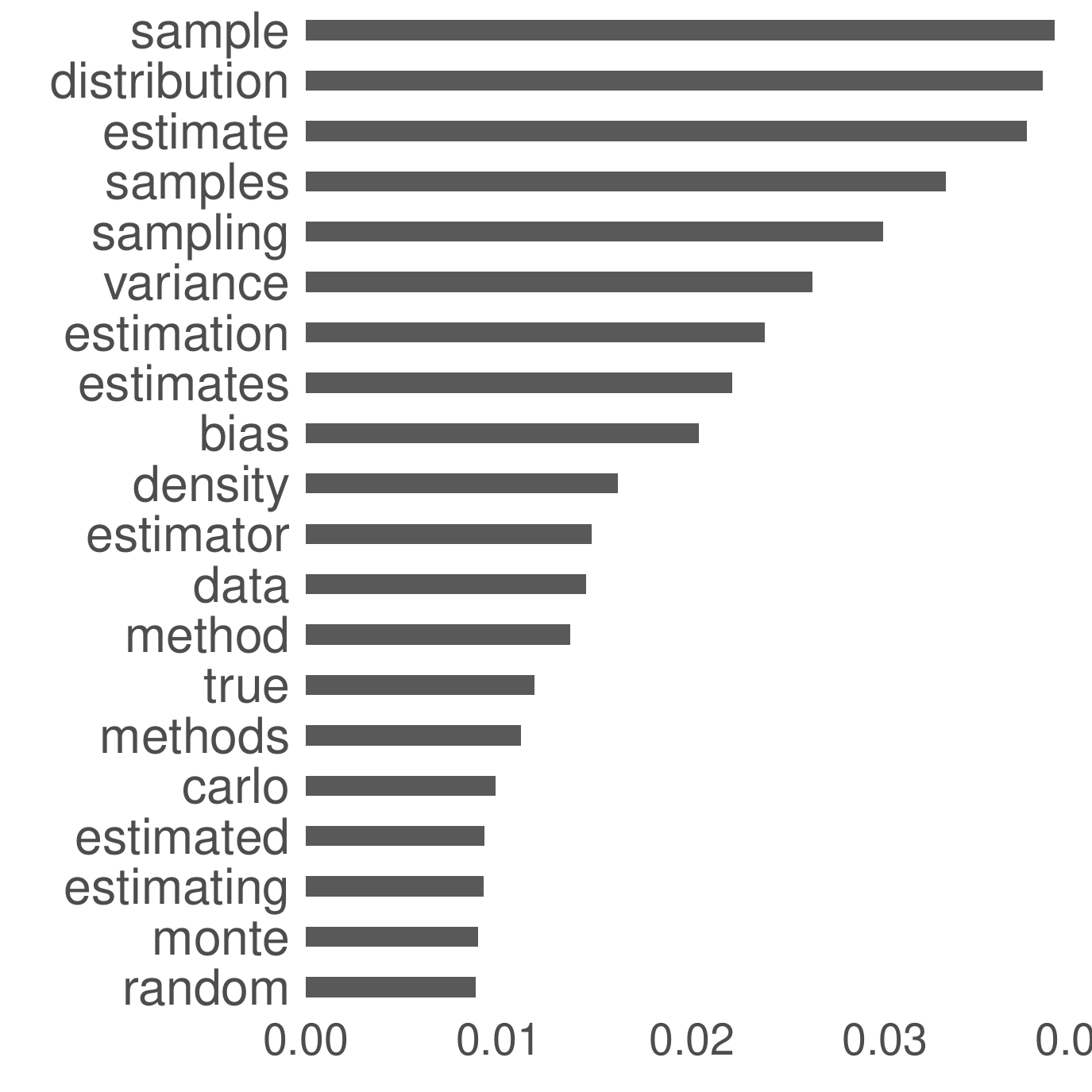}}\hfill%
	\subfloat[cLDA: t$85$ (Bayesian models)\label{fig:nips-clda-lda-topics:c}]{
		\includegraphics[width=.3\linewidth]
		{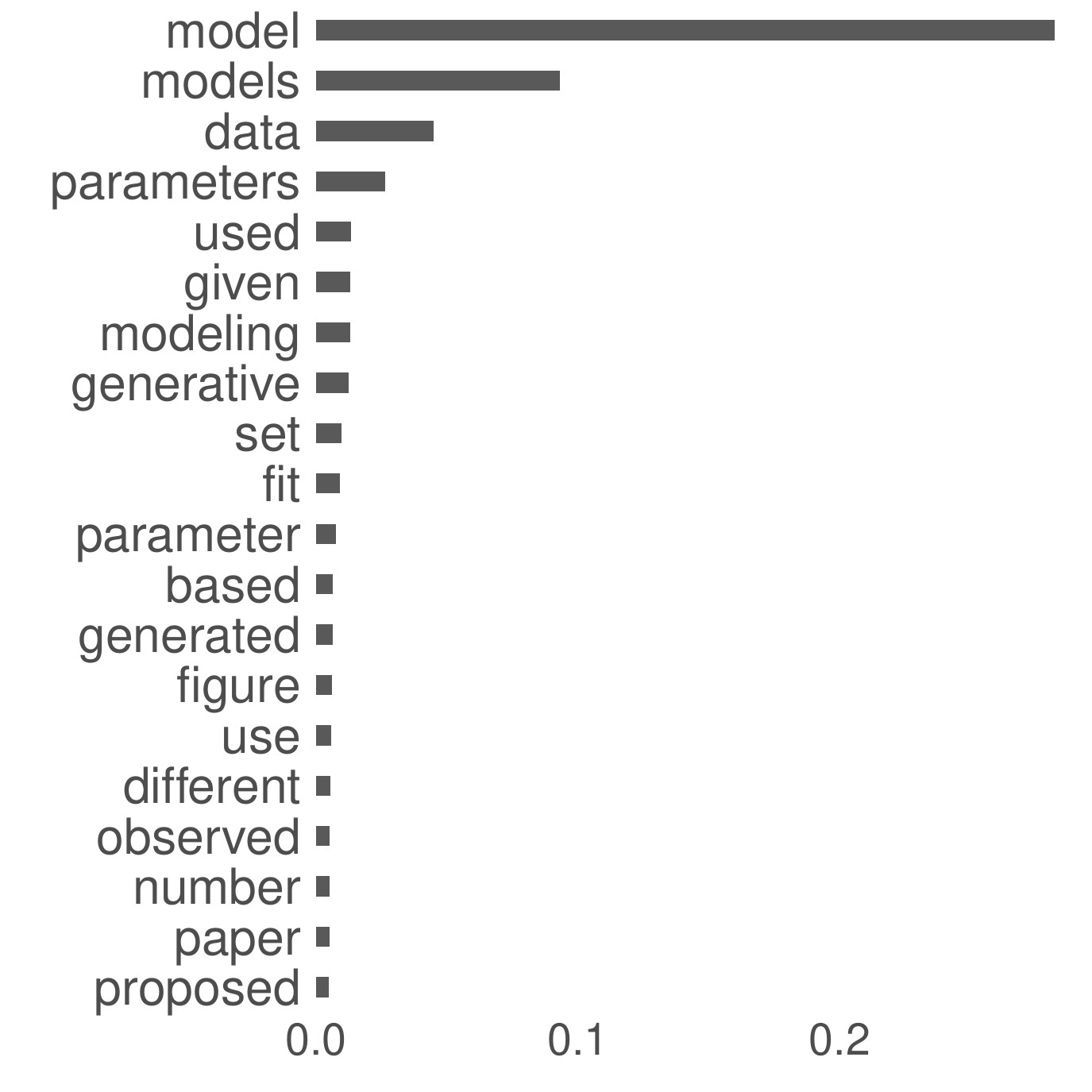}}\\
	\subfloat[LDA: t$8$ (ANN)\label{fig:nips-clda-lda-topics:d}]{
		\includegraphics[width=.3\linewidth]
		{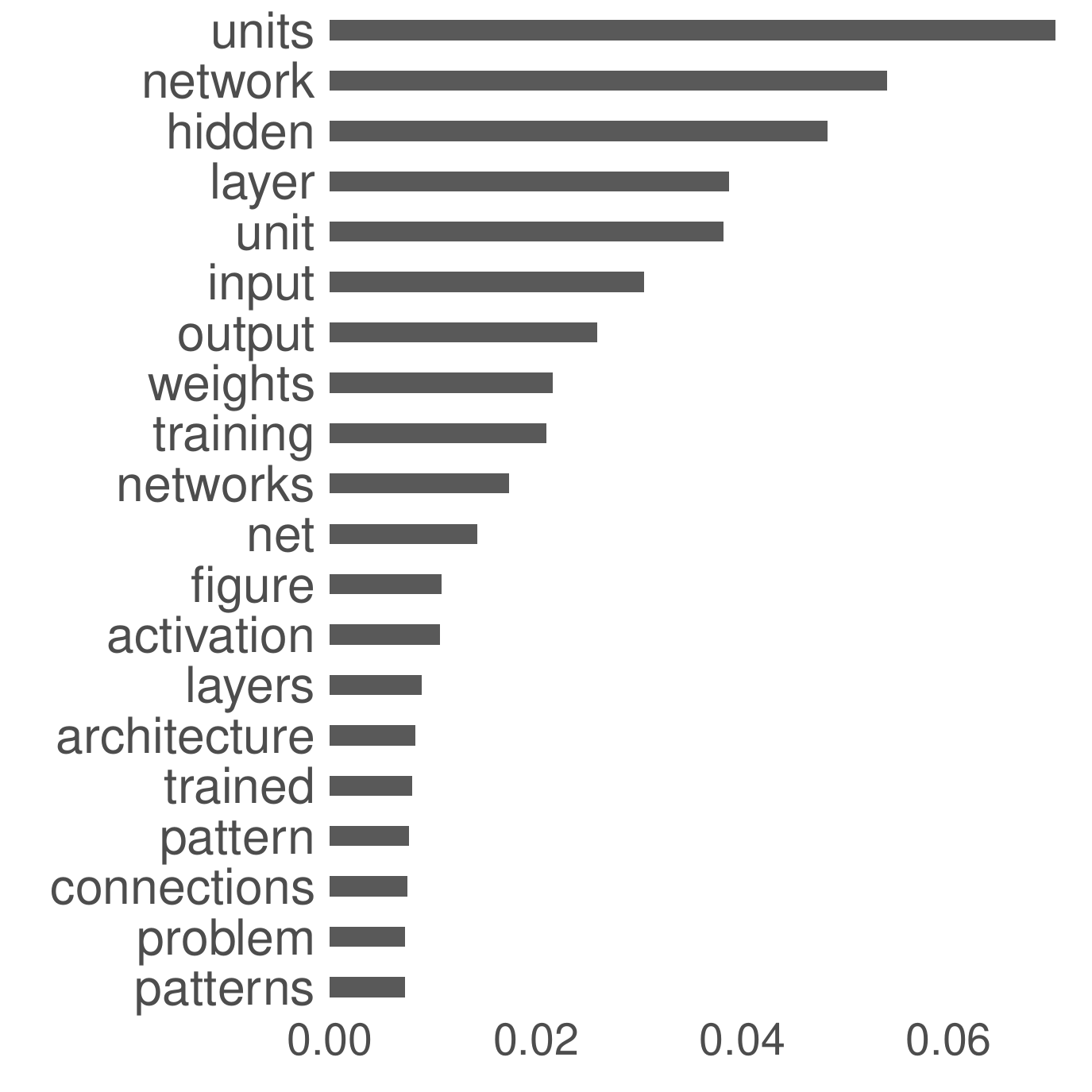}}\hfill%
	\subfloat[LDA: t$57$ (MCMC)\label{fig:nips-clda-lda-topics:e}]{
		\includegraphics[width=.3\linewidth]
		{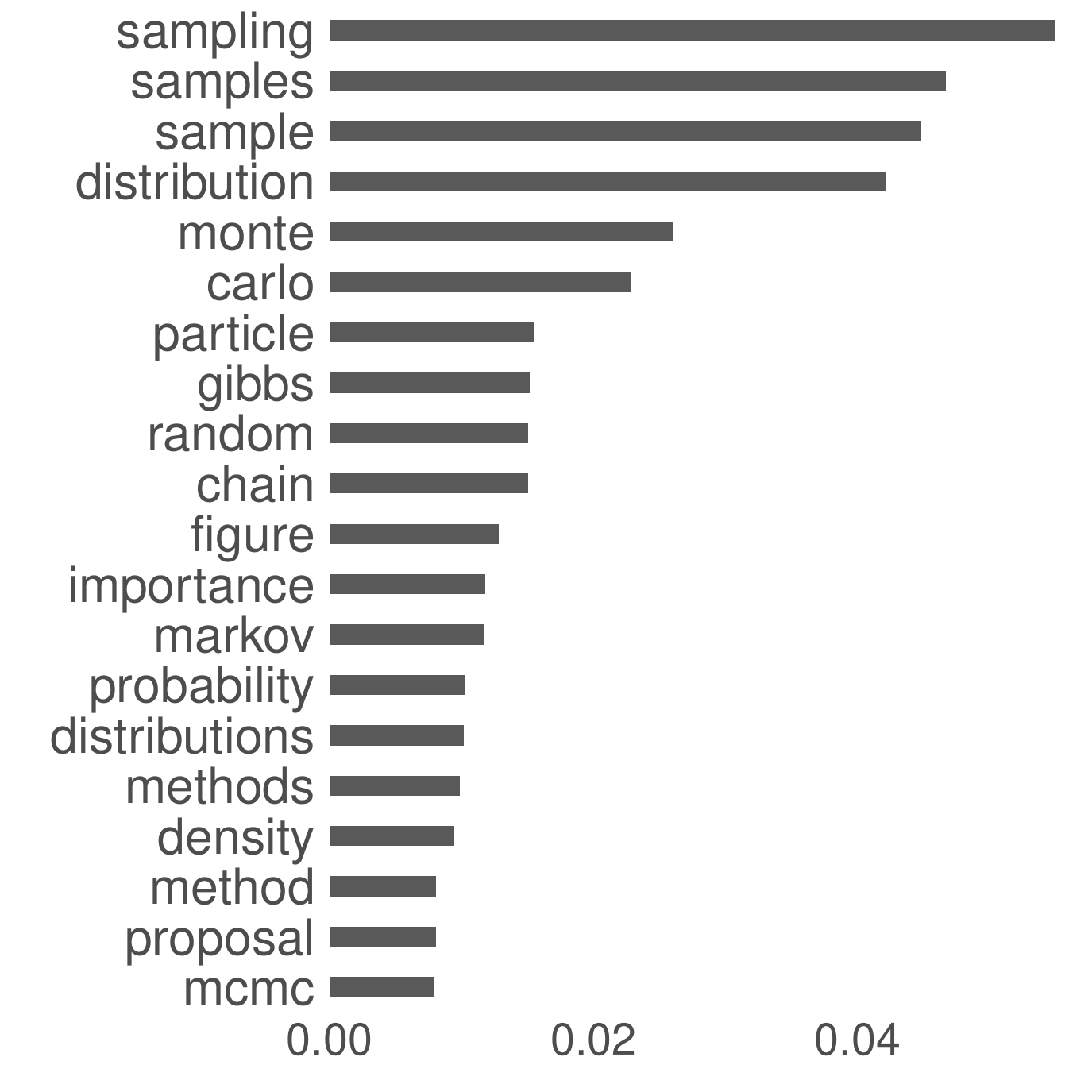}}\hfill%
	\subfloat[LDA: t$85$ (Bayesian models)\label{fig:nips-clda-lda-topics:f}]{
		\includegraphics[width=.3\linewidth]
		{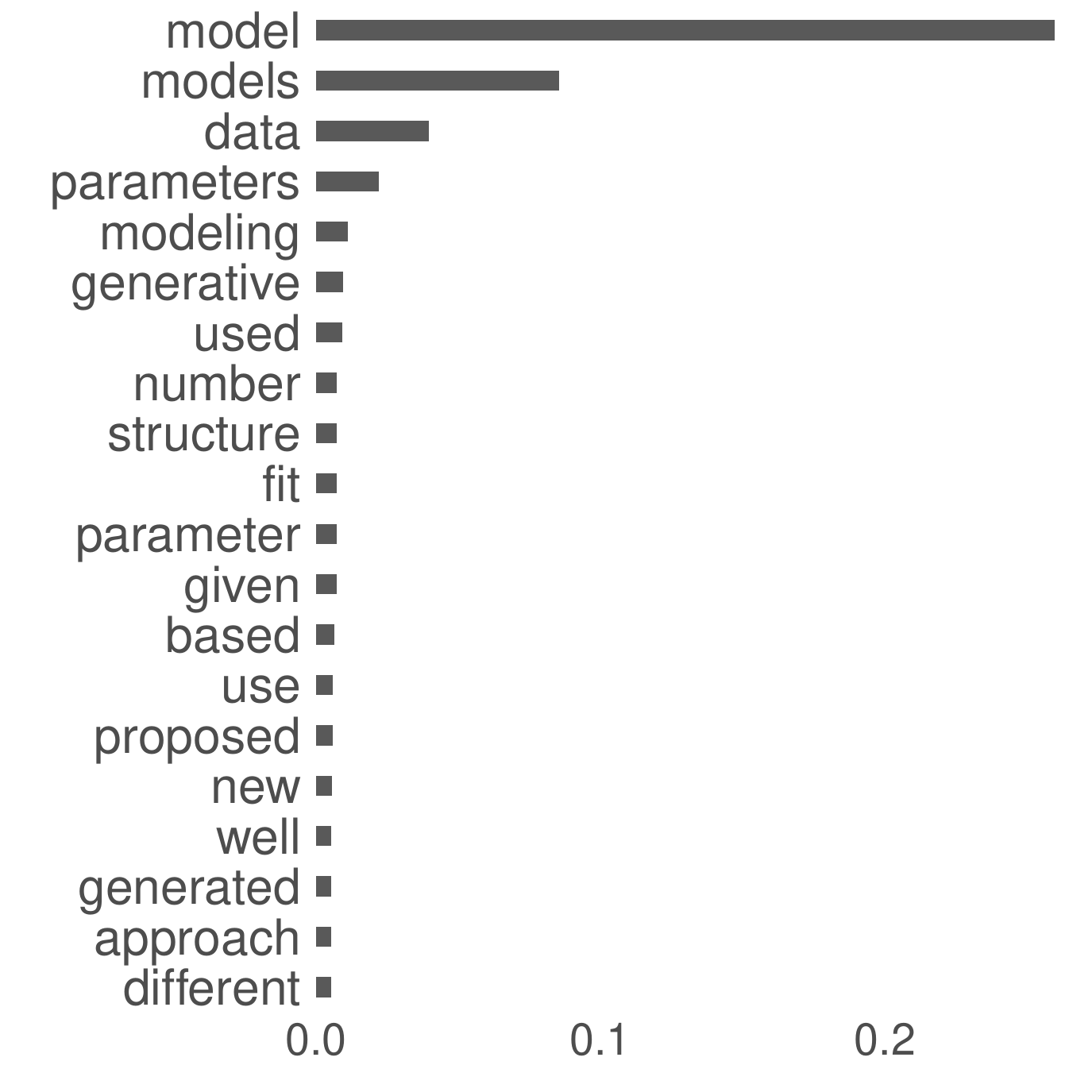}}
	\caption{$20$ most probable words for topics from $90$-topic cLDA (top row) and LDA (bottom row) 
	models trained on corpus NIPS $00$-$18$. Bars represent the corresponding (estimated) 
	probabilities of words given a topic.}
	\label{fig:nips-clda-lda-topics}
\end{figure*}

In text mining, one can use trained cLDA models for tasks such 
as classifying text documents and summarizing document collections 
and corpora. cLDA also gives excellent options to visualize and 
browse documents. To illustrate some of these, we provide a 
usability study for the proposed cLDA model next. We are interested 
in two aspects of a learned cLDA model: (a) interpretability of the 
learned topics and topic structures, and (b) visualizing corpora in 
terms of learned topics. 

cLDA infers collection-level and document-level topic distributions 
based on a common set of corpus-level topics. So it is natural to 
check whether cLDA's topics are as interpretable as LDA's topics. 
We present three examples of LDA and cLDA topics t$8$, t$57$, and 
t$85$ in \figurename~\ref{fig:nips-clda-lda-topics} to furnish a 
qualitative comparison of these models, for corpus NIPS $00$-$18$. 
For each topic $k$, we show $20$ most probable words based on 
the non-decreasing order of the $\beta_{kv}$ values over words 
$v$'s (denoted by bars in each plot). Note that we did any 
necessary alignments for the set of topics from each of these 
models to ease comparison. We labeled topics based on their most 
probable words. (We do these post processing steps for this section 
and all the following.) Note that cLDA topics are meaningful, 
easily interpretable, and comparable with LDA topics.

A typical question of interest in topic modeling is to identify 
topics that evolve over time. cLDA provides a way to model time 
evolving corpus. On the other hand, LDA has some limitations due to 
several issues such as the required alignment of topics and 
information loss of segmentation, as mentioned in  
Section~\ref{sec:introduction}. To illustrate some of these issues 
involved, we study corpus NIPS $00$-$18$---consists of four 
collections based on document timestamps---using both cLDA and LDA 
models. Once we fit a cLDA model (e.g. via AGS chain) for the 
corpus, we can use each estimated collection-level topic 
distribution for a time period as its natural topic allocation. 
Recall that LDA does not have collection-level topic mixtures by 
the model construction. However, one can estimate them via each 
collection's word topic vector  $\bz_{j}$ (e.g. sampled from the 
CGS chain). 

\begin{figure}[t!] 
\centering 
\subfloat[M$1$: AGS run on the corpus with four collections 
\label{fig:nips-sel-pi:a}]{
	\includegraphics[width=.7\linewidth]
	{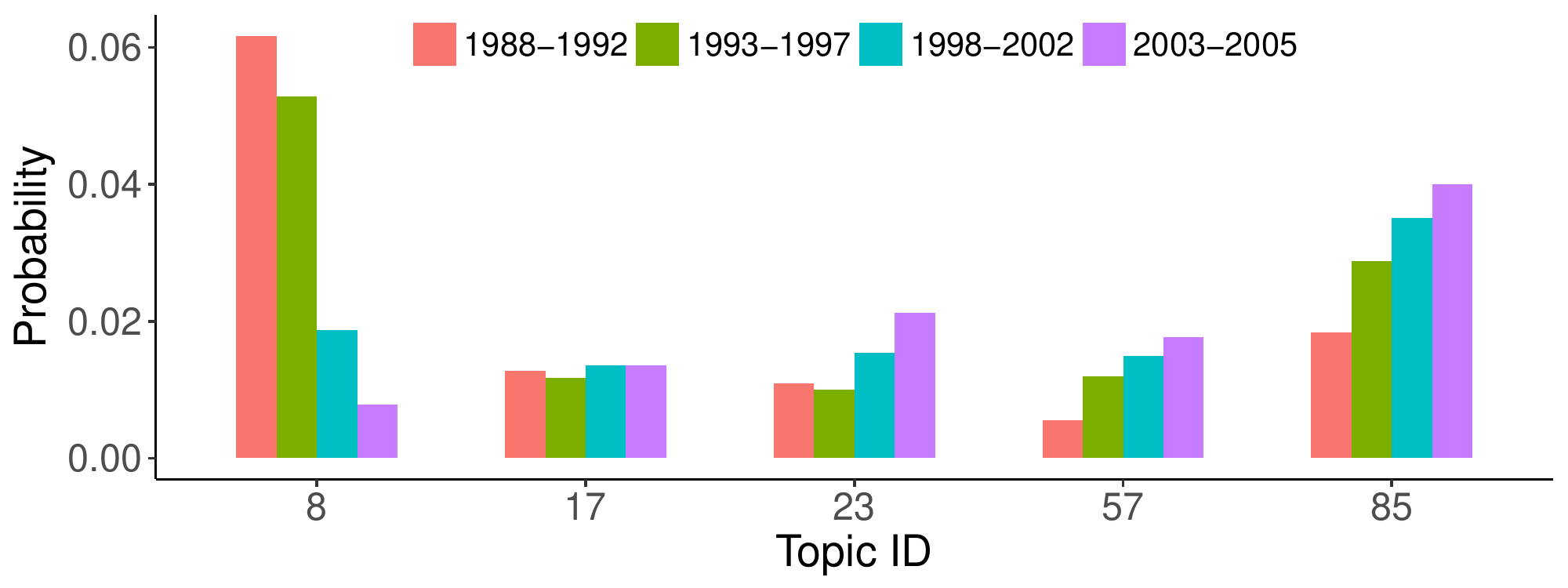}}\\%
\subfloat[M$2$: Single CGS run on the whole 
corpus\label{fig:nips-sel-pi:b}]{
	\includegraphics[width=.7\linewidth]
	{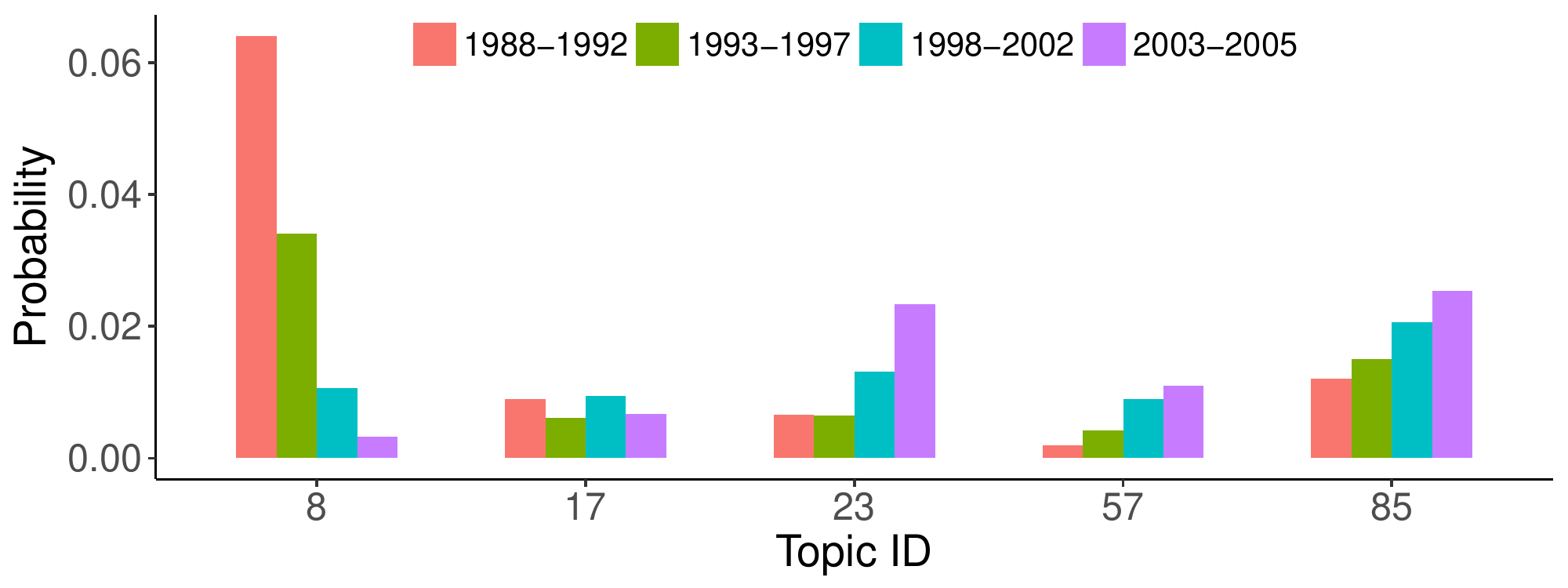}}\\%
\subfloat[M$3$: Single CGS run on each of the four partitions of 
the 
corpus\label{fig:nips-sel-pi:c}]{
	\includegraphics[width=.7\linewidth]
	{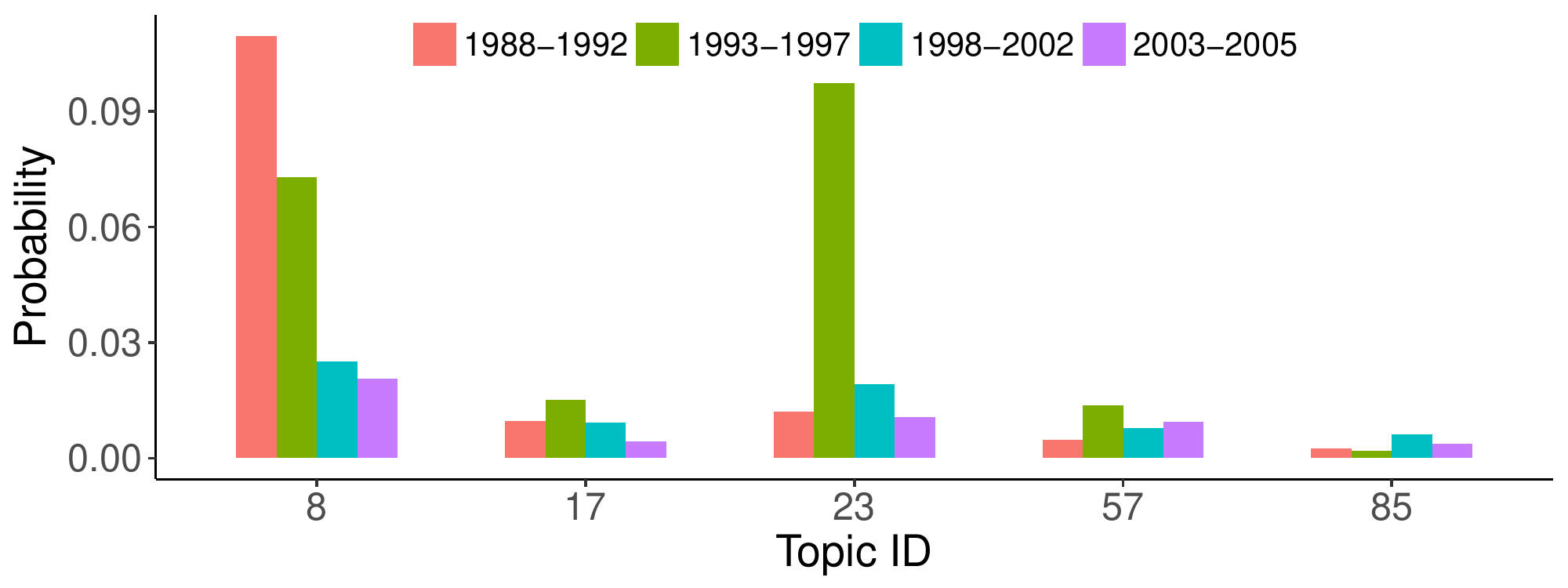}}
\caption{Estimates of topic distributions (of five randomly 
	selected topics) for four timespans $1988$-$1992$, 
	$1993$-$1997$, 
	$1998$-$2002$, and $2003$-$2005$ of the NIPS conference 
	proceedings 
	via three topic modeling approaches. Clearly, cLDA identifies 
	interpretable topic patterns compared to other two LDA 
	approaches. 
	See discussion in the text. 
}
\label{fig:nips-sel-pi}
\end{figure}

\figurename~\ref{fig:nips-sel-pi} shows estimates of four  
collection-level topic distributions $\bpi_1, \bpi_2, \bpi_3$, and 
$\bpi_4$ for corpus NIPS $00$-$18$ based on three different topic 
modeling approaches, which we will describe below. In 
\figurename~\ref{fig:nips-sel-pi:a}, we applied the cLDA AGS 
algorithm on the whole corpus with four collections, i.e., $J$ = 4 
(denoted by M$1$). Each barplot represents the value of a topic 
element in the last sample $\bpi^{(1000)}$ from the AGS chain. (We 
show values for a few selected topics here; see an extended list of 
topics in the Appendix, 
\figurename~\ref{fig:nips-clda-pi-samples}.)
\figurename~\ref{fig:nips-sel-pi:b} is based on the LDA CGS 
algorithm on the whole corpus (denoted by M$2$). To estimate 
collection $j$'s topic mixture, we used the last sample 
$\bz_j^{(1000)}$ from the CGS chain. In 
\figurename~\ref{fig:nips-sel-pi:c}, we considered each 
collection in the corpus as a separate corpus. We then ran 
the LDA CGS algorithm on each sub-corpus (denoted by M$3$) 
one by one. We used the last sample $\bz_j^{(1000)}$ from 
collection $j$'s CGS chain to estimate its topic mixture. 

One can infer interesting patterns from the estimates of  
collection-level topic mixtures via method M$1$, as shown in 
\figurename~\ref{fig:nips-sel-pi:a}. Several research topics got 
increased and decreased interest, and some topics were relatively 
constant in popularity over the time span $1988$-$2005$ of NIPS. As 
we would have expected, topic t$57$, a topic on MCMC and 
inference, and topic t$85$, a topic on generative models got 
increased popularity over the time span, which is interesting to 
watch. One topic that lost popularity is t$8$, which is about 
neural networks and related topics. 
(Additional details are given 
in Section~\ref{app:nips-results}.) We can also see 
that the estimates via method M$1$ are superior to the estimates 
based on other two methods M$2$ and M$3$. Note that for method 
M$3$, one must solve the non-trivial topic alignment problem for 
each LDA model learned for every sub-corpus. Additionally, we 
notice that method M$2$ gives better estimates than method M$3$. We 
believe this is due to the lack of information sharing among 
the partitions in method M$3$ that might have affected the quality 
of the learned LDA models. On the other hand, cLDA considers the 
corpus as a whole to incorporate the organization hierarchy of 
documents into the model, eliminating the issues discussed.

\begin{figure}[t!] 
	\centering 
	\subfloat[M$1$: AGS run on the corpus with 
	five collections 
	\label{fig:yelp-clda-pi}]{
		\includegraphics[width=.8\linewidth]
		{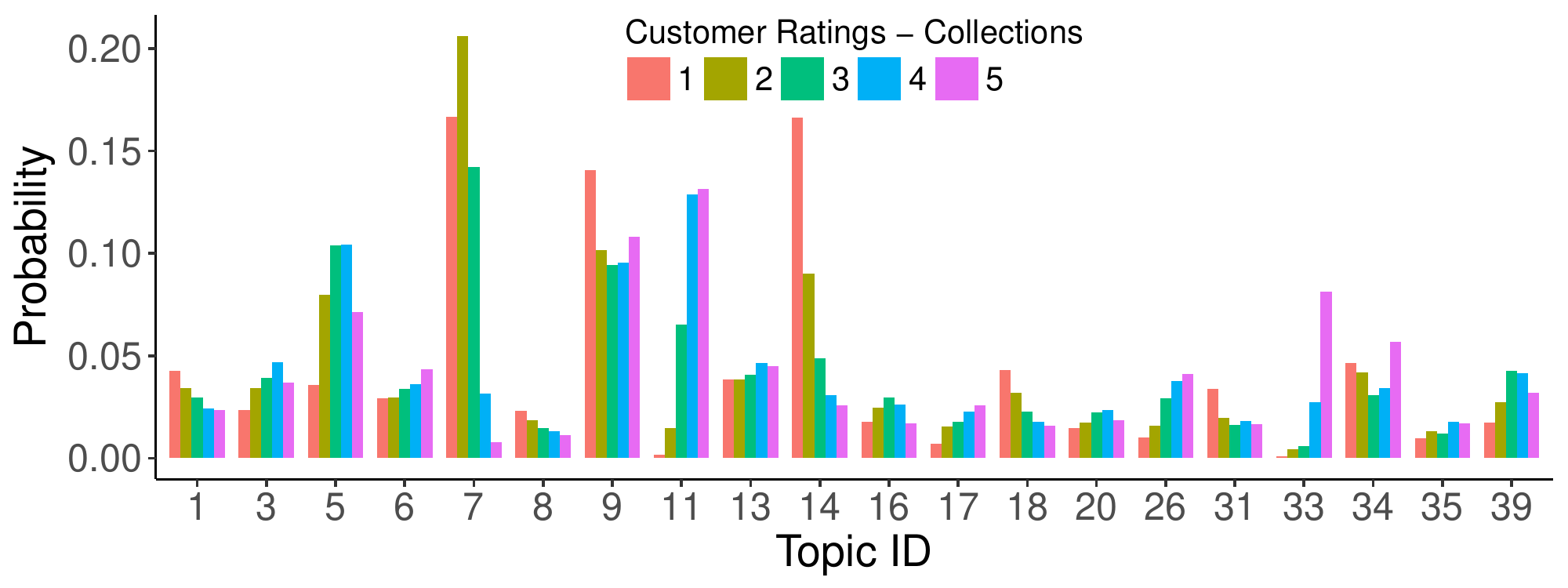}}
	\\%
	\subfloat[M$2$: Single CGS run on the whole corpus
	\label{fig:yelp-lda-pi}]{
		\includegraphics[width=.8\linewidth]
		{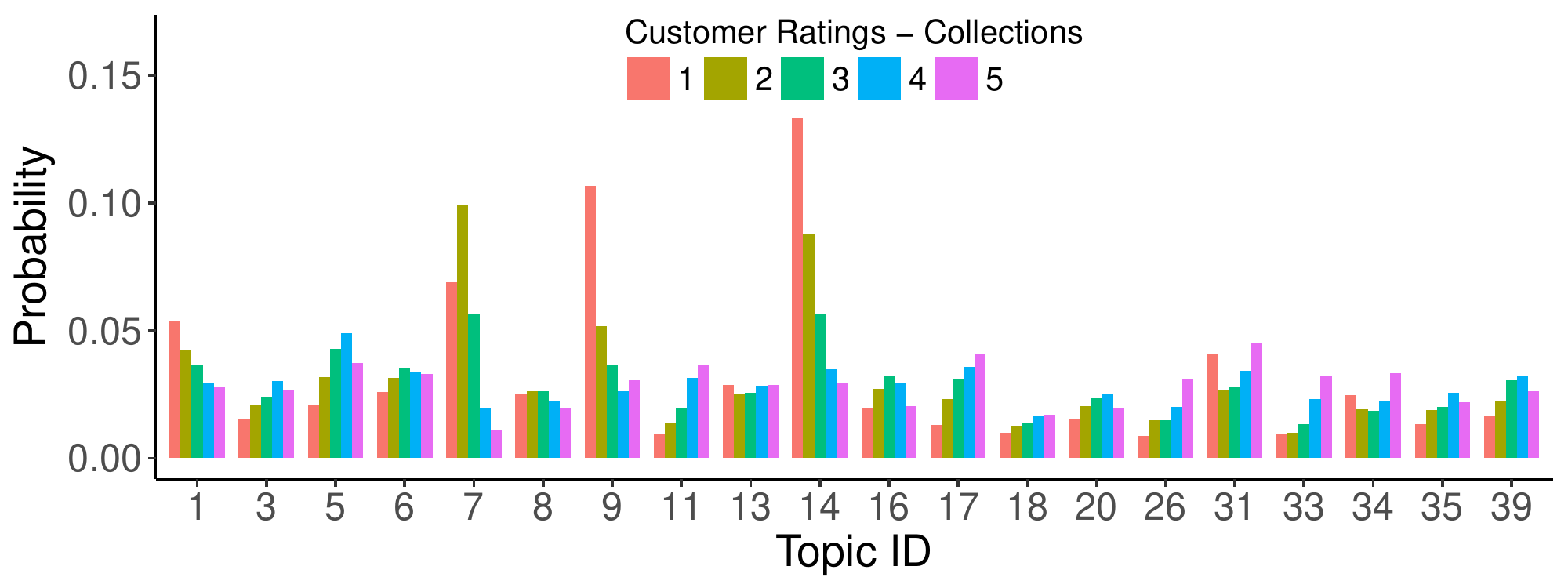}}
	\caption{Estimates of topic proportions for five customer 
		ratings $1$, $2$, $3$, $4$, and $5$ via algorithms AGS and 
		CGS 
		using the Yelp corpus. Table~\ref{tab:yelp-topic-desc} 
		gives 
		additional details about the listed topics. Clearly, cLDA 
		provides meaningful and interpretable patterns of topics in 
		the corpus for different customer ratings, compared to its  
		non-hierarchical precursor, LDA. In both of these models, 
		topics are aligned to ease comparison, based on most 
		probable 
		words given a topic. See description in the text.     
	}
	\label{fig:yelp-sel-pi}	
\end{figure}

Modeling customer behaviors is the core interest in operations 
management community and market research. Analyzing patterns of 
topics that pervade through customer reviews can give interesting 
insights about customer needs. 
We wish to study topic patterns for different customer ratings, and 
we thus use corpus Yelp.  A natural choice is to partition the 
corpus into collections based on customer review ratings on the 
scale $1$-$5$, and fit a cLDA model. 
Note that the two other corpora used in this paper have collections 
that occurred more naturally; but, for this corpus, we follow this 
partitioning scheme just for experimentation. 
\figurename~\ref{fig:yelp-clda-pi} shows estimated 
collection-level topic distributions of a $40$-topic cLDA model 
trained on corpus Yelp via AGS chain. Specifically, each barplot  
represents the value of a topic element for a collection from the 
last sample $\bpi^{(1000)}$ of the AGS chain. We can see a gradual 
increase of probabilities for some topics (e.g. t$3$, t$11$, t$26$) 
and gradual decrease of probabilities for some topics (e.g. t$7$, 
t$14$, t$18$), for rating from $1$ to $5$. We noticed that the 
topics with gradual increase in probabilities correspond to 
positive reviews, and the topics with gradual decrease in 
probabilities correspond to negative reviews. 

Table~\ref{tab:yelp-topic-desc} gives additional details of these 
topics and their manually assigned sentiment. In our experience, 
people do not give detail reviews about the place, ambiance, food, 
or customer service, if they really like them (e.g. reviews with 
rating $5$). If they partially like a place (e.g. reviews with 
rating $3$ and $4$), they provide comments with all the details. 
Additionally, topics that show little or relatively small 
variability for different ratings are general topics about food 
styles. 
For completeness, we include the collection-level topic 
distributions estimated via a single CGS chain (i.e. via method 
M$2$) for the same Yelp corpus with $K = 40$. Note that LDA is 
inferior in giving any interpretable analyses from the results, as 
shown in \figurename~\ref{fig:yelp-lda-pi}.    

%

\begin{table}[t!] 
	\centering
	\begin{threeparttable}[b] 
		\caption{A subset of topics from a 
			$40$-topic cLDA model for corpus Yelp}
		\label{tab:yelp-topic-desc}
		\begin{tabular}{c c l}
			\toprule
			Topic ID\tnote{a} & Sentiment 
			& 
			Short Description \\ 
			\midrule
			3 & positive & food \\  
			11 & positive & food, place \\ 
			26 & positive & food, ambiance, group dining \\ 
			33 & positive & food, atmosphere \\ 
			5 & moderately positive & food, service \\ 
			6 & moderately positive & food \\ 
			13 & moderately positive & food, waiting time \\ 
			35 & moderately positive & food, sandwich, salad \\ 
			39 & moderately positive & place, ambiance \\ 
			8 & neutral & Mexican food\\ 
			9 & neutral & place, food \\ 
			16 & neutral & bar, food, atmosphere \\ 
			20 & neutral & location, parking\\ 
			34 & neutral & location, ambiance \\ 
			1 & moderately negative &  ambiance, night-life \\ 
			7 & moderately negative &  food, service \\ 
			31 & moderately negative & food, meal quantity \\  
			14 & negative & food, service \\ 
			18 & negative & place, food \\   
			\bottomrule
		\end{tabular}
		\begin{tablenotes} 
			\item [a] The topics are ordered in the 
			non-decreasing order of the positive 
			sentiment or polarity of the most probable 
			words given a topic.
		\end{tablenotes} 
	\end{threeparttable} 
\end{table}

\begin{figure}[t!] 
	\centering
	\includegraphics[width=.8\linewidth]
	{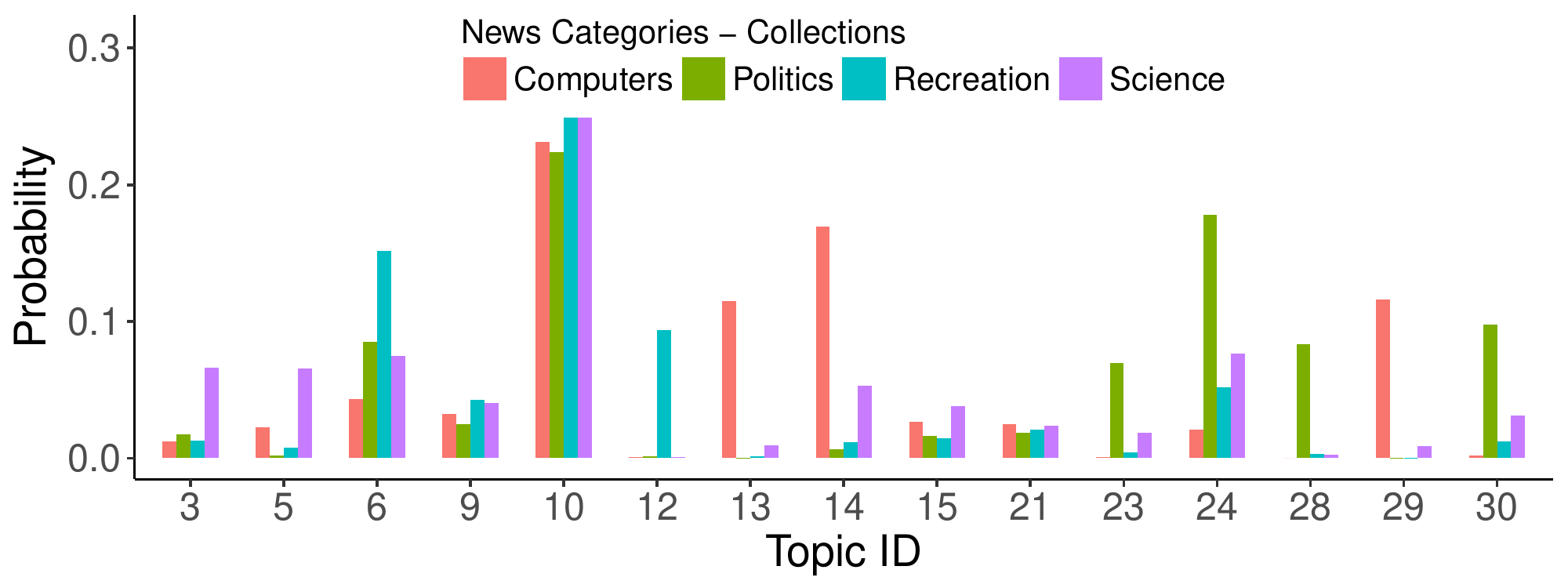}
	\caption{Estimates of topic distributions for collections Computers, 
		Politics, Recreation, and Science in corpus $16$newsgroups.}
	\label{fig:16news-clda-pi-samples-new}
\end{figure}

A key feature of the cLDA model, compared to its non-hierarchical 
alternative, LDA, is that it provides collection level topics, 
which summarize themes or topics of individual collections in a 
corpus. We perform a study on the applicability of this feature 
using corpus $16$newsgroups. 
Recall that corpus $16$newsgroups is partitioned based on four 
subject groups. It is natural to check whether each collection's 
topic distribution exhibits proper weights to the topics that 
are related to the subject matter of the collection. 
\figurename~\ref{fig:16news-clda-pi-samples-new} 
shows estimated collection-level topic distributions, i.e., 
$\bpi_j, \, j = 1, \ldots, 4$, from a learned $30$-topic cLDA model 
via the AGS algorithm. (We found that topic $10$, which 
is dominant in all collections, is a potential collection of 
stop-words in the corpus. \figurename~\ref{fig:16news-topics-new} 
provides examples of topics 
for reference.) 
The results show that cLDA enables us to summarize topics of 
individual collections and perform a meaningful comparison of 
topics among collections, an aspect that is not well exploited in a 
flat modeling framework such as LDA. We also notice sparse 
allocations for topics in each collection and no major sharing of 
topics among collections in 
\figurename~\ref{fig:16news-clda-pi-samples-new}: this conveys the 
fact that corpus $16$newsgroups may be surely separable via the 
subject matter of newsgroups. 

%% file: discussion.tex
\section{Summary}
\label{sec:discussion}

In this paper, we developed a topic model, compound latent  
Dirichlet allocation (cLDA), that can incorporate several 
characteristics of a corpus including the organization hierarchy of 
documents. One can employ cLDA model to (a) explore research topics 
across multiple/consecutive proceedings or workshops of a 
conference, (b) discover cultural differences in blogs and forums 
from different countries,  or (c) compare nuanced review 
summarization on various customer ratings (for customer relations 
management). We proposed posterior inference methods for cLDA (e.g. 
AGS, MGS, VEM), which can be used for analyzing collections of 
documents and we recommend algorithm AGS. We also discussed 
guidelines for selecting $K$ and hyperparameters in cLDA, which 
perform well empirically. Additionally, we have seen that proposed 
parametric hierarchy in cLDA adds only a little computational 
burden compared to the non-hierarchical alternative, LDA. cLDA 
model makes several natural assumptions about the data generating 
framework, which we believe helped us getting a much simpler 
inference scheme, compared to nonparametric alternatives such as 
Hierarchical Dirichlet Process \citep{TehEtal:2006}. We have also  
provided an empirical comparison of cLDA with two-level DP.

Lastly, we studied the applicability and performance of cLDA model 
in both synthetic and real-world corpora. cLDA is 
quite useful for analyzing topic distributions of collections in a 
corpus, and it can better understand the underlying thematic 
structure of the corpus. cLDA provides relevant insights about 
topics that evolve over time, as noted in the experiments on NIPS 
conference proceedings. This model also presents some ideas to 
employ metadata such as customer ratings for ``soft'' segmentation 
while analyzing the thematic structure of customer reviews. 
Although such partitions do not emerge naturally, together with 
cLDA, they may provide diverse perspectives of customer behaviors.
 
\begin{notes-to-stay}
We also wish to study the applicability of this model for a corpus 
where documents exhibit multiple collection memberships. For example, 
a corpus built from the Wikipedia articles, where an article can be 
assigned to one or more Wikipedia categories.     
\end{notes-to-stay}

%% file: app-posterior.tex
\section{Expressions for the Prior, Likelihood, and Posterior}
\label{app:posterior}
This section derives expressions of the prior $p_h(\bpsi)$, 
likelihood $\ell_{\bw} (\bpsi)$, and the posterior 
$p_{h,\bw}(\bpsi)$. From the hierarchical model 
\eqref{eq:cLDA-e}--\eqref{eq:cLDA-a} given in Section 
\ref{sec:hierarchical-model}, we can write the prior $p_h(\bpsi)$ 
as 

\begin{equation*}
p_h (\bz \given \btheta, \bpi, \bbeta) 
p_h (\btheta \given \bpi) p_h (\bpi) p_h (\bbeta), 
\label{eq:prior-psi}
\end{equation*}
where the individual density functions are given by 
\eqref{eq:cLDA-e}--\eqref{eq:cLDA-b} of the model. 
Let $n_{jdk} = \sum_{i=1}^{n_{jd}} z_{jdik}$, i.e.\ $n_{jdk}$ is
the number of words in document $d$ in collection $j$ that are 
assigned to topic $k$, and let $n_{j.k} = \sum_{d=1}^{D_j} \sum_{i=1}^
{n_{jd}} z_{jdik}$, i.e.\ $n_{j.k}$ is the number of words in 
all documents in collection $j$ that are assigned to topic $k$.
Using the Dirichlet and multinomial distributions specified in
\eqref{eq:cLDA-e}--\eqref{eq:cLDA-b}, we obtain

\begin{eqnarray}
p_h(\bpsi) &=& 
  \prod_{j=1}^J 
  \left(\frac{\Gamma\bigl( K \alpha \bigr)}
  {\Gamma(\alpha)^K} 
  \prod_{k=1}^K \pi_{jk}^{\alpha - 1} 
  \right)
	\Biggl[ 
	\prod_{j=1}^J \prod_{d=1}^{D_j} \prod_{k=1}^K 
	\theta_{jdk}^{n_{jdk}} 
	\Biggr] 
	\nonumber \\ 
& &	\Biggl[ 
		\prod_{j=1}^J \prod_{d=1}^{D_j} 
		\Biggl(\frac{\Gamma\bigl( \gamma \bigr)}
								{\prod_{k=1}^K {\Gamma(\gamma \pi_{jk})}} 
					 \prod_{k=1}^K \theta_{jdk}^{\gamma \pi_{jk} - 1} 
		\Biggr)
  \Biggr] \nonumber \\ 
	& & \Biggl[ \prod_{k=1}^K \biggl( 
	\frac{\Gamma(V\eta)}{\Gamma(\eta)^V} \prod_{v=1}^V \beta^{\eta 
	- 1}_{kv} \biggr)
  \Biggr].
\label{eq:dens-of-btz}
\end{eqnarray}

For $j = 1, \ldots, 
J$, $d = 1, \ldots, D_j$ and $k = 1, \ldots, K$, let $S_{jdk} = \{ 
i: 1 \leq i \leq n_{jd} \text{ and } z_{jdik} = 1 \}$, which is the 
set of indices of all words in document $d$ in collection $j$ whose 
latent topic variable is $k$.  With this notation, Equation 
\eqref{eq:cLDA-b} induces the likelihood function 
$\ell_{\bw} (\bpsi) := p(\bw \given \bz, \btheta, \bpi, \bbeta)$  
\begin{equation}
  \label{eq:lda-likelihood}
  \begin{split}
    p (\bw \given \bz, \btheta, \bpi, \bbeta) & = 
    \prod_{j=1}^{J} \prod_{d=1}^{D_j} \prod_{i=1}^{n_{jd}}
                \prod_{k:z_{jdik}=1} \prod_{v=1}^V
                \beta_{kv}^{w_{jdiv}} \\
            & = \prod_{j=1}^{J} \prod_{d=1}^{D_j} \prod_{k=1}^K \prod_{v=1}^V
                \prod_{i \in S_{jdk}} \beta_{kv}^{w_{jdiv}} \\
            & = \prod_{j=1}^{J} \prod_{d=1}^{D_j} \prod_{k=1}^K \prod_{v=1}^V
                \beta_{kv}^{\sum_{ i \in S_{jdk} } w_{jdiv}} \\
            & = \prod_{j=1}^{J} \prod_{d=1}^{D_j} \prod_{k=1}^K \prod_{v=1}^V
                \beta_{kv}^{m_{jdkv}},
  \end{split}
\end{equation}
where $m_{jdkv} = \sum_{i \in S_{jdk}} w_{jdiv}$ counts the number of
words in document $d$ in collection $j$ for which the latent topic is $k$ and the
index of the word in the vocabulary is $v$. Recalling the definition
of $n_{jdk}$ given just before~\eqref{eq:dens-of-btz}, and noting
that $\sum_{i \in S_{jdk}} w_{jdiv} = \sum_{i=1}^{n_{jd}} z_{jdik}
w_{jdiv}$, we see that $m_{jdkv} = \sum_{i=1}^{n_{jd}} z_{jdik} 
w_{jdiv}$ and $\sum_{v = 1}^V m_{jdkv} = n_{jdk}$ 

Plugging the likelihood~\eqref{eq:lda-likelihood} and the
prior~\eqref{eq:dens-of-btz} into the expression for the Bayes rule 
\eqref{eq:posterior-psi}, and 
absorbing constants in Dirichlet normalizing constants into an 
overall constant of proportionality, we have the posterior density 
\begin{eqnarray} 
	\begin{split}
  p_{h,\bw}(\bpsi) 
	& \propto   
	\Biggl[ \prod_{k=1}^K \prod_{v=1}^V \beta_{kv}^{\sum_{j=1}^{J} 
	\sum_{d=1}^{D_j} m_{jdkv} + \eta - 1} \Biggr] \\ 
	& \hspace{4.7mm}
	\Biggl[ \prod_{j=1}^J \prod_{d=1}^{D_j} \frac{\prod_{k=1}^K 
	\theta_{jdk}^{n_{jdk} + \gamma \pi_{jk} - 1}}{\prod_{k=1}^K 
	{\Gamma(\gamma \pi_{jk})}}   \Biggr]  
	\Biggl[ \prod_{j=1}^J \prod_{k=1}^K \pi_{jk}^{\alpha - 1} \Biggr]
	\end{split} \nonumber 
\end{eqnarray}

\section{Conditional posterior for $z_{jdi}$}
\label{app:posterior-z}

This section derives a closed form expression for the 
conditional posterior $p_{h, \bw} \left(z_{jdik} = 1 \given 
\bz^{(-jdi)}, 
\bpi \right)$ for $k = 1, \ldots, K$. The vector $\bz^{(-jdi)}$ 
contains 
topic assignments of all words in the corpus except for word 
$w_{jdi}$. We can then write  
\begin{eqnarray}
\begin{split}
p_{h, \bw} \left(z_{jdik} = 1 \given \bz^{(-jdi)}, \bpi \right) 
& \propto p \left(z_{jdik} = 1, w_{jdiv} = 1 \given \bz^{(-jdi)}, 
\bw^{(-jdi)}, \bpi\right)  \\ 
& = \frac{p \left(z_{jdik} = 1, w_{jdiv} = 1, \bz^{(-jdi)} \given 
	\bw^{(-jdi)}, \bpi \right)}
{p \left(\bz^{(-jdi)} \given \bw^{(-jdi)}, \bpi \right)} \\ 
& \propto  
\frac{\gamma \pi_{jk} + n^{(-jdi)}_{jdk}}
{\gamma + n^{(-jdi)}_{jd.}} \,
\frac{\eta  + m^{(-jdi)}_{..kv}}
{V\eta + m^{(-jdi)}_{..k.}}	
\end{split}
\label{eq:posterior-z-derivation}
\end{eqnarray}
Here, we used the fact that $p 
\left(\bz^{(-jdi)} \given \bw^{(-jdi)}, \bpi \right) \propto 
p_{h,\bw}(\bpi, \bz)^{(-jdi)}$ given by \eqref{eq:pi-z-posterior}
and $\Gamma(x + 1) = x \Gamma(x)$. The superscript $(-jdi)$ means 
that we discard the contribution of word $w_{jdi}$ in count 
statistics $n_{jdk}$, $n_{jd.}$, $m_{..kv}$, and $m_{..k.}$.     
This development is motivated by the LDA collapsed Gibbs sampling  
algorithm~\citep[CGS]{GriffithsSteyvers:2004}.

%% file: app-mmala.tex
\section{Langevin Monte Carlo}
\label{app:mmala}

The Metropolis Adjusted Langevin Algorithm 
~\cite[MALA]{Girolami:2011}, as described in Section~\ref{sec:mgs}, 
is given the following steps 
\begin{enumerate}
	\item Propose $\bpi^{(*)}$ via Langevin dynamics 
	\eqref{eq:langevin-discrete-eq}
	\item Calculate the MH acceptance ratio 
	\begin{equation}
	a (\bpi^{(t)}, \bpi^{(*)}) = \frac{p(\bpi^{(*)})}
	{p(\bpi^{(t)})}
	\frac{\exp \bigl(-\frac{1}{2\varepsilon^2} {\| \bpi^{(t)} - 
			\bmu(\bpi^{(*)}, \varepsilon) \|}^2  \bigr)}
	{\exp \bigl(-\frac{1}{2\varepsilon^2} {\| \bpi^{(*)} - 
			\bmu(\bpi^{(t)}, \varepsilon) \|}^2 \bigr)}
	\end{equation}
	and set $\bpi^{(t+1)} = \bpi^{(*)}$ with probability 
	min$(1, a(\bpi^{(t)}, \bpi^{(*)}))$, and set $\bpi^{(t+1)} = 
	\bpi^{(t)}$ with the remaining probability. 
\end{enumerate}

\subsection{Riemann Manifold Metropolis Adjusted Langevin 
Algorithm}


Girolami et al. \cite[Section 5]{Girolami:2011} suggests the 
preconditioning matrix ${\bG}(\bpi)$ as an arbitrary metric tensor 
on a Riemannian manifold induced by the parameter space of a 
statistical model. 
We write the stochastic 
differential equation for the Langevin diffusion on the Riemannian 
manifold as 
\begin{equation}
\text{d}\bpi(t) = \frac{1}{2} \widetilde{\nabla}_{\bpi} {\cal L} ( \bpi^{(t)} ) \text{d}t + \text{d}\tilde{\bb}(t)   
\label{eq:reimann-langevin}
\end{equation} 
where the natural gradient is 
\begin{equation}
\widetilde{\nabla}_{\bpi} {\cal L}( \bpi^ {(t)} ) = \bG\{ \bpi (t) \}^{-1} \nabla_{\bpi} {\cal L}(\bpi^{(t)})
\end{equation} 
and the Brownian motion is 
\begin{equation}
\text{d}\tilde{\bb}_k(t) = {| \bG\{ \bpi (t) \} |}^{-1/2} \sum_{k^\prime = 1}^{K} \frac{\partial }{\partial \pi_{k^\prime}} \Bigl [ \bG\{ \bpi (t) \}^{-1}_{kk^\prime} \, {| \bG\{ \bpi (t) \} |}^{1/2} \Bigl ] \text{d}t 
+ {\Bigl [ {\bG\{ \bpi (t) \}}^{-1/2} \text{d}\bb(t)  \Bigl ]}_k
\label{eq:reimann-langevin-brownian}
\end{equation} 
where the subscript $k$ indicates $k^\text{th}$ element in vector 
$\text{d}\tilde{\bb}(t)$. 
Note that in the Euclidean space the metric tensor ${\bG}(\bpi)$ is 
an identity matrix, and thus \eqref{eq:reimann-langevin} will be 
reduced to the standard Langevin SDE.  

By expanding the gradients in~\eqref{eq:reimann-langevin-brownian} 
and discretizing \eqref{eq:reimann-langevin} via the 
first-order Euler integration, we get the proposal for 
the Riemann Manifold Metropolis Adjusted Langevin Algorithm (MMALA):
\begin{equation}
\bpi^{(*)} = \bmu(\bpi^{(t)}, \varepsilon) + \varepsilon  {\bG(\bpi^{(t)})}^{-1/2} \bxi^{(t)},
\label{eq:riemann-langevin-euler}
\end{equation}
where $k^\text{th}$ element in vector $\bmu(\bpi^{(t)}, \varepsilon)$ is 
given by 
\begin{equation}
\begin{split}
\bmu(\bpi^{(t)}, \varepsilon)_k &= \pi_k^{(t)} 
+ \frac{\varepsilon^2}{2} \Bigl \{ {\bG ( \bpi^{(t)} )}^{-1} 
\nabla_{\bpi} {\cal L}(\bpi^{(t)}) \Bigr \}_k \\ 
&\hspace{5mm}- \varepsilon^2 \sum_{k^\prime = 1}^{K} \Bigl \{ 
\bG(\bpi^{(t)} )^{-1} \frac{\partial \bG(\bpi^{(t)} )}{\partial 
\pi_{k^\prime}} \bG(\bpi^{(t)} )^{-1} \Bigr \}_{kk^\prime} \\ 
&\hspace{5mm}+ \frac{\varepsilon^2}{2} \sum_{k^\prime = 1}^{K} \bigl \{\bG(\bpi^{(t)} )^{-1} \bigl \}_{kk^\prime} \text{tr} \Bigl \{ \bG(\bpi^{(t)} )^{-1} \frac{\partial \bG(\bpi^{(t)} )}{\partial \pi_{k^\prime}}  \Bigr \} 
\end{split}
\end{equation}
The corresponding proposal density is given by 
\begin{equation}
q(\bpi^{(*)} \leftarrow \bpi^{(t)}) = 
\mathcal{N}_K(\bpi^{(*)} \given \bmu(\bpi^{(t)}, \varepsilon), \, 
\varepsilon^2 \bG(\bpi^{(t)} )^{-1})
\end{equation}

\subsection{Langevin Updates on Probability Simplices: Boundary 
Considerations}

\eat{
To perform valid moves (e.g. via MALA or MMALA) of $\bpi$ that 
lies on the probability simplex, one needs to consider boundary 
conditions. One natural solution to handle boundaries is to 
re-parameterize $\bpi$~\cite{Patterson:2013}. Here, we revisit the 
original parameter $\bpi_j \in \simplex_K$, $j = 1, 2, \ldots, J$ 
of interest and the unnormalized conditional density 
\eqref{eq:posterior-pi}.  Let $\bvarphi_j 
= (\varphi_{j1}, \ldots, \varphi_{jK}) \in \Real^K$. We take the 
prior on $\bvarphi_j$ as a product of i.i.d. Gamma random variables 
as 
\begin{equation}
\begin{split}
|\varphi_{jk}| &\iid \text{Gamma} \Bigl (\alpha, 1 \Bigr ), \\
p(\bvarphi_j) &\propto \prod_{k = 1}^{K} {|\varphi_{jk}|}^{\alpha - 1} e^{-|\varphi_{jk}|}. 
\end{split}
\end{equation}
We define $|\varphi_{j.}| := \sum_{k=1}^K |\varphi_{jk}|$. Let  
$\pi_{jk}$ be $|\varphi_{jk}| / |\varphi_{j.}|$, for each $k = 1, 
2, \ldots, K$. This choice keeps the prior on $\bpi_j$ a    
Dirichlet density. We can now re-write the unnormalized conditional 
density~\eqref{eq:posterior-pi} on $\bpi_j$ as 
}

We first re-write the unnormalized conditional 
density~\eqref{eq:posterior-pi} on $\bpi_j$ as 
\begin{equation}
\tilde{p}(\bvarphi_j \given \bz_j, \bw_j) \propto  
\prod_{d=1}^{D_j} \prod_{k=1}^K
\Biggl(\frac{ \Gamma(\gamma \frac{|\varphi_{jk}|}{|\varphi_{j.}|} + n_{jdk})}
						{\Gamma(\gamma \frac{|\varphi_{jk}|}{|\varphi_{j.}|})}
\Biggr)
\prod_{k=1}^K {|\varphi_{jk}|}^{\alpha - 1} e^{-|\varphi_{jk}|}
\label{eq:posterior-pi-re-parametrization}
\end{equation}
The MMALA update for the new parametrization is given by 
\begin{equation}
\bvarphi_j^{(*)} = \bmu(\bvarphi_j^{(t)}, \varepsilon) + \varepsilon  {\bG(\bvarphi_j^{(t)})}^{-1/2} \bxi^{(t)}
\label{eq:riemann-langevin-euler-reparam}
\end{equation}
where for $k = 1, \ldots, K$, we have  
\begin{equation}
\begin{split}
\bmu(\bvarphi_j^{(t)}, \varepsilon)_k &= \varphi_{jk}^{(t)}  
+ \frac{\varepsilon^2}{2} \Bigl \{ {\bG ( \bvarphi_j^{(t)} )}^{-1} \nabla_{\bvarphi_j} {\cal L}(\bvarphi_j^{(t)}) \Bigr \}_k \\ 
&\hspace{5mm}- \varepsilon^2 \sum_{k^\prime = 1}^{K} \Bigl \{ \bG(\bvarphi_j^{(t)} )^{-1} \frac{\partial \bG(\bvarphi_j^{(t)} )}{\partial \varphi_{jk^\prime}} \bG(\bvarphi_j^{(t)} )^{-1} \Bigr \}_{kk^\prime} \\ 
&\hspace{5mm}+ \frac{\varepsilon^2}{2} \sum_{k^\prime = 1}^{K} \bigl \{\bG(\bvarphi_j^{(t)} )^{-1} \bigl \}_{kk^\prime} \text{tr} \Bigl \{ \bG(\bvarphi_j^{(t)} )^{-1} \frac{\partial \bG(\bvarphi_j^{(t)} )}{\partial \varphi_{jk^\prime}}  \Bigr \}
\end{split} 
\end{equation}
The proposal density of this diffusion process is given by 
\begin{equation}
\begin{split}
q(\bvarphi_j^{(*)} \leftarrow \bvarphi_j^{(t)}) &= 
\mathcal{N}_K(\bvarphi_j^{(*)} \given \bmu(\bvarphi_j^{(t)}, \varepsilon), \, \varepsilon^2 \bG(\bvarphi_j^{(t)} )^{-1}) \\ 
&\propto {| \bG(\bvarphi_j^{(t)})^{-1} |}^{-1/2} 
\exp \Bigl \{ -\frac{1}{2 \varepsilon^2} {(\bvarphi_j^{(*)} - \bmu(\bvarphi_j^{(t)}, \varepsilon))}^\text{T} 
\bG(\bvarphi_j^{(t)}) {(\bvarphi_j^{(*)} - \bmu(\bvarphi_j^{(t)}, \varepsilon))}
\Bigr\} \\
\end{split}
\label{eq:pi-transition-density-re-parameterization}
\end{equation}
We take the metric tensor $\bG ( \bvarphi^{(t)}_j )$ 
as $\diag(|\bvarphi^{(t)}_j|)^{-1}$ 
(\cite{Patterson:2013} suggested this choice in a different 
context.), as it gives simplified expressions for  
\begin{equation}
\bmu(\bvarphi_j^{(t)}, \varepsilon)_k = \varphi_{jk}^{(t)} 
+ \frac{\varepsilon^2}{2} \, \Bigl \{ \diag(|\bvarphi_j^{(t)}|) \, \nabla_{\bvarphi_j} {\cal L}(\bvarphi_j^{(t)}) \Bigr \}_k
+  \frac{\varepsilon^2}{2} \, \sign(\varphi^{(t)}_{jk}) 
\end{equation}
and
\begin{equation}
q(\bvarphi_j^{(*)} \leftarrow \bvarphi_j^{(t)}) 
\propto \Bigl [ \prod_{k=1}^{K} {|\varphi_{jk}^{(t)}|}^{-1/2} \Bigr ]
\exp \Bigl \{ -\frac{1}{2 \varepsilon^2} {(\bvarphi_j^{(*)} - \bmu(\bvarphi_j^{(t)}, \varepsilon))}^\text{T} 
\diag({|\bvarphi_j^{(t)}|})^{-1} {(\bvarphi_j^{(*)} - \bmu(\bvarphi_j^{(t)}, \varepsilon))}
\Bigr\} 
\end{equation}
To derive $\nabla_{\bvarphi_j} 
{\cal L}(\bvarphi_j^{(t)})$, we first write $\tilde{p}(\bvarphi_j 
\given \bz_j, \bw_j)$ in a convenient logarithmic form: 
\begin{equation}
\log \tilde{p}(\bvarphi_j \given \bz_j, \bw_j) \propto  \sum_{k = 1}^{K} \Bigl [ \Bigl \{ (\alpha - 1) \log |\varphi_{jk}| \Bigr \} - |\varphi_{jk}| \Bigr ]
+ \sum_{d=1}^{D_j} \sum_{k=1}^K \Bigl [ \lgamma(\gamma \frac{|\varphi_{jk}|}{|\varphi_{j.}|} + n_{jdk}) - \lgamma(\gamma \frac{|\varphi_{jk}|}{|\varphi_{j.}|})  \Bigr ] \nonumber 
\end{equation}
We then define $\nabla_{\bvarphi_j} {\cal L}(\bvarphi_j^{(t)})$ 
by the partial derivatives  
\begin{equation}
\begin{split}
\frac{\partial \log \tilde{p}(\bvarphi_j \given \bz_j, \bw_j)}{\partial \varphi_{jk^*}}  
&= \sign (\varphi_{jk^*}) \Bigl \{ 
\frac{\gamma}{{|\varphi_{j.}|}^2}  \sum_{d=1}^{D_j} \sum_{k = 1}^K  {|\varphi_{jk}|} 
\Bigl [  \varPsi(\gamma \frac{{|\varphi_{jk}|}}{{|\varphi_{j.}|}}) - \varPsi (\gamma \frac{{|\varphi_{jk}|}}{{|\varphi_{j.}|}} + n_{jdk}) \Bigr ] \\
&\hspace{4mm}- \frac{\gamma}{{|\varphi_{j.}|}} \sum_{d=1}^{D_j} 
\Bigl [  \varPsi(\gamma \frac{{|\varphi_{jk^*}|}}{{|\varphi_{j.}|}}) - \varPsi (\gamma \frac{{|\varphi_{jk^*}|}}{{|\varphi_{j.}|}} + n_{jdk^*}) \Bigr ] 
+ \Bigl [ \frac{(\alpha - 1)}{{|\varphi_{jk^*}|}} - 1 \Bigr ] \Bigr \} \\
\end{split}
\end{equation}

\noindent
Similar to the AGS chain, in this scheme, we implement a Markov 
chain on $(\bpi, \bz)$ via MMALA updates within Gibbs 
sampling (MGS), as in Algorithm \ref{alg:gibbs-mmala-pi-z}. 

%% file: app-est-eta-gamma.tex
\section[Estimating hyperparameters eta and gamma]
{Estimating hyperparameters $\eta$ and $\gamma$}
\label{sec:est-eta-gamma}

In this section, we derive expressions for the fixed point 
iterations of hyperparameters $\eta$ and $\gamma$ in the cLDA 
model. We use the following two bounds by \cite[Appendix 
B]{minka2000estimating} in our development. 

\begin{eqnarray}
\frac{\Gamma (x)}{\Gamma(n + x)} &\geq& 
\frac{\Gamma (\hat{x}) \exp\left((\hat{x} - x)b\right)}
{\Gamma(n + \hat{x})} \label{eq:minka-bound1}\\ 
b &=& \Psi(n + \hat{x}) - \Psi(\hat{x}) \nonumber \\[1em] 
\frac{\Gamma(n + x)}{\Gamma (x)} &\geq& cx^a \: \:
\text{ if } n \geq 1, x \geq 1 \label{eq:minka-bound2}\\ 
a &=& \left(\Psi(n + \hat{x}) - \Psi(\hat{x})\right) \hat{x} 
\nonumber\\
c &=&  \frac{\Gamma(n + \hat{x})}{\Gamma (\hat{x})} \hat{x}^{-a}
\nonumber
\end{eqnarray}

From the cLDA hierarchical model 
\eqref{eq:cLDA-e}--\eqref{eq:cLDA-a}, 
after integrating out $\bbeta$'s, we get the marginal posterior of 
$(\bpi, \bz)$, given $\bw$ and $\eta$ as:  
\begin{equation}
p_{\eta, \bw}(\bpi, \bz) \propto \prod_{k = 1}^K 
\left [ \frac{\Gamma(V\eta)}{\Gamma(m_k + V\eta)} 
\prod_{v = 1}^V \frac{\Gamma(m_{kv} + \eta)} {\Gamma(\eta)}
\right]
\end{equation}
Using \eqref{eq:minka-bound1} and \eqref{eq:minka-bound2}, we write 
it as: 
\begin{eqnarray}
p_{\eta, \bw}(\bpi, \bz) &\geq& \prod_{k = 1}^K 
\left [ 
\frac{\Gamma(V\eta_0) \exp\left(V(\eta_0 - \eta)b_k\right)}
{\Gamma(m_k + V\eta_0)} 
\prod_{v = 1}^V c_{kv} \eta^{a_{kv}}
\right] \\ 
a_{kv} &=& \eta_0 \Big( \Psi(m_{kv} + \eta_0) - \Psi (\eta_0) 
\Big) \\ 
b_k &=& \Psi(m_k + V\eta_0) - \Psi(V\eta_0)\\ 
c_{kv} &=& \frac{\Gamma(m_{kv} + \eta_0)}{\Gamma(\eta_0)} 
\eta_0^{-a_{kv}} 
\end{eqnarray}
Here, $\eta_0$ represents the value of $\eta$ at the current 
state. We now denote the lower bound on $\log p_{\eta, \bw} (\bpi, 
\bz)$ as ${\cal L}_\eta$, which we write   
\begin{equation}
{\cal L}_\eta = \sum_{k=1}^K \sum_{v = 1}^V 
\Big( \log c_{kv} + a_{kv} \log \eta \Big) + 
\sum_{k=1}^K \Big( \log \Gamma (V\eta_0) - 
\log \Gamma(m_k + V\eta_0) + V (\eta_0 - \eta) b_k \Big).
\end{equation}
Taking derivatives with respect to $\eta$, 
\begin{equation}
\frac{\partial {\cal L}_\eta}{\partial \eta} = 
\frac{1}{\eta} \sum_{k=1}^K \sum_{v = 1}^V a_{kv} - 
V \sum_{k=1}^K b_k 
\end{equation}
and setting it to zero, we get 
\begin{equation}
\eta = \frac{\sum_{k=1}^K \sum_{v = 1}^V a_{kv}}{V\sum_{k=1}^K b_k}
= \frac{\eta_0}{V} 
\frac{\sum_{k=1}^K \sum_{v = 1}^V 
	\Big[ \Psi(m_{kv} + \eta_0) - \Psi(\eta_0)\Big]} 
{\sum_{k=1}^K \Big[ \Psi(m_k + V\eta_0) - \Psi(V\eta_0) \Big]}
\end{equation}
We can compute the maximum via the fixed point iteration 
\citep{Minka:2000}. 

Similarly, from the cLDA hierarchical model  
\eqref{eq:cLDA-e}--\eqref{eq:cLDA-a}, 
after integrating out $\btheta$'s, we get the marginal posterior of 
$(\bpi, \bz)$, given $\bw$ and $\gamma$ as:  
\begin{equation}
p_{\gamma, \bw}(\bpi, \bz) \propto \prod_{j = 1}^J \prod_{d = 
1}^{D_j}   
\left [ \frac{\Gamma(\gamma)}{\Gamma(n_{jd} + \gamma)} 
\prod_{k = 1}^K 
\frac{\Gamma(n_{jdk} + \gamma \pi_{jk})} {\Gamma(\gamma \pi_{jk})}
\right]
\end{equation}
Using \eqref{eq:minka-bound1} and \eqref{eq:minka-bound2}, we  
write it as: 
\begin{eqnarray}
p_{\gamma, \bw}(\bpi, \bz) &\geq& \prod_{j = 1}^{J} \prod_{d = 
1}^{D_j} 
\left [ 
\frac{\Gamma(\gamma_0) \exp\left((\gamma_0 - \gamma)b_{jd}\right)}
{\Gamma(n_{jd} + \gamma_0)} 
\prod_{k = 1}^K c_{jdk} (\gamma \pi_{jk})^{a_{jdk}}
\right] \\ 
a_{jdk} &=& \gamma_0 \pi_{jk} \Big( \Psi(n_{jdk} + \gamma_0 
\pi_{jk}) - \Psi (\gamma_0 \pi_{jk}) \Big) \\ 
b_k &=& \Psi(n_{jdk} + \gamma_0) - \Psi(\gamma_0) \\ 
c_{jdk} &=& \frac{\Gamma(n_{jdk} + \gamma_0 \pi_{jk})}
{\Gamma(\gamma_0 \pi_{jk})}{(\gamma_0 \pi_{jk})}^{-a_{jdk}} 
\end{eqnarray}
Here, $\gamma_0$ represents the value of $\gamma$ at the current 
state. We now denote the lower bound on $\log p_{\gamma, \bw} 
(\bpi, \bz)$ as ${\cal L}_\gamma$, which we write   
\begin{eqnarray}
{\cal L}_\gamma &=& \sum_{j = 1}^J \sum_{d = 1}^{D_j} \sum_{k = 1}^K
\Big[ \log c_{jdk} + a_{jdk} \log \gamma + a_{jdk} \log \pi_{jk} 
\Big] + \\ 
&\hspace{1cm}& \sum_{j = 1}^J \sum_{d = 1}^{D_j} 
\Big[ \log \Gamma (\gamma_0) - \log \Gamma(n_{jd} + \gamma_0) 
+ (\gamma_0 - \gamma) b_{jd} \Big] \nonumber.
\end{eqnarray}
Taking derivatives with respect to $\gamma$, 
\begin{equation}
\frac{\partial {\cal L}_\gamma}{\partial \gamma} = 
-\sum_{j = 1}^{J} \sum_{d = 1}^{D_j} b_{jd} + 
\frac{1}{\gamma} \sum_{j = 1}^{J} \sum_{d = 1}^{D_j} 
\sum_{k = 1}^{K} a_{jdk} 
\end{equation} 
and setting it to zero, we get 
\begin{eqnarray}
\gamma &=& \frac{\sum_{j = 1}^{J} \sum_{d = 1}^{D_j} \sum_{k = 
1}^{K} a_{jdk}}{\sum_{j = 1}^{J} \sum_{d = 1}^{D_j} b_{jd}} 
\nonumber \\
&=& \gamma_0 \frac{\sum_{j = 1}^{J} \sum_{d = 1}^{D_j} 
\sum_{k = 1}^{K} \pi_{jk} \Big[ 
\Psi (n_{jdk} + \gamma_0 \pi_{jk}) - \Psi (\gamma_0 \pi_{jk}) \Big]}
{\sum_{j = 1}^{J} \sum_{d = 1}^{D_j} \Big[ \Psi (n_{jd} + \gamma_0) 
- \Psi (\gamma_0) \Big]} 
\end{eqnarray}
We can compute the maximum via the fixed point iteration 
\citep{Minka:2000}. 

%% file: app-vem.tex
\eject 

\section{Variational Inference}
\label{app:vem}

We now develop variational methods \citep{Jordan:1999} to 
approximate the  intractable posterior $p_{h, \bw} (\bpsi)$ in the 
cLDA model. Our approach can be viewed as an extension of 
\cite{BleiNgJordan:2003}'s inference scheme in the LDA model. 
Briefly, in variational methods, one considers a restricted family 
of distributions instead of working on the intractable posterior, 
and then seeks the member of the family that is ``closest'' to the 
posterior~\citep{Jordan:1999, Bishop:2006}. One way to restrict the 
family of approximating distributions is to use a parametric 
distribution (i.e., variational distribution) that is governed by a 
set of parameters (i.e., variational parameters). Typically, this 
parametric distribution is much simpler to work with than the 
original posterior by assuming independence between respective 
variables.  The goal is then to identify the parameters which give 
the tightest lower-bound with in the family.

Let $p_{h} (\bpsi, \bw)$ be the joint probability of $\bpsi = 
(\bbeta, \bpi, \btheta, \bz)$ and $\bw$ based on the cLDA model. 
Suppose $q(\bpsi)$ is any parametric distribution over latent 
variables $\bpsi$. 
We can then write the log marginal probability of the data $\bw$ as 
\citep{Bishop:2006}
\begin{equation}
	\log m(h) = {\cal L}(q, p_{h}) + \text{KL}(q, p_{h, \bw})
	\label{eq:log-marginal-prob}
\end{equation}
where\footnote{The summation $\sum_{\bz}$ represents the
summation over all $z_{jdi}$s. We use summation 
instead of an integral because $z_{jdi}$s are discrete.}
\begin{equation}
{\cal L}(q, p_{h}) = \int \sum_{\bz}
q(\bpsi) \log \left \{ \frac{p_{h}(\bpsi, \bw)}{q(\bpsi)} \right \}
d\bbeta d\bpi d\btheta  
\label{eq:variational-lowerbound-original}
\end{equation}
and
\begin{equation}
	\text{KL}(q, p_{h, \bw}) = -\int
	\sum_{\bz} q(\bpsi) \log \left \{
		\frac{p_{h, \bw} (\bpsi)}{q(\bpsi)} \right \} d\bbeta d\bpi 
d\btheta.
	\label{eq:kl-divergence}
\end{equation}
Note that ${\cal L}(q, p_{h})$ 
in~\eqref{eq:variational-lowerbound-original} is a 
functional of the distribution $q(\bpsi)$ and a function of the 
hyperparameters $h$. The Kullback-Leibler (KL) divergence specified
in~\eqref{eq:kl-divergence} satisfies $\text{KL}(q,
p_{h, \bw}) \geq 0$---by the positivity of the KL
divergence, with equality if, and only if, $q(\bpsi)$
equals the posterior $p_{h, \bw}(\bpsi)$.  
Following~\eqref{eq:log-marginal-prob}, ${\cal L}(q, p_{h})$ is a 
lower-bound for the log marginal probability. We can maximize the 
lower-bound ${\cal L}(q, p_{h})$ with respect to
$q(\bpsi)$, which is also equivalent to minimizing $\text{KL}(q, 
p_{h, \bw})$. The tightest lower-bound occurs when the KL 
divergence vanishes, i.e., when $q(\bpsi)$ equals the posterior 
distribution (but it is intractable to work with).  Thus, in 
variational methods, one considers a restricted family of 
distributions $q(\bpsi)$ instead of working on the intractable 
posterior, and then seeks the member of the family for which the 
lower-bound ${\cal L}(q, p_{h})$ is maximized.

For  cLDA, we define a 
fully factorized variational distribution with the variational 
parameters $\bpsi^{\prime} = (\blambda, \btau, \brho, \bphi)$ as    
\begin{equation}
q(\bpsi \given \bpsi^{\prime}) = 
\Bigl [ \prod_{k = 1}^{K} q(\bbeta_k \given \blambda_k) \Bigr ]  
\Bigl [ \prod_{j = 1}^{J} q(\bpi_j \given \btau_j) 
\Bigl [ \prod_{d = 1}^{D_j} q(\btheta_d \given \brho_d) 
\Bigl (
\prod_{i = 1}^{n_{jd}} q(\bz_{di} \given \bphi_{jdi}  ) 
\Bigr )
\Bigr ] 
\Bigr ]  
\label{eq:variational-posterior}
\end{equation}
We take its independent component distributions 
\cite{BleiNgJordan:2003} are  
\begin{equation}
\begin{split}
\bbeta_k &\sim \Dir_{V}(\blambda_k) \\ 
\bpi_j &\sim \Dir_{K}(\btau_j) \\ 
\btheta_{jd} &\sim \Dir_{K}(\brho_{jd}) \\ 
\bz_{jdi} &\sim \Mult_{K}(\bphi_{jdi}) \\ 
\end{split}
\label{eq:variational-distributions}
\end{equation}
Note that the lower-bound 
\eqref{eq:variational-lowerbound-original}, which is an expectation 
with respect to \eqref{eq:variational-posterior}, is intractable 
due to the non-conjugate relationships between 
$\bpi_j$'s and $\btheta_{jd}$'s. 
We will also see that estimation of variational parameters 
$(\blambda, \brho, \bphi)$ follows closely to  
\cite{BleiNgJordan:2003}'s scheme, but updating 
parameter $\btau$ does not have a closed form expression. 
\cite{Kim:2013} proposed a solution to an expectation of 
similar form in a different context, which we employ here. 
The corresponding variational expectation maximization 
scheme for cLDA is described in Algorithm \ref{alg:vem} and is 
denoted by the acronym VEM.

\begin{algorithm}[t]
	\KwData{Observed words $\bw$ and document metadata}
	\KwResult{Optimal variational parameters $(\blambda, \btau, 
	\brho, 
		\bphi)$} \vspace{.3cm}
	
	initialize $(\blambda^{(0)}, \btau^{(0)})$\;
	
	\While{not converged}{ 
		
		\tcp{Step $1$: Expectation} 
		
		\vspace{.1cm}
		
		\For{document $d = 1, \ldots, D_j$, $j = 1, \ldots, J$}{
			
			initialize $\rho_{jdk}^{(0)} = \frac{\gamma \tau_{jk}}
			{\tau_{j.}} + \frac{n_{jd}}{K}, \, k = 1, \ldots, K$\;
			
			\tcp{Variational updates for each document}
			
			\While{not converged}{ 
				
				\For{word $w_{jdi}$, $i = 1, \ldots, n_{jd}$}{
					variational Multinomial update for 
					$\bphi_{jdi}$ 
					via~\eqref{eq:var-multinomial-phi}\; 
				}
				variational Dirichlet update for $\brho_{jd}$ via 
				\eqref{eq:var-dirichlet-rho}\; 
				
			}
			
		}
		
		\tcp{Variational updates for each topic}
		
		variational Dirichlet update for $\blambda_{k}, \, k = 1, 
		\ldots, K$ via~\eqref{eq:var-dirichlet-lambda}\; 
		
		\vspace{.2cm}
		
		\tcp{Step $2$: Maximization}
		
		\vspace{.1cm}
		
		\tcp{Constraint Newton updates for collection-level topic 
			mixtures}
		
		\For{collection $j = 1, \ldots, J$}{
			initialize $(a^{(0)}_j, \bomega^{(0)}_j)$ based on the 
			current $\btau_j$\; 
			\While{not converged}{ 
				constraint Newton update for $\bomega_{j}$ via 
				\eqref{eq:constraint-newton-step}\; 
				Newton update for $a_j$ 
				via~\eqref{eq:newton-step-a-j}\; 
			}
			set $\btau_j = a^{(\text{final})}_j * 
			\bomega^{(\text{final})}_j$ 
		}
		optimize hyperparameter $h = (\alpha, \gamma, \eta)$\; 
		
	}
	\caption{Variational expectation maximization (VEM)}
	\label{alg:vem}
\end{algorithm}

We now describe a way to estimate the variational parameters 
$(\blambda, \btau, \brho, \bphi)$ via minimizing the KL divergence 
between the posterior $p_{h,\bw}(\bpsi)$ and the variational 
distribution $q(\bpsi \given \bpsi^{\prime})$. We first write down 
the variational 
lower-bound~\eqref{eq:variational-lowerbound-original}  as 
follows.  
\begin{equation}
\begin{split}
{\cal L}(q, p_{h}) &= 
\sum_{k = 1}^{K} \Exp_{q_k}[ \log p_{\eta}(\bbeta_k) ] 
+ \sum_{j = 1}^{J} \Exp_{q_j}[ \log p_{\alpha}(\bpi) ] 
+ \sum_{j = 1}^{J} \sum_{d = 1}^{D_j} \Exp_{q_{jd}}[ \log 
p_{\gamma}(\btheta_{jd} \given \bpi_{j}) ] \\ 
&+ \sum_{j = 1}^{J} \sum_{d = 1}^{D_j} \sum_{i = 1}^{n_{dj}} 
\Exp_{q_{jdi}}[ \log p(z_{jdi} \given \btheta_{jd}) ] 
+ \sum_{j = 1}^{J} \sum_{d = 1}^{D_j} \sum_{i = 1}^{n_{dj}} 
\Exp_{q_{jdi}}[ \log p(w_{jdi} \given z_{jdi}, \bbeta) ] \\ 
&- \sum_{k = 1}^{K} \Exp_{q_k}[ \log q(\bbeta_k \given \blambda_k) 
] 
- \sum_{j = 1}^{J} \Exp_{q_j}[ \log q(\bpi_j \given \btau_j)  ] \\
&- \sum_{j = 1}^{J} \sum_{d = 1}^{D_j} \Exp_{q_{jd}}[ \log 
q(\btheta_{jd} \given \brho_{jd}) ]
- \sum_{j = 1}^{J} \sum_{d = 1}^{D_j} \sum_{i = 1}^{n_{dj}} 
\Exp_{q_{jdi}}[ \log q(\bz_{jdi} \given \bphi_{jdi})  ] \\ 
\end{split}
\label{eq:variational-lowerbound-full}
\end{equation}

To ease notation, we denote $\beta_{k.} = \sum_{v = 1}^{V} 
\beta_{kv}$. Let $\btheta \sim \Dir_L (\brho)$. We evaluate 
the expectation of the log of a single probability component 
$\theta_k$ analytically, using \cite[Appendix 
A.1]{BleiNgJordan:2003}: 
\[\Exp_{q}[ \log \theta_k \given \rho_k ] = \varPsi(\rho_{k}) - \varPsi(\rho_{.})\] 
Similarly, we evaluate the following expectations analytically:   
\begin{equation}
\begin{split}
\Exp_{q}[ \log \beta_{kv} \given \lambda_{kv} ] &= \varPsi(\lambda_{kv}) - \varPsi(\lambda_{k.}) \\ 
\Exp_{q}[ \log \pi_{jk} \given \tau_{jk} ] &= \varPsi(\tau_{jk}) - \varPsi(\tau_{j.}) \\ 
\Exp_q[\log \theta_{jdk} \given \rho_{jdk}  ] &= \varPsi(\rho_{jdk}) - \varPsi(\rho_{jd.}) \\ 
\end{split}
\end{equation}

\noindent
Let $\btheta \sim \Dir_L (\brho)$, 
$\Exp [\theta_k] = \rho_k / \rho_{.}$, and $\alpha \in [0, \infty)$. 
We can expand the intractable expectation $\Exp[ \lgamma (\alpha 
\theta_k)]$ as~\citep[Theorem 3.1]{Kim:2013} 
\begin{equation}
\Exp[ \lgamma (\alpha \theta_k)] \leq 
\lgamma (\alpha \Exp[\theta_k]) + 
\frac{\alpha}{\rho_{.}}(1 - \Exp[\theta_k]) + 
(1 - \alpha\Exp[\theta_k]) \Bigl[ \log \Exp[\theta_k] + \varPsi(\rho_{.}) - \varPsi(\rho_{k}) \Bigr] 
\label{eq:kim-lemma}
\end{equation}    

\noindent
We then write the individual expectations as follows:  
\begin{equation}
\begin{split}
\Exp_{q_k}[ \log p_{\eta}(\bbeta_k) ] &=
\lgamma (V \eta) - 
V \lgamma (\eta) + 
\sum_{v = 1}^{V} (\eta - 1) \bigl [ \varPsi(\lambda_{kv}) - \varPsi(\lambda_{k.}) \bigr ] \\ 
\Exp_{q_j}[ \log p_{\alpha}(\bpi) ] &=
\lgamma (K \alpha) - K \lgamma (\alpha) + 
\sum_{k = 1}^{K} (\alpha - 1) \bigl [ \varPsi(\tau_{jk}) - \varPsi(\tau_{j.}) \bigr ] \\ 
\Exp_{q_{jd}}[ \log p_{\gamma}(\btheta_{jd} \given \bpi_{j}) ] &= 
\Exp_q[ \lgamma (\gamma)] - \sum_{k - 1}^{K}  \Exp_q[\lgamma (\gamma \pi_{jk})]  
+ \sum_{k = 1}^{K} \Exp_q[ (\gamma \pi_{jk} - 1) \log \theta_{jdk} ] \\ 
&\geq \lgamma (\gamma) - \sum_{k = 1}^{K} 
\Bigl [ 
\lgamma (\gamma \Exp_q[\pi_{jk}] ) + \frac{\gamma}{\tau_{j.}} (1 - \Exp_q[\pi_{jk}]) \\ 
&\hspace{3.2cm} + (1 - \gamma \Exp_q[\pi_{jk}]) \bigl [ \log \Exp_q[\pi_{jk}] + \varPsi(\tau_{j.}) - \varPsi(\tau_{jk}) \bigr ] 
\Bigr ] \\ 
&\hspace{2cm} + \sum_{k = 1}^{K} 
\Bigl [ 
\gamma \Exp_q[\pi_{jk}] \Exp_q[\log \theta_{jdk}] - \Exp_q[\log \theta_{jdk}] 
\Bigr ] \\ 
&\geq \lgamma (\gamma) - \frac{\gamma}{\tau_{j.}} (K - 1) 
- (\gamma - K) \Bigl [
\log \tau_{j.} - \varPsi(\tau_{j.}) + \varPsi(\rho_{jd.})  
\Bigr ] \\
&\hspace{2cm} - \sum_{k = 1}^{K} 
\Bigl [  
\lgamma (\frac{\gamma \tau_{jk}}{\tau_{j.}}) + (1 - \frac{\gamma \tau_{jk}}{\tau_{j.}}) \bigl [ \log (\tau_{jk}) - \varPsi(\tau_{jk}) + \varPsi(\rho_{jdk}) \bigr ] 
\Bigr ] \\ 
\Exp_{q_{jdi}}[ \log p(z_{jdi} \given \btheta_{jd}) ] &=
\sum_{k = 1}^{K} \phi_{jdik} \bigl [ \varPsi(\rho_{jdk}) - \varPsi(\rho_{jd.}) \bigr ] \\ 
\Exp_{q_{jdi}}[ \log p(w_{jdi} \given z_{jdi}, \bbeta) ] &=
\sum_{k = 1}^{K} \sum_{v = 1}^{V} \phi_{jdik} w_{jdiv} 
\bigl [ \varPsi(\lambda_{kv}) - \varPsi(\lambda_{k.}) \bigr ] \\ 
\Exp_{q_k}[ \log q(\bbeta_k \given \blambda_k) ] &=
\lgamma (\lambda_{k.}) - \sum_{v = 1}^{V} \lgamma (\lambda_{kv}) + 
\sum_{v = 1}^{V} (\lambda_{kv} - 1) \bigl [ \varPsi(\lambda_{kv}) - \varPsi(\lambda_{k.}) \bigr ] \\ 
\Exp_{q_j}[ \log q(\bpi_j \given \btau_j)  ]  &=
\lgamma (\tau_{j.}) - \sum_{k = 1}^{K} \lgamma (\tau_{jk}) + 
\sum_{k = 1}^{K} (\tau_{jk} - 1) \bigl [ \varPsi(\tau_{jk}) - \varPsi(\tau_{j.}) \bigr ] \\ 
\Exp_{q_{jd}}[ \log q(\btheta_{jd} \given \brho_{jd})  ] &= 
\lgamma (\rho_{jd.}) - \sum_{k = 1}^{K} \lgamma (\rho_{jdk}) + 
\sum_{k = 1}^{K} (\rho_{jdk} - 1) \bigl [ \varPsi(\rho_{jdk}) - \varPsi(\rho_{jd.}) \bigr ] \\ 
\Exp_{q_{jdi}}[ \log q(\bz_{jdi} \given \bphi_{jdi}) ] &=  
\sum_{k = 1}^{K} \phi_{jdik} \log \phi_{jdik} \\ 
\end{split}
\label{eq:var-exp}
\end{equation} 
For $\Exp_{q_{jd}}[ \log p_{\gamma}(\btheta_{jd} \given \bpi_{jd})]$, 
the second step uses~\eqref{eq:kim-lemma} and the independence 
assumption of the variational distribution, 
$\Exp_q[ \pi_{jk} \log \theta_{jdk} ] = \Exp_q[ \pi_{jk} ] 
\Exp_q[ \log \theta_{jdk} ]$, and the third step uses the result 
$\Exp_q[ \pi_{jk} ] = \tau_{jk} / \tau_{j.}$.  

We have the variational parameters $(\blambda, \btau, \brho, \bphi)$
for the latent variables $(\bbeta, \bpi, \btheta, \bz)$ in the cLDA 
model. We will see in the following subsections that the updates 
for the variational parameters, except for $\btau_j$'s, 
follow closely that of the variational Dirichlet and Multinomial 
updates of the LDA model~\citep[Appendix]{BleiNgJordan:2003}.

\subsection{Variational Dirichlet Update for Collections} 
\label{sec:var-dir-tau}

Grouping the expectations that contain $\btau_j$ from the 
lower-bound ${\cal L}$\footnote{We ignore the arguments of ${\cal 
L}$ to ease notation.}, we get: 
\begin{equation}\nonumber 
{\cal L}_{[\tau_j]} = \Exp_{q_j}[ \log p_{\alpha}(\bpi) ] 
- \Exp_{q_j}[ \log q(\bpi_j \given \btau_j)  ] 
+ \sum_{d = 1}^{D_j} \Exp_{q_{jd}}[ \log p_{\gamma}(\btheta_{jd} \given \bpi_{j}) ] 
\end{equation}
This lower-bound does not produce a closed form expression for   
updating $\tau_j$'s. \cite{Kim:2013} suggested to use a Newton's 
update with equality constraints for update in a similar 
hierarchical modeling context. A similar approach is followed here. We first 
break down each Dirichlet parameter $\btau_j$ into a scale  
parameter $a_j$ and a base measure $\bomega_j$ that satisfies the 
equality constraint $\sum_{k = 1}^{K} \omega_{jk} = 1$. The 
corresponding variational distribution for $\bpi_j$ is redefined as 
$\bpi_j \sim \Dir_{K}(a_j \bomega_j)$. We will see 
that this decomposition will enable us to perform a Newton's 
update with equality constraints.    
Utilizing the equality constraint and collecting terms that contain 
$a_j$ and $\bomega_{j}$, we get
\begin{equation} 
\begin{split}
{\cal L}_{[a_j \bomega_j]} &= 
\sum_{k = 1}^{K} \Bigl[ 
(\alpha - a_j \omega_{jk})  \bigl[ \varPsi(a_j \omega_{jk}) - \varPsi(a_j) \bigr]  + 
\lgamma (a_j \omega_{jk}) 
\Bigr] - 
\lgamma (a_j ) \\ 
&\hspace{.4cm} - \sum_{d = 1}^{D_j} \Bigl[ 
\frac{\gamma}{a_j} (K - 1) 
+ (\gamma - K) \bigl [
\log a_j - \varPsi(a_j) + \varPsi(\rho_{jd.})  
\bigr ] \Bigr]\\
&\hspace{.4cm} - \sum_{d = 1}^{D_j} \sum_{k = 1}^{K} \Bigl[ 
\lgamma (\gamma \omega_{jk}) + (1 - \gamma \omega_{jk}) \bigl [ \log (a_j \omega_{jk}) - \varPsi(a_j \omega_{jk}) + \varPsi(\rho_{jdk}) \bigr ] 
\Bigr]
\end{split}
\end{equation}
To maximize ${\cal L}_{[a_j \bomega_j]}$ with respect to $\omega_{jk}$, 
we first form the objective function for $\omega_{jk}$ by collecting 
terms as 
\begin{equation} 
\begin{split}
{\cal L}_{[\omega_{jk}]} &= 
\varPsi(a_j \omega_{jk}) \Bigl[ \alpha + D_j - a_j \omega_{jk} - \gamma D_j \omega_{jk} \Bigr]  
+ \gamma \omega_{jk} \sum_{d = 1}^{D_j} \varPsi(\rho_{jdk}) \\ 
&\hspace{.4cm} + \lgamma (a_{j} \omega_{jk}) 
- D_j \Bigl[ \lgamma (\gamma \omega_{jk}) + (1 - \gamma \omega_{jk}) \log (a_{j} \omega_{jk}) \Bigr]  
\end{split}
\end{equation} 
Its first and second derivatives, denoted by $g_{jk}$ and $h_{jk}$, 
are: 
\begin{eqnarray} 
\frac{\partial}{\partial \omega_{jk}} {\cal L}_{[\omega_{jk}]} 
&=& a_j\varPsi^{\prime}(a_j \omega_{jk}) \Bigl[ \alpha + D_j - a_j \omega_{jk} - \gamma D_j \omega_{jk} \Bigr]  
-  \gamma D_j \varPsi(a_j \omega_{jk}) + \gamma \sum_{d = 1}^{D_j} \varPsi(\rho_{jdk}) \nonumber \\ 
&& - \: D_j \Bigl[ \gamma \varPsi(\gamma \omega_{jk}) + \frac{1}{\omega_{jk}}  - \gamma - \gamma \log (a_{j} \omega_{jk}) \Bigr]  
\label{eq:omega-grad} \\ 
\frac{\partial^{2}}{\partial \omega_{jk}} {\cal L}_{[\omega_{jk}]} 
&=& a_j^2\varPsi^{\prime\prime}(a_j \omega_{jk}) \Bigl[ \alpha + D_j - a_j \omega_{jk} - \gamma D_j \omega_{jk} \Bigr] 
- a_{j}\varPsi^{\prime}(a_j \omega_{jk}) \Bigl[ a_j + 2 \gamma D_j \Bigr] \nonumber \\ 
&& - \: D_j \Bigl[ \gamma^2 \varPsi^{\prime}(\gamma \omega_{jk}) - \frac{1}{\omega^2_{jk}} - \frac{\gamma}{\omega_{jk}}  \Bigr]  
\label{eq:omega-hess}
\end{eqnarray}
We can see that the hessian given by 
\eqref{eq:omega-hess} is diagonal. We use $u$ to denote the dual 
variable for the sums to one constraint. We then form  
the constraint Newton step $\Delta \omega_{jk}$ by solving the set 
of linear equations   
\begin{equation} 
\begin{bmatrix} \text{diag}(\bm{h}) & \mathbf{1} \\ {\mathbf{1}}^{\text{T}} & 0 \end{bmatrix}
\left[ \begin{array}{c} \Delta \omega_{jk} \\ u \end{array} \right] =  
\left[ \begin{array}{c} -\bm{g} \\ 0 \end{array} \right],
\label{eq:constraint-newton}
\end{equation}
that yields  
\begin{equation} 
\Delta \omega_{jk} = \Bigl\{ \frac{ \sum_{k=1}^{K} 
\frac{g_{jk}}{h_{jk}} }{ \sum_{k=1}^{K} \frac{1}{h_{jk}} } \Bigr\}   
\left[ 
\begin{array}{c} 
\frac{1}{h_{j1}} \\ 
\cdots \\ 
\frac{1}{h_{jK}} 
\end{array} 
\right] - 
\left[ 
\begin{array}{c} 
\frac{g_{j1}}{h_{j1}} \\ 
\cdots \\ 
\frac{g_{jK}}{h_{jK}}
\end{array} 
\right]
\label{eq:constraint-newton-step}
\end{equation}
By construction, this update satisfies $\sum_{k=1}^{K} \Delta 
\omega_{jk} = 0$ and preserves the sums to one constraint of 
the variational parameter $\omega_{jk}$~\citep{Kim:2013}. 

To maximize ${\cal L}_{[a_j \bomega_j]}$ with respect to $a_j$, 
we first form the objective function for $a_j$, by collecting 
terms as 
\begin{equation} 
\begin{split}
{\cal L}_{[a_j]} &= 
\sum_{k = 1}^{K} \bigl[ \varPsi(a_j \omega_{jk}) - \varPsi(a_j) \bigr]
\Bigl( \alpha + D_j - a_j \omega_{jk} - \gamma D_j \omega_{jk} \Bigr)  
+ \sum_{k = 1}^{K} \lgamma (a_j \omega_{jk}) \\ 
&\hspace{.4cm}  - \lgamma (a_j) - \frac{\gamma D_j (K - 1)}{a_j}
\end{split}
\end{equation} 
Its first and second derivatives, denoted by $g_{j}$ and $h_{j}$, 
are: 
\begin{eqnarray} 
\frac{\partial}{\partial a_j} {\cal L}_{[a_j]} 
&=& \sum_{k = 1}^{K} \bigl[ \omega_{jk} \varPsi^{\prime}(a_j \omega_{jk}) - \varPsi^{\prime}(a_j) \bigr]
\Bigl( \alpha + D_j - a_j \omega_{jk} - \gamma D_j \omega_{jk} \Bigr) \nonumber \\ 
&& + \: (K - 1) \gamma D_j a_j^{-2} 
\label{eq:a-grad} \\ 
\frac{\partial^{2}}{\partial a_j} {\cal L}_{[a_j]} 
&=& \sum_{k = 1}^{K} \bigl[ \omega^2_{jk} \varPsi^{\prime\prime}(a_j \omega_{jk}) - \varPsi^{\prime\prime}(a_j) \bigr]
\Bigl( \alpha + D_j - a_j \omega_{jk} - \gamma D_j \omega_{jk} \Bigr) \nonumber \\ 
&& - \: \sum_{k = 1}^{K}  \omega_{jk} \bigl[ \omega_{jk} \varPsi^{\prime}(a_j \omega_{jk}) - \varPsi^{\prime}(a_j) \bigr]
- 2(K - 1) \gamma D_j a_j^{-3}
\label{eq:a-hess}
\end{eqnarray}
The Newton update step for $a_j$ is then given by 
\begin{equation}
\Delta a_j = - h^{-1}_{j} g_{j}. 
\label{eq:newton-step-a-j}
\end{equation}
We alternately maximize ${\cal L}_{[a_j \bomega_j]}$ with 
respect to $a_j$ and $\bomega_j$ until convergence.

\subsection{Variational Multinomial Update for Words} %
\label{sec:var-dir-phi}

We derive the expression for updating the variational parameter 
$\phi_{jdik}$---the probability that $jdi^{\text{th}}$ word is 
generated by topic $k$---via maximizing the lower-bound ${\cal L}$ 
with respect to the constraint $\sum_{k=1}^{K} \phi_{jdik} = 1$.   
We first form the Lagrangian by collecting the terms that contain 
$\phi_{jdik}$ from~\eqref{eq:var-exp}, and applying 
Lagrangian multipliers as
\begin{equation} 
\begin{split}
{\cal L}_{[\phi_{jdik}]} 
&= \phi_{jdik} \Bigl[ [\varPsi(\rho_{jdk}) - \varPsi(\rho_{jd.})] + 
 [\varPsi(\lambda_{kv}) - \varPsi(\lambda_{k.})] - \log \phi_{jdik} \Bigr] \\ 
&+ \mu_{jdi} \Bigl[ \sum_{k=1}^{K} \phi_{jdik} - 1 \Bigr]
\end{split}
\end{equation}
Taking derivatives with respect to $\phi_{jdik}$, we get 
\begin{equation} 
\frac{\partial}{\partial \phi_{jdik}}{\cal L}_{[\phi_{jdik}]} 
= [\varPsi(\rho_{jdk}) - \varPsi(\rho_{jd.})] + 
 [\varPsi(\lambda_{kv}) - \varPsi(\lambda_{k.})] - \log \phi_{jdik} 
- 1 + \mu_{jdi} 
\end{equation}
Setting this to zero yields the maximum value for $\phi_{jdik}$ 
\begin{equation} 
\phi_{jdik} 
\propto \exp \Bigl( [\varPsi(\rho_{jdk}) - \varPsi(\rho_{jd.})] + 
 [\varPsi(\lambda_{kv}) - \varPsi(\lambda_{k.})] \Bigr)
\label{eq:var-multinomial-phi}
\end{equation}
  
\subsection{Variational Dirichlet Update for Documents}  %
\label{sec:var-dir-rho}

We maximize the lower-bound ${\cal L}$ with respect to $\rho_{jdk}$. 
Collecting terms that contain $\rho_{jd}$ from~\eqref{eq:var-exp}, 
we get 
\begin{equation} 
\begin{split}
{\cal L}_{[\rho_{jd}]} 
&= \sum_{k = 1}^{K} \Bigl( \frac{\gamma \tau_{jk}}{\tau_{j.}} + \sum_{i = 1}^{n_{jd}}\phi_{jdik} - \rho_{jdk} \Bigr)
 \bigl [ \varPsi(\rho_{jdk}) - \varPsi(\rho_{jd.}) \bigr ] \\ 
&- \lgamma (\rho_{jd.}) + \sum_{k = 1}^{K} \lgamma (\rho_{jdk})
\end{split}
\end{equation}
Taking derivatives with respect to $\rho_{jdk}$, we get 
\begin{equation} 
\frac{\partial}{\partial \rho_{jdk}}{\cal L}_{[\rho_{jd}]}
=  \bigl [ \varPsi^{\prime}(\rho_{jdk}) - \varPsi^{\prime}(\rho_{jd.}) \bigr ] \Bigl( \frac{\gamma \tau_{jk}}{\tau_{j.}} + \sum_{i = 1}^{n_{jd}}\phi_{jdik} - \rho_{jdk} \Bigr)
\end{equation}
Setting this to zero yields the maximum value for $\rho_{jdk}$
\begin{equation} 
\rho_{jdk} = \frac{\gamma \tau_{jk}}{\tau_{j.}} + \sum_{i = 1}^{n_{jd}}\phi_{jdik}
\label{eq:var-dirichlet-rho}
\end{equation}

\subsection{Variational Dirichlet Update for Topics} 
\label{sec:var-dir-lambda}

We maximize the lower-bound ${\cal L}$ with respect to 
$\lambda_{kv}$. Collecting terms that contain $\blambda_{k}$ from 
\eqref{eq:var-exp}, we get 
\begin{equation} 
\begin{split}
{\cal L}_{[\blambda_{k}]} 
&= \sum_{k = 1}^{K} \sum_{v = 1}^{V} \Bigl( \eta - \lambda_{kv} + \sum_{j = 1}^{J} \sum_{d = 1}^{D_j} \sum_{i = 1}^{n_{jd}}\phi_{jdik} w_{jdiv} \Bigr)
 \bigl [ \varPsi(\lambda_{kv}) - \varPsi(\lambda_{k.}) \bigr ] \\ 
&- \sum_{k = 1}^{K} \bigl [ \lgamma (\lambda_{k.}) - \sum_{v = 1}^{V} \lgamma (\lambda_{kv}) \bigr ]
\end{split}
\end{equation}
Taking derivatives with respect to $\lambda_{kv}$, we get 
\begin{equation} 
\frac{\partial}{\partial \lambda_{kv}}{\cal L}_{[\blambda_{k}]}
=  \bigl [ \varPsi^{\prime}(\lambda_{kv}) - \varPsi^{\prime}(\lambda_{k.}) \bigr ] \Bigl( \eta - \lambda_{kv} + \sum_{j = 1}^{J} \sum_{d = 1}^{D_j} \sum_{i = 1}^{n_{jd}}\phi_{jdik} \Bigr)
\end{equation}
Setting this to zero yields the maximum value for $\lambda_{kv}$
\begin{equation} 
\lambda_{kv} = \eta + \sum_{j = 1}^{J} \sum_{d = 1}^{D_j} \sum_{i = 1}^{n_{jd}}\phi_{jdik} 
\label{eq:var-dirichlet-lambda}
\end{equation}

\subsection{Optimize Hyperparameters} 
\label{sec:opt-h}

As in the variational EM algorithm of 
LDA~\citep{BleiNgJordan:2003}, 
the E-step of the cLDA VEM algorithm updates the variational 
parameters based on the expressions provided above (see 
Algorithm~\ref{alg:vem}). We can use the optimal lower-bound ${\cal 
L}(q^*, p_h)$ as the tractable approximation for the log marginal 
likelihood $\log m(h)$. In the M-step of VEM, we can then update the 
hyperparameters $h = (\alpha, \gamma, \eta)$ by maximizing the 
optimal lower-bound with respect to $h$. We collect terms that 
contain each hyperparameter, separately, as follows: 
\begin{eqnarray} 
{\cal L}_{[\alpha]} &=& J \lgamma (K \alpha) - JK \lgamma (\alpha) + 
\sum_{j = 1}^{J} \sum_{k = 1}^{K} \alpha \bigl [ \varPsi(\tau_{jk}) - 
\varPsi(\tau_{j.}) \bigr ] \\ 
{\cal L}_{[\eta]} &=& K\lgamma (V \eta) - KV \lgamma (\eta) + 
\sum_{k = 1}^{K} \sum_{v = 1}^{V} \eta \bigl [ \varPsi(\lambda_{kv}) - 
\varPsi(\lambda_{k.}) \bigr ]  \\ 
{\cal L}_{[\gamma]} &=& \sum_{j = 1}^{J} D_j [\lgamma (\gamma) - \frac{\gamma}{\tau_{j.}} 
(K - 1)] - \gamma \sum_{j = 1}^{J} \sum_{d = 1}^{D_j} \Bigl [
\log \tau_{j.} - \varPsi(\tau_{j.}) + \varPsi(\rho_{jd.})  
\Bigr ] \nonumber \\
&& - \sum_{j = 1}^{J} \sum_{d = 1}^{D_j} \sum_{k = 1}^{K} 
\Bigl [  
\lgamma (\frac{\gamma \tau_{jk}}{\tau_{j.}}) - 
\frac{\gamma \tau_{jk}}{\tau_{j.}} \bigl [ \log (\tau_{jk}) - 
\varPsi(\tau_{jk}) + \varPsi(\rho_{jdk}) \bigr ] 
\Bigr ]
\end{eqnarray}
The first and second derivatives are the following: 
\begin{eqnarray} 
\frac{\partial}{\partial \alpha}{\cal L}_{[\alpha]} &=& JK \bigl [ 
\varPsi (K \alpha) - \varPsi (\alpha) \bigr ] + 
\sum_{j = 1}^{J} \sum_{k = 1}^{K} \bigl [ \varPsi(\tau_{jk}) - 
\varPsi(\tau_{j.}) \bigr ] \\ 
\frac{\partial^2}{\partial \alpha}{\cal L}_{[\alpha]} &=& JK^2 
\varPsi^{\prime} (K \alpha) - JK \varPsi^{\prime} (\alpha) \\ 
\frac{\partial}{\partial \eta}{\cal L}_{[\eta]} &=& KV \bigl [ 
\varPsi (V \eta) - \varPsi (\eta) \bigr ] + 
\sum_{k = 1}^{K} \sum_{v = 1}^{V} \bigl [ \varPsi(\lambda_{kv}) - 
\varPsi(\lambda_{k.}) \bigr ]  \\ 
\frac{\partial^2}{\partial \eta}{\cal L}_{[\eta]} &=& KV^2
\varPsi^{\prime} (V \eta) - KV \varPsi^{\prime} (\eta) \\ 
\frac{\partial}{\partial \gamma}{\cal L}_{[\gamma]} &=& 
\sum_{j = 1}^{J} D_j [\varPsi (\gamma) - \frac{1}{\tau_{j.}} 
(K - 1)] - \sum_{j = 1}^{J} \sum_{d = 1}^{D_j} \Bigl [
\log \tau_{j.} - \varPsi(\tau_{j.}) + \varPsi(\rho_{jd.})  
\Bigr ] \nonumber \\
&& - \sum_{j = 1}^{J} \sum_{d = 1}^{D_j} \sum_{k = 1}^{K} 
\frac{\tau_{jk}}{\tau_{j.}}
\Bigl [  
\varPsi (\frac{\gamma \tau_{jk}}{\tau_{j.}}) - 
\bigl [ \log (\tau_{jk}) - 
\varPsi(\tau_{jk}) + \varPsi(\rho_{jdk}) \bigr ] 
\Bigr ] \\ 
\frac{\partial^2}{\partial \gamma}{\cal L}_{[\gamma]} &=& 
\sum_{j = 1}^{J} D_j \varPsi^{\prime} (\gamma) 
 - \sum_{j = 1}^{J} \sum_{d = 1}^{D_j} \sum_{k = 1}^{K} 
\frac{\tau^2_{jk}}{\tau^2_{j.}}
\varPsi^{\prime} (\frac{\gamma \tau_{jk}}{\tau_{j.}})
\end{eqnarray}  
Using these derivatives, one can update $\alpha$, $\gamma$, $\eta$ 
via a Newton method~\citep{BleiNgJordan:2003, Minka:2000}.  

%% file: supp-compare-ags-mgs-vem.tex
\section{Comparison of AGS, MGS, and VEM on a Synthetic Corpus}
\label{sec:algorithm-illustration}

\begin{figure*}[t!] 
\centering 
\subfloat[AGS: $\bpi^{(2000)}_{1} = (\epsilon, .996, .003)$]{
\includegraphics[width=.45\linewidth]
{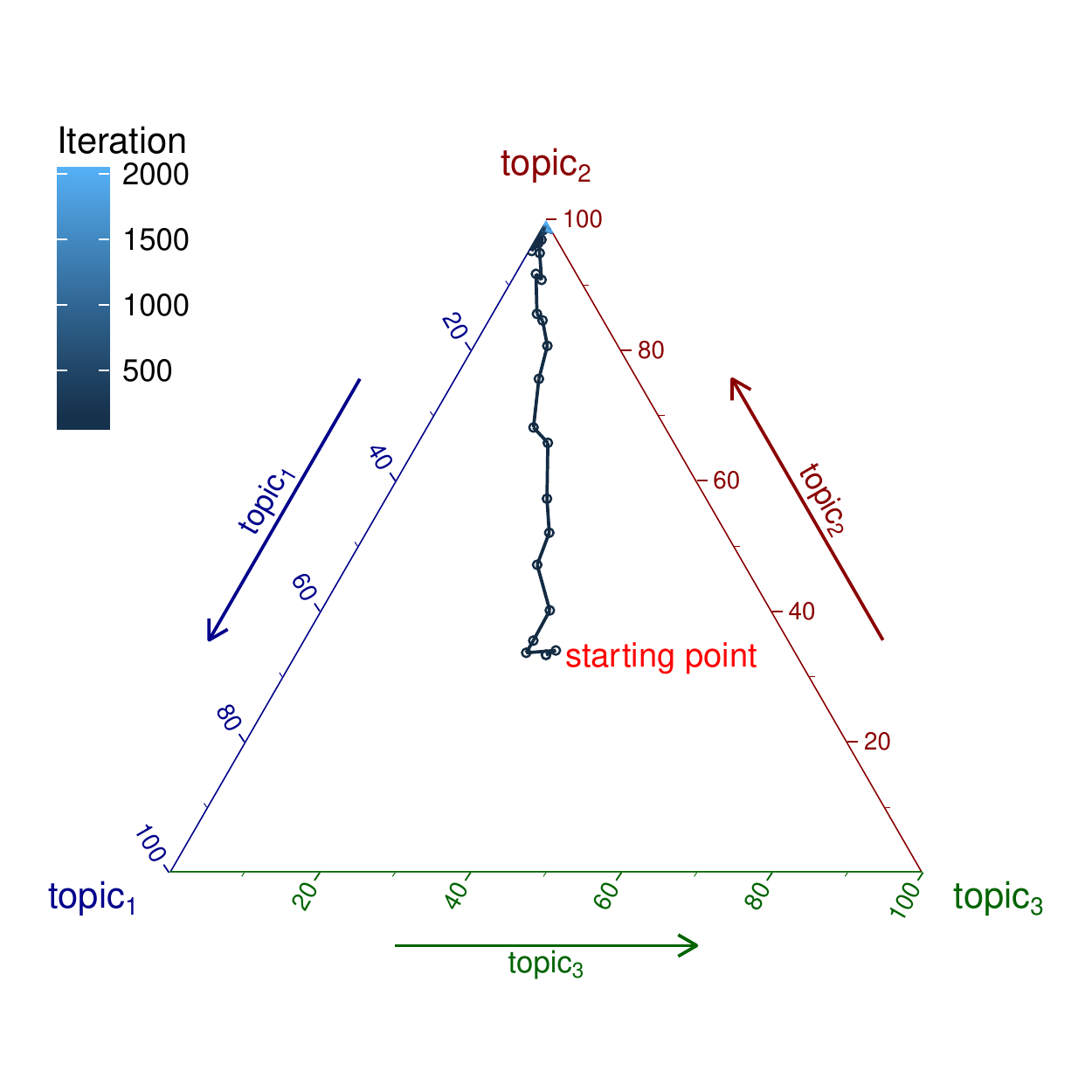}}
\subfloat[MGS: $\bpi^{(2000)}_{1} = (.001, .997, \epsilon)$]{
\includegraphics[width=.45\linewidth]
{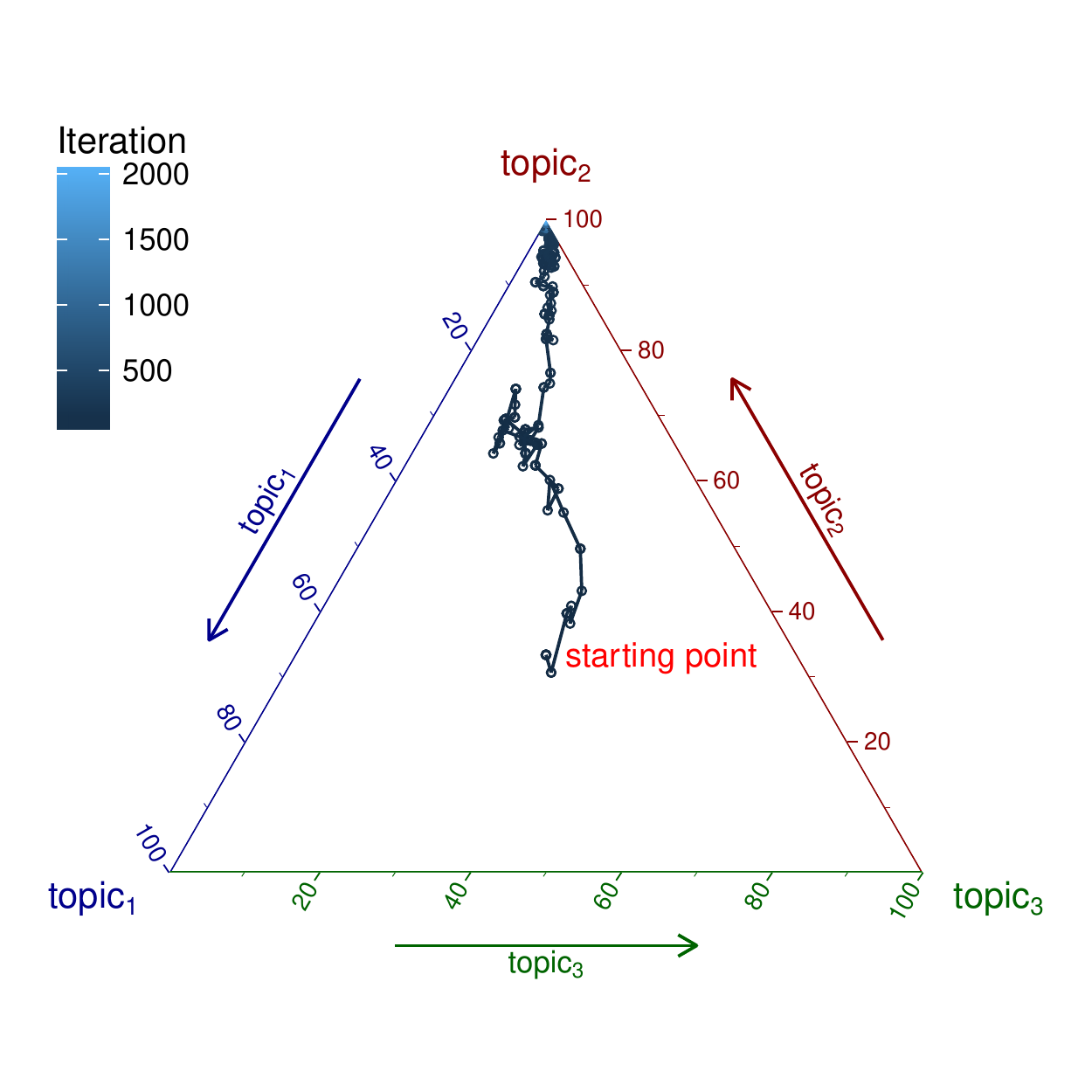}}
\\ \subfloat[VEM: $\bpi^{(45)}_{1} = (.057, .935, .006)$]{
\includegraphics[width=.45\linewidth]
{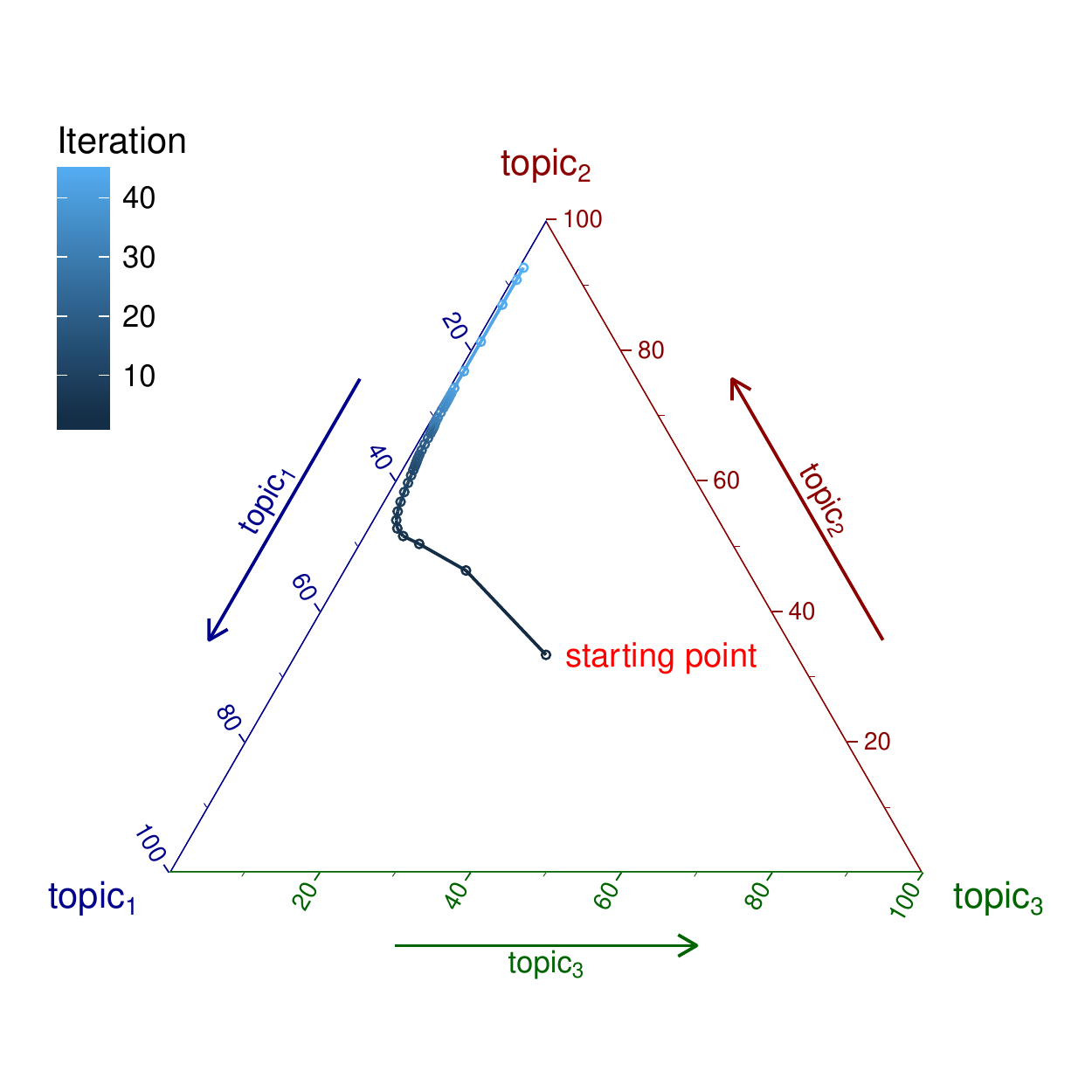}}
 
\caption{Plots of values of $\bpi_{1}$ via algorithms AGS, MGS, and 
VEM. Here, the variable $\epsilon$ denotes a small number. With 
approximately $42$ iterations the AGS chain reached the optimal 
region, i.e., $.003$ from the true value $\bpi^{\text{true}}_{1} = 
(.002, \epsilon, .997)$, but the MGS chain took $248$ iterations to 
reach there. Algorithm VEM never reached the optimal 
regions: at convergence, VEM hit points that are $0.08$ far 
from $\bpi^{\text{true}}_{1}$. See discussion in the text.}
\label{fig:pi-1-random-walk}
\end{figure*}

This section gives additional details for the comparative study 
given in Section~\ref{sec:pi-sampling-illustration} of the main 
paper. 
\begin{notes-to-stay}
Here we compare the performance of three approximate inference 
methods, AGS, MGS, and VEM, for the cLDA model. For this purpose, 
we 
created a synthetic corpus by simulating lines $1$-$5$ of the cLDA 
hierarchical model with the following configurations. We 
took the number of collections $J = 2$ and the number of topics 
$K = 3$. We did this solely so that we can visualize the results of 
the algorithms. Besides, we took the vocabulary size $V = 40$, the 
number of documents in each collection $D_j = 100$, and the 
hyperparameters $h_{\text{true}} = (\alpha, \gamma, \eta) = (.1, 1, 
.25)$. The collection-level Dirichlet sampling 
with hyperparameter $\alpha$ (via line~$2$ of the hierarchical 
model) 
produced two topic distributions $\bpi^{\text{true}}_{1} 
= (.002, .997, \epsilon)$ and $\bpi^{\text{true}}_{2} = 
(.584, .030, .386)$, where $\epsilon$ denotes a small number. 
The data $\bw$ is generated by simulating line $5$.     
  
We first study the ability of these algorithms to recover 
parameters 
$\bpi^{\text{true}}_{1}$ and $\bpi^{\text{true}}_{2}$. We do this 
by 
comparing samples of $\bpi_j$ from the AGS and MGS chains on 
$(\bpi, 
\bz)$ with variational estimates of $\bpi_j$ from the VEM algorithm 
iterations\footnote{An implementation of all algorithms and 
datasets 
discussed in this paper is available as an R package at 
\url{https://github.com/clintpgeorge/clda}}. We initialized 
$\bpi^{(0)}_{1} = \bpi^{(0)}_{2} = (.33, .33, .33)$ for all three 
algorithms. Using the data $\bw$, we ran both AGS and MGS chains 
for 
$2000$ iterations, and the VEM algorithm converged after $45$ EM 
iterations. 
\end{notes-to-stay} 
\figurename~\ref{fig:pi-1-random-walk} gives  
trace plots of values of $\bpi_{1}$ and $\bpi_{2}$ on the 
$2$-simplex for all three algorithms. 
Consider any of the choices of hyperparameters 
$h$ and the number of topics $K$. Markov chains AGS and MGS 
induces a chain on $\bpsi$ with invariant distribution 
$p_{h,\bw}(\bpsi)$ by using the conditional distribution of 
$(\bbeta, \btheta)$, given by~\eqref{eq:theta-beta-posterior} 
(see Section \ref{sec:inference}). They essentially give us a 
sequence $(\bbeta^{(1)}, \bpi^{(1)}, \btheta^{(1)}, \bz^{(1)}), 
\ldots, (\bbeta^{(S)}, \bpi^{(S)}, \btheta^{(S)}, \bz^{(S)})$. 
Consider the component $\bpi_j^{(s)}$ of $\bpi^{(s)}$.  
While both $\bpi_j^{(s)}$ and $\bpi_j^{\text{true}}$ are points in 
the $K$-$1$ simplex, their interpretations are different: 
$\bpi_j^{(s)}$ is a distribution on the $K$ topics 
$\beta_1^{(s)}, \ldots, \beta_K^{(s)}$, while 
$\bpi_j^{\text{true}}$ is a distribution on the $K$ topics
$\beta_1^{\text{true}}, \ldots, \beta_K^{\text{true}}$, and 
these are different sets of topics. This is also the case with 
$K$ components of $\btheta_d^{(s)}$ and $\btheta_d^{\text{true}}$ 
(see, e.g., \cite{GriffithsSteyvers:2004,George:2015}). The results 
of VEM described in the paper hold a similar case.  
Thus, before making any comparison between the values of 
$\bpi_j^{(s)}$ and $\bpi_j^{\text{true}}$, one needs to align 
the corresponding sets of topics. For simple corpora such as the 
one used in our empirical evaluation, one can trivially
re-align topics to compare $\bpi_j^{\text{true}}$ and 
$\bpi_2^{(s)}$s (see Table \ref{tab:ags-mgs-vem}).
%
%

\begin{notes-to-stay} From the plots, we can see that both AGS and 
MGS chains were able to recover the values $\bpi^{\text{true}}_{1}$ 
and $\bpi^{\text{true}}_{2}$ reasonably well, even though the AGS 
chain has an edge. Converging to optimal regions 
$\|\bpi^{\text{true}}_1 - \bpi^{(s)}_1 \| < 10^{-2}$ and 
$\| \bpi^{\text{true}}_2 - \bpi^{(s)}_2 \| < 10^{-2}$, 
the AGS chain  took cycles $42$ and $23$ only; but, the MGS chain 
required cycles $248$ and $139$, respectively. ($\|.\|$ denotes 
L-$1$ norm on the simplex $\simplex_K$.) Lastly, the 
VEM algorithm started off nicely, but never reached the optimal 
regions; at convergence, VEM hit points that are $0.1432$ far 
from $\bpi^{\text{true}}_{1}$ and $0.2507$ far from 
$\bpi^{\text{true}}_{2}$.
\end{notes-to-stay} 

\begin{figure*}[t!] 
  \centering 
  \subfloat[$\pi_{11}$: AGS]{
  \includegraphics[width=.45\linewidth]
  {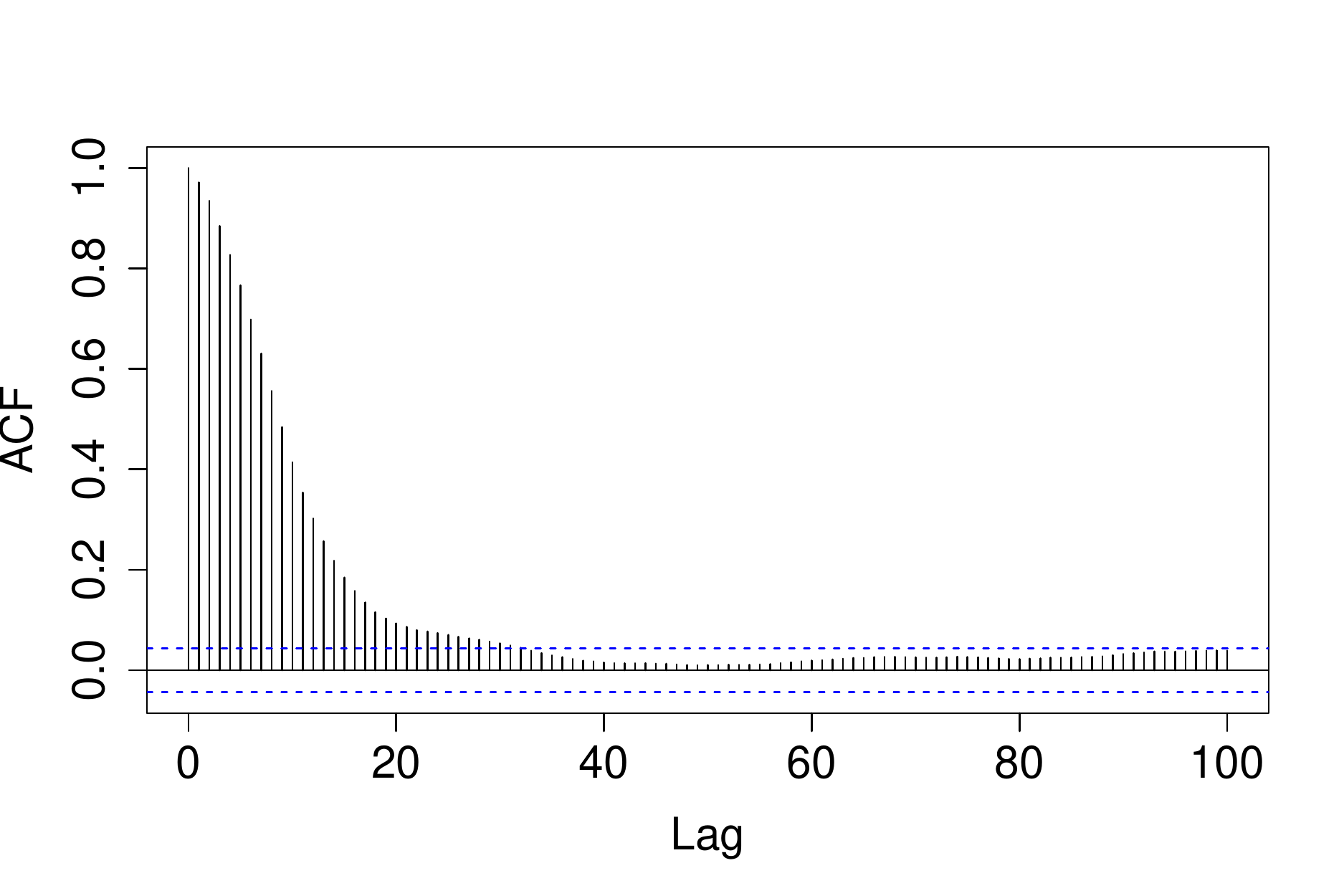}}
  \subfloat[$\pi_{22}$: AGS]{
  \includegraphics[width=.45\linewidth]
  {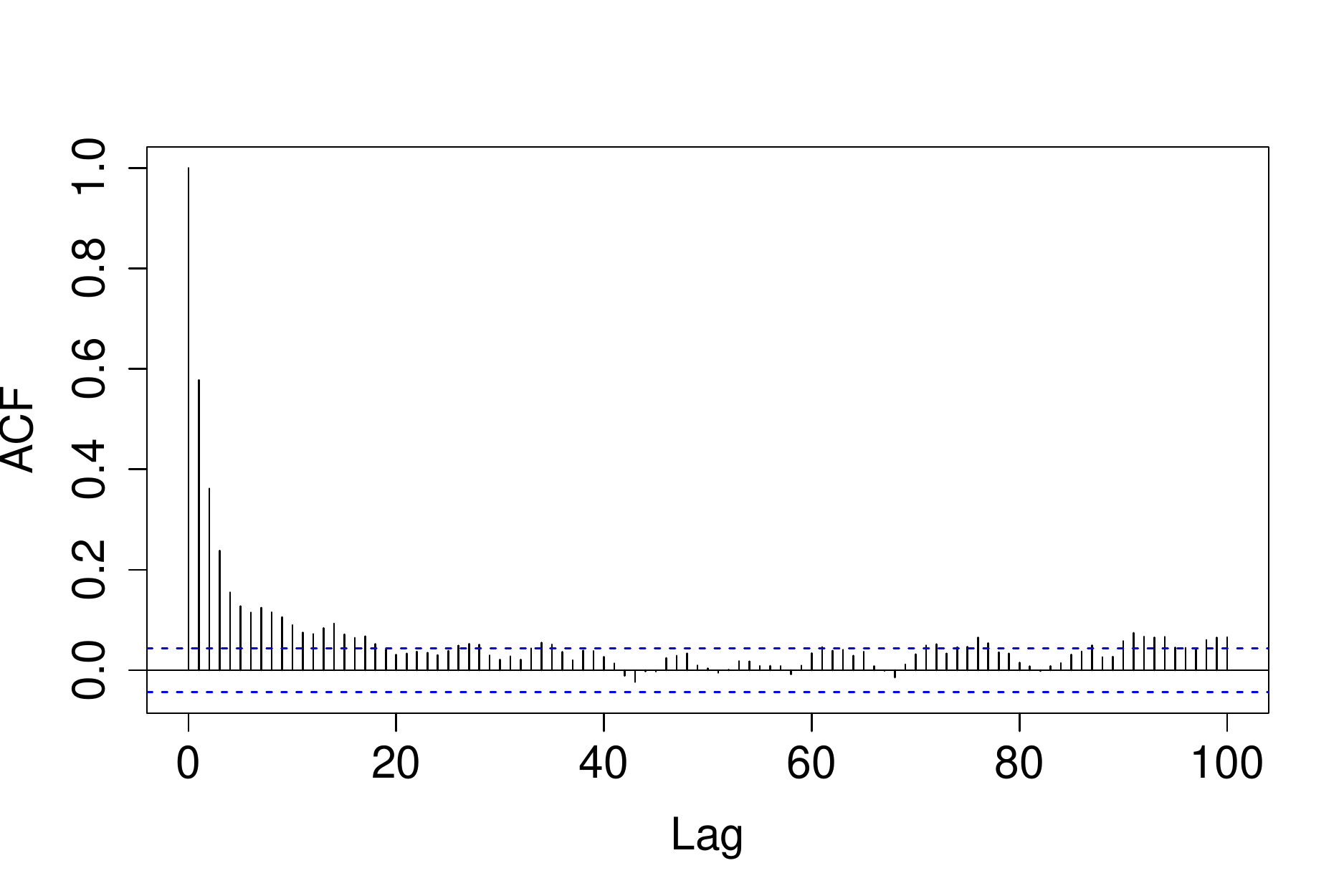}}
  \\ \subfloat[$\pi_{11}$: MGS]{
  \includegraphics[width=.45\linewidth]
  {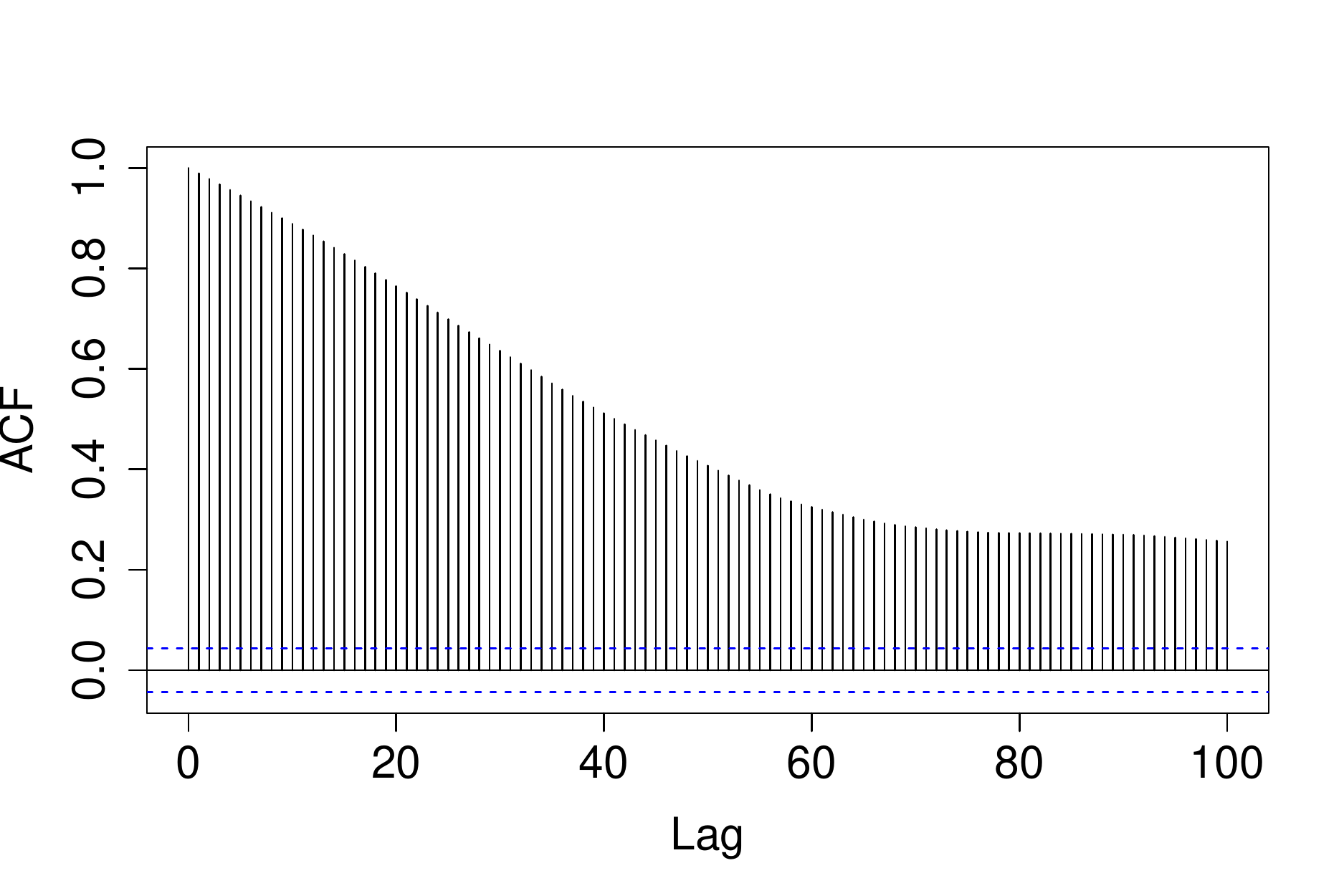}}
  \subfloat[$\pi_{22}$: MGS]{
  \includegraphics[width=.45\linewidth]
  {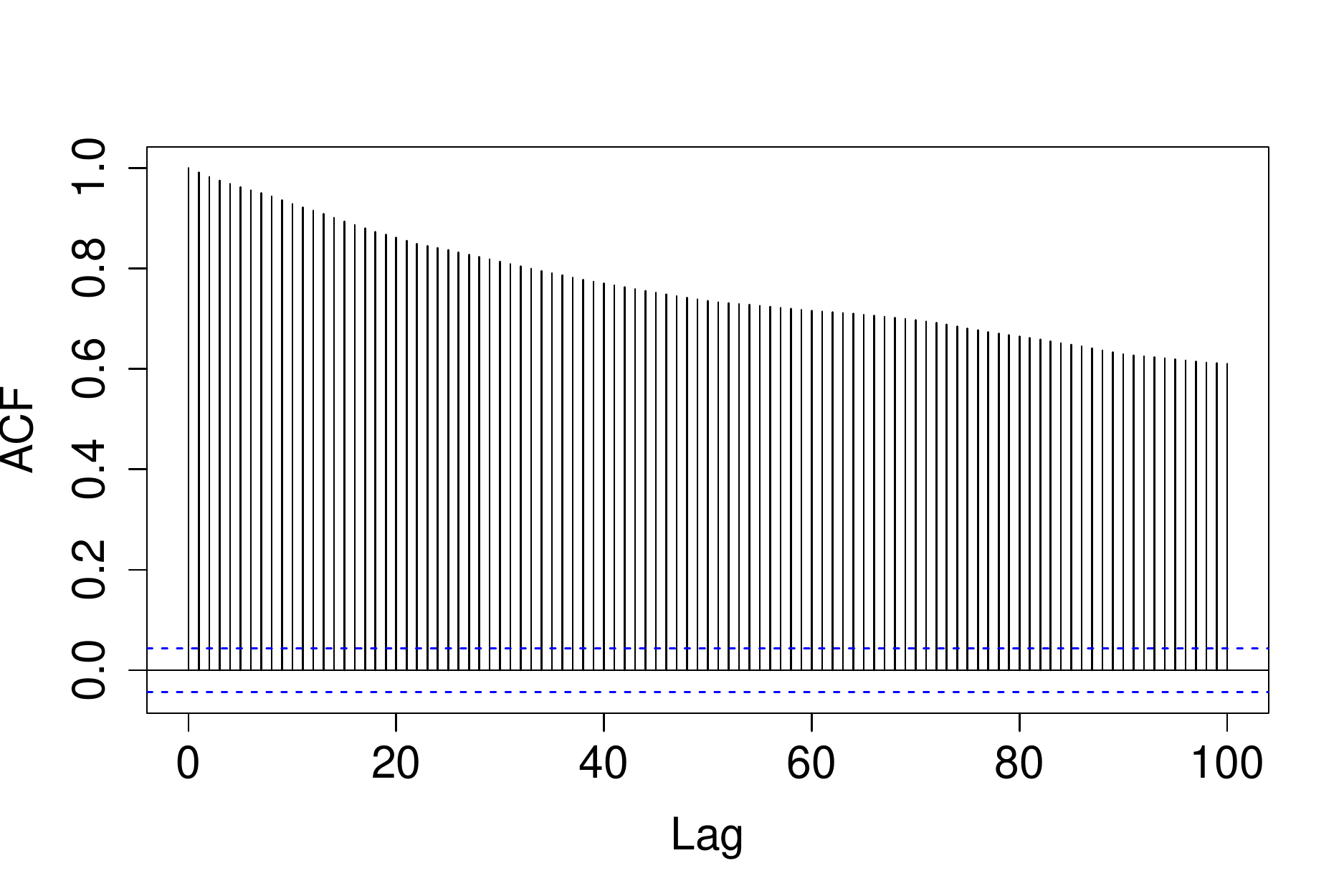}}
  	
  \caption{Plots of ACF for samples $\pi_{11}$ and $\pi_{22}$  
	from the Markov chains MGS and AGS.}
	\label{fig:aux-mmala-acf}
\end{figure*}

We now compare the mixing rates of the two chains AGS and 
MGS. People often use diagnostics such as trace plots and 
auto-correlation function (ACF) plots for this purpose. Although 
both 
chains appear to converge in reasonable cycles, the AGS chain mixes 
faster as shown in  \figurename~\ref{fig:pi-1-random-walk} 
and~\ref{fig:pi-2-random-walk}. \figurename~\ref{fig:aux-mmala-acf} 
further supports this fact: it gives plots of ACF's for two random 
elements $\pi_{11}$ and $\pi_{22}$ of the $\bpi$ matrix, from each 
iteration of the chains AGS and MGS. These plots suggests that the 
AGS chain mixes faster, iterations separated by a lag of $25$ or 
$30$ 
are essentially uncorrelated. Note that even though the VEM 
algorithm 
did not reach the optimal regions here, in our experience, it  
converges relatively quickly. For modeling corpora with large 
document collections, we still recommend the reader to use the VEM 
algorithm as a practical alternative, considering its speed gains 
and 
parallelization capabilities. On the other hand, AGS gives more 
accurate results; hence, we use AGS for our future analysis.  


%

Note that sampling $\bz$'s in both AGS and MGS chains is quite 
similar to sampling $\bz$'s in the LDA CGS 
\citep{GriffithsSteyvers:2004} chain. The CGS chain is a 
well-studied chain, see, e.g.,~\cite{George:2015}; so, we do not 
report diagnostics for samples of $\bz$ from the chains AGS and MGS 
here.

%% file: app-perplexity.tex
\section{Perplexity Calculation for cLDA and LDA}
\label{app:perplexity}

In this section, we derive expressions for Perplexity (defined in 
Section~\ref{sec:eval-performance}) for both cLDA and LDA 
models. To compute the perplexity score~\eqref{eq:c-lda-perplexity}, 
we first need an expression for the predictive 
likelihood $p (w_{jdi} \given \bw^{\text{train}})$. We obtain this 
by marginalizing the likelihood over $\bbeta, \bpi, \btheta$ and 
$z_{jdi}$. There is a slight abuse of notation here: the variables 
$\bbeta, \bpi, \btheta$ are all dependent on the training 
data $\bw^{\text{train}}$ only, but we ignore this in the notation.  
From the hierarchical 
model, we can write the predictive likelihood for word $w_{jdi}$ as 
\begin{equation}
\ell_{w_{jdi}} (\bbeta, \btheta_{jd}, z_{jdi}) 
= \prod_{k = 1}^K \prod_{v = 1}^V \beta_{kv}^{z_{jdik} w_{jdiv}} 
= \sum_{k = 1}^K z_{jdik} \prod_{v = 1}^V \beta_{kv}^{w_{jdiv}}
\end{equation} 
where we used the property that $\prod_{k = 1}^K \prod_{v = 1}^V 
\beta_{kv}^{z_{jdik} w_{jdiv}} = \prod_{v = 1}^V 
\beta_{k^{\prime}v}^{w_{jdiv}}$ for some $k^{\prime}$ such that 
$z_{jdik^{\prime}} = 1$. The 
dependence of the likelihood on $\bpi_j$'s is ignored in the  
notation as it's included in $\btheta_{jd}$'s.   
Since $z_{jdi} \sim \Mult_K(\btheta_{jd})$, $E (z_{jdi}) = \theta_{jdk}$,
for $j = 1, \ldots, J$, $d = 1, \ldots, D_j$, $i = 1, \ldots, n_{jd}$ 
and $k = 1, \ldots, K$. We then have 
\begin{equation}
\begin{split}
\ell_{w_{jdi}} (\bbeta, \btheta_{jd}) 
&= E \Bigl [ \sum_{k = 1}^K z_{jdik} \prod_{v = 1}^V 
\beta_{kv}^{w_{jdiv}} \Bigl ] 
= \sum_{k = 1}^K \Bigl [ E(z_{jdik}) \prod_{v = 1}^V 
\beta_{kv}^{w_{jdiv}} \Bigl ] \\ 
&= \sum_{k = 1}^K \Bigl [ \theta_{jdk} \prod_{v = 1}^V 
\beta_{kv}^{w_{jdiv}} \Bigl ] \\ 
\end{split}
\label{eq:c-lda-perplexity-pred-ll}
\end{equation}      
Let $(\bbeta^{[1]}, \btheta^{[1]}), \ldots, (\bbeta^{[S]}, 
\btheta^{[S]})$ be a Markov chain with invariant distribution 
$\nu_{h, \bw^{\text{train}}}(\bbeta, \btheta)$. One can then 
estimate the marginal likelihood $p (w_{jdi} \given 
\bw^{\text{train}})$ in~\eqref{eq:c-lda-perplexity} by the Monte 
Carlo average 
\begin{equation}
\frac{1}{S} \sum_{s = 1}^S \sum_{k = 1}^K \Bigl [ \theta^{[s]}_{jdk} \prod_{v = 1}^V {\beta^{[s]}}_{kv}^{w_{jdiv}} \Bigl ]
\label{eq:c-lda-perplexity-est}
\end{equation}   
We can use any of the augmented chains on $(\bbeta, \bpi, \btheta, 
\bz)$ described in Section~\ref{sec:mcmc} to compute this 
average.   
An elegant alternative is to substitute the following estimates of 
$\theta^{[s]}_{jdk}$ and ${\beta^{[s]}}_{kv}$ in 
\eqref{eq:c-lda-perplexity-est}  
\begin{equation}
\begin{split}
\hat{\theta}^{[s]}_{jdk} 
&= E_{\bpi_j, \bz_d, \bw_d} \left (\theta^{[s]}_{jdk} \given \bz_d, 
\bpi_j, \bw_d \right) = \frac{n_{jdk}^{\text{train}} + \gamma 
\bpi^{[s]}_{j}}{n_{jd.}^{\text{train}} + \gamma} \\ 
\hat{\beta}^{[s]}_{kv} 
&= E_{\bz, \bw} \left (\beta^{[s]}_{kv} \given \bz, \bw \right) =  
\frac{m_{..kv}^{\text{train}} + \eta}{m_{..k.}^{\text{train}} + V 
\eta} \\ 
\end{split}
\end{equation}
which can be computed for every sample in a chain on $(\bpi, \bz)$. 
The resulting likelihood estimate dominates the original estimate 
\eqref{eq:c-lda-perplexity-est} in terms of variance. This approach 
is sometimes called as Rao-Blackwellization, see, e.g., 
\cite[Chapter 4]{RobertCasella:2005}. To compute these estimates, 
we only need a Markov chain on $(\bpi, \bz)$, which has reduced 
computational cost compared to a Markov chain on $(\bbeta, \bpi, 
\btheta, \bz)$.  
Note that one can plug in variational estimates of $\theta_{jdk}$ 
and $\beta_{kv}$ via the cLDA variational EM algorithm (see 
Section~\ref{app:vem}) into~\eqref{eq:c-lda-perplexity-pred-ll} to 
estimate the marginal likelihood $p (w_{jdi} \given 
\bw^{\text{train}})$.   

Using similar arguments, we can derive an expression for the 
estimate of the marginal likelihood $p (w_{di} \given 
\bw^{\text{train}})$ for the LDA model. Given the CGS chain 
$z^{[1]}, \ldots, z^{[S]}$ \citep{GriffithsSteyvers:2004}, we have 
\begin{equation}
\hat{p} (w_{di} \given \bw^{\text{train}})
= \frac{1}{S} \sum_{s = 1}^S \sum_{k = 1}^K \Bigl [ 
\hat{\theta}^{[s]}_{dk} \prod_{v = 1}^V 
{\hat{\beta}^{[s]}}_{kv}^{w_{div}} \Bigl ], 
\label{eq:lda-perplexity}
\end{equation} 
where 
\begin{equation}
\begin{split}
\hat{\theta}^{[s]}_{dk} 
&= \frac{n_{dk}^{\text{train}} + \alpha}{n_{d.}^{\text{train}} + 
K\alpha} \\ 
\hat{\beta}^{[s]}_{kv} 
&= \frac{m_{.kv}^{\text{train}} + \eta}{m_{.k.}^{\text{train}} + 
	V\eta}. 
\end{split}
\end{equation}

%% file: app-nips-results.tex
\section{Additional Experimental Results}
\label{app:nips-results}
This section gives additional results of experiments on real world  
corpora that are discussed in Section~\ref{sec:experiments}. 

\begin{figure}[htbp!]
	\centering
	\subfloat[cLDA topics: \texttt{hclust} dendrogram]
	{\includegraphics[width=.55\textwidth]
		{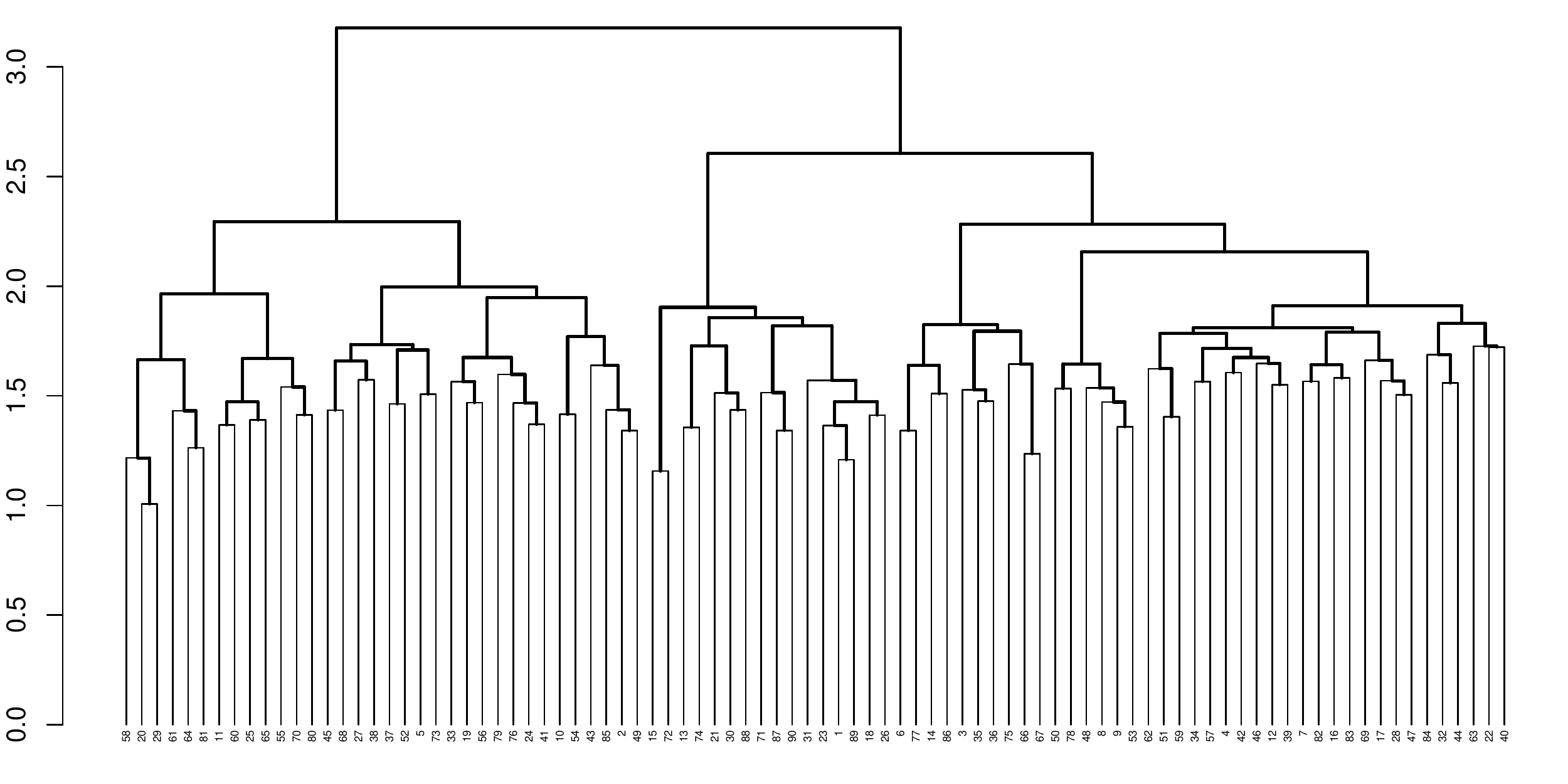}} 
	\subfloat[cLDA topics' size (log scale)]
	{\includegraphics[width=.44\textwidth]
		{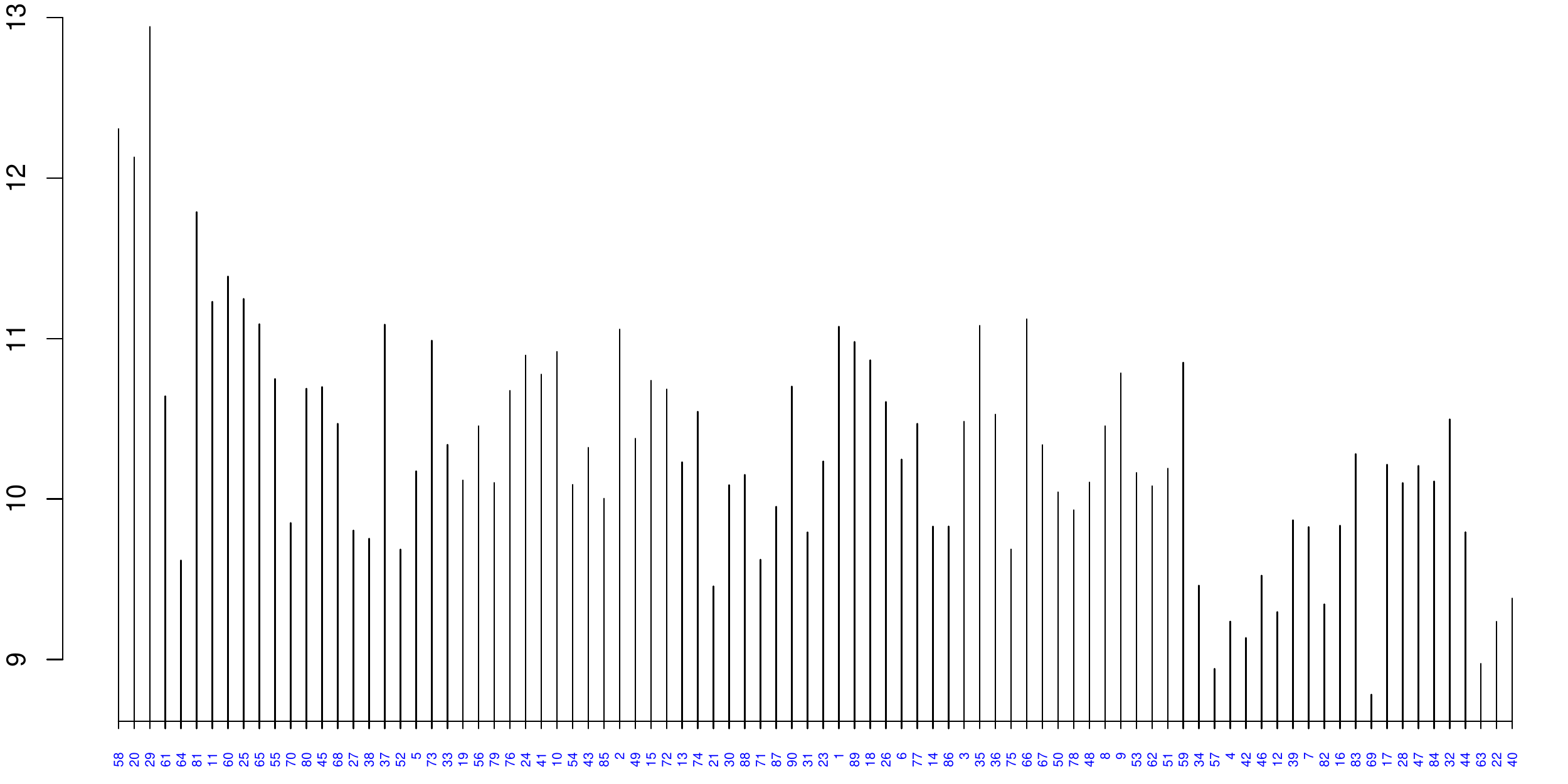}} \\ 
	\subfloat[LDA topics: \texttt{hclust} dendrogram]
	{\includegraphics[width=.55\textwidth]
		{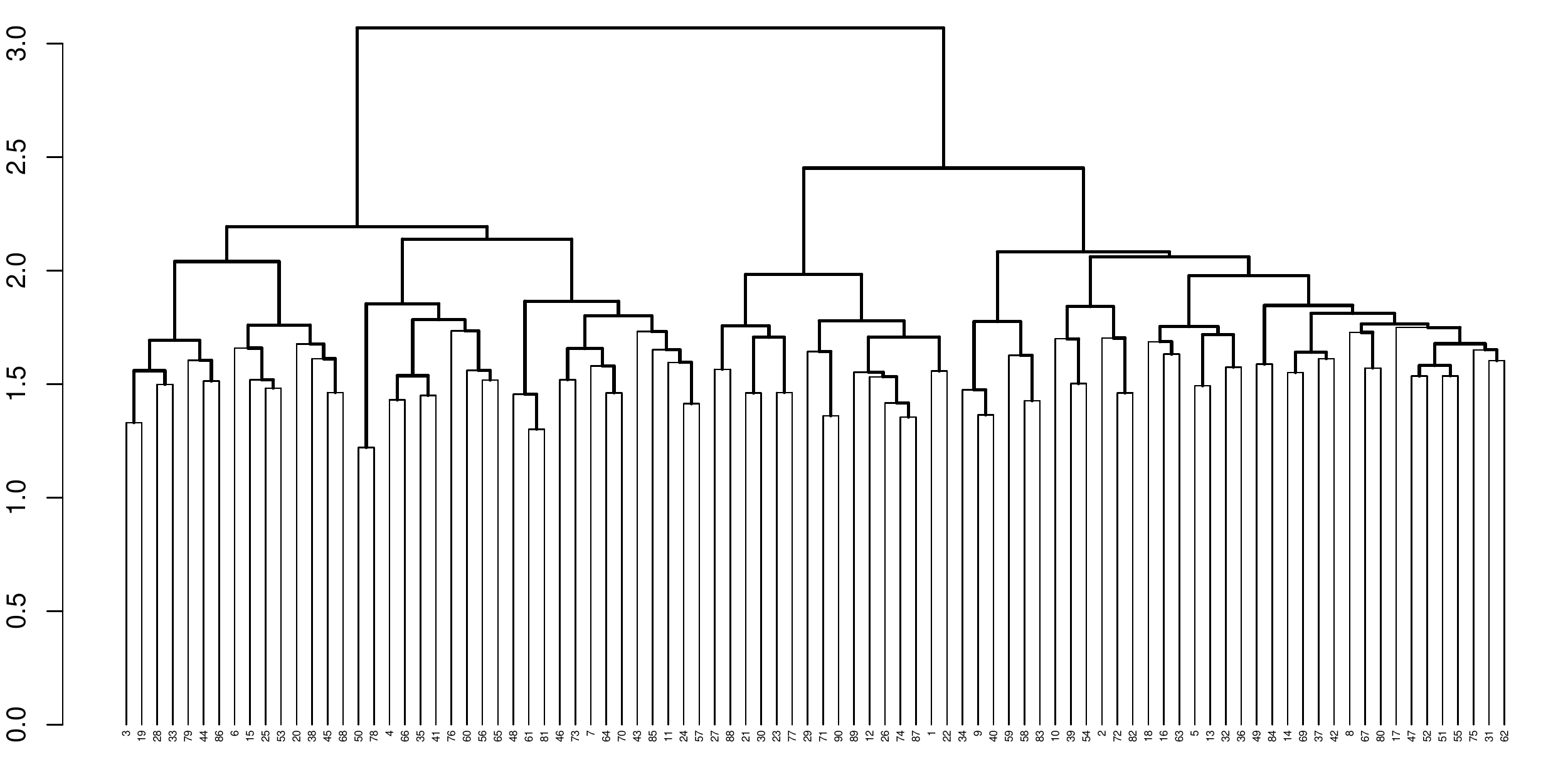}}
	\subfloat[LDA topics' size (log scale)]
	{\includegraphics[width=.44\textwidth]
		{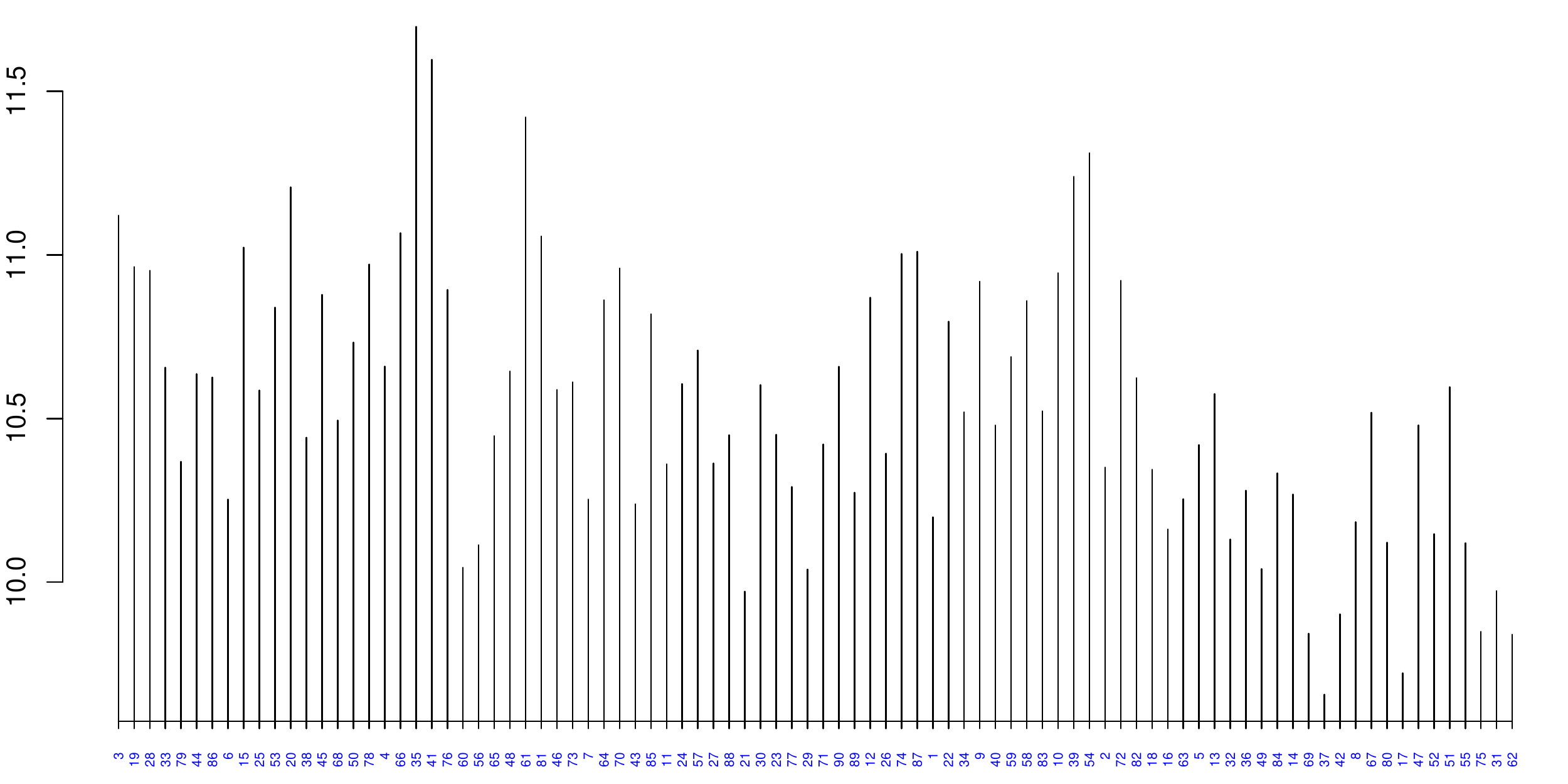}} \\ 
	\subfloat[HDP topics: \texttt{hclust} dendrogram]
	{\includegraphics[width=.55\textwidth]
		{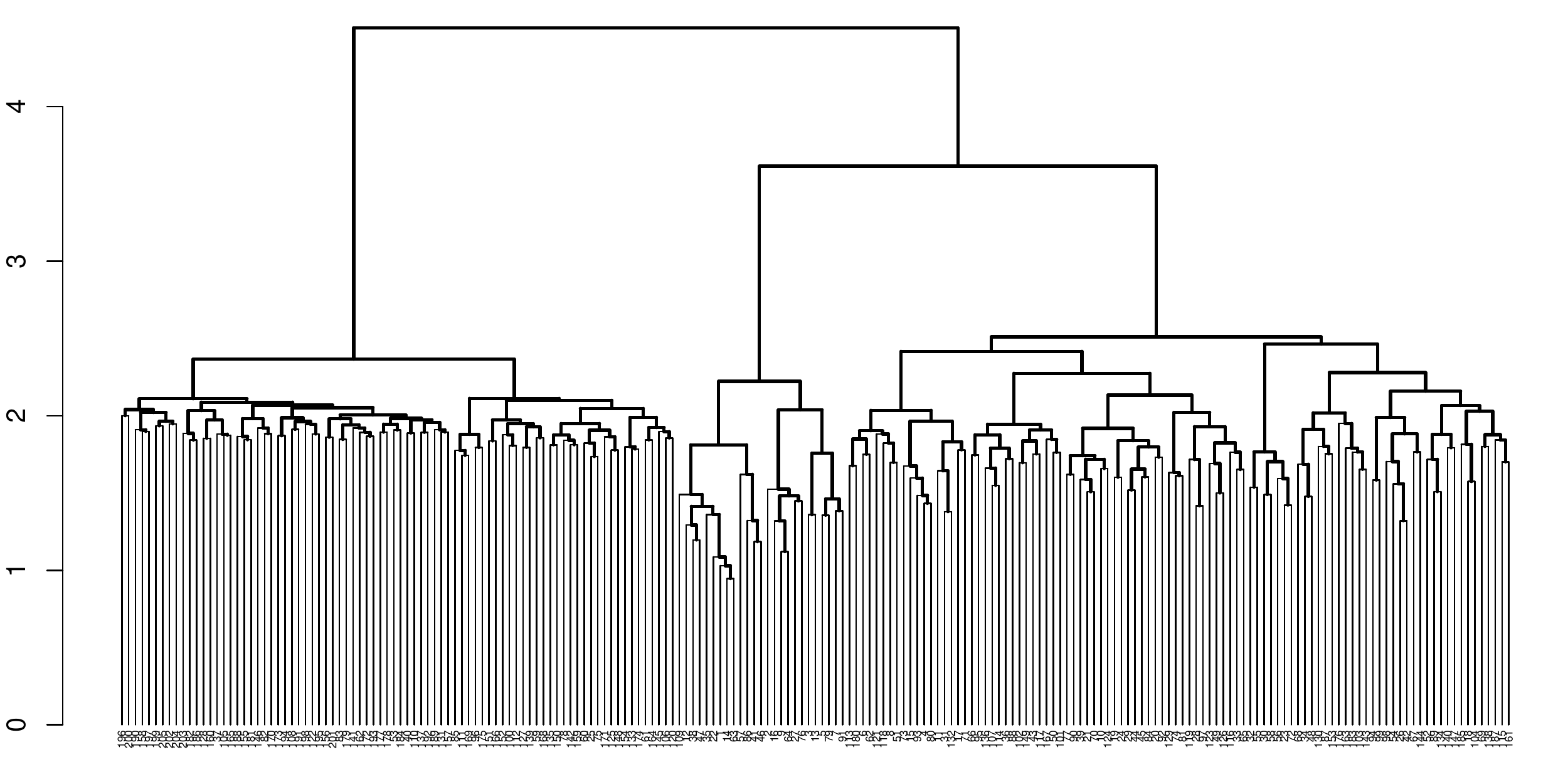}}
	\subfloat[HDP topics' size (log scale)]
	{\includegraphics[width=.44\textwidth]
		{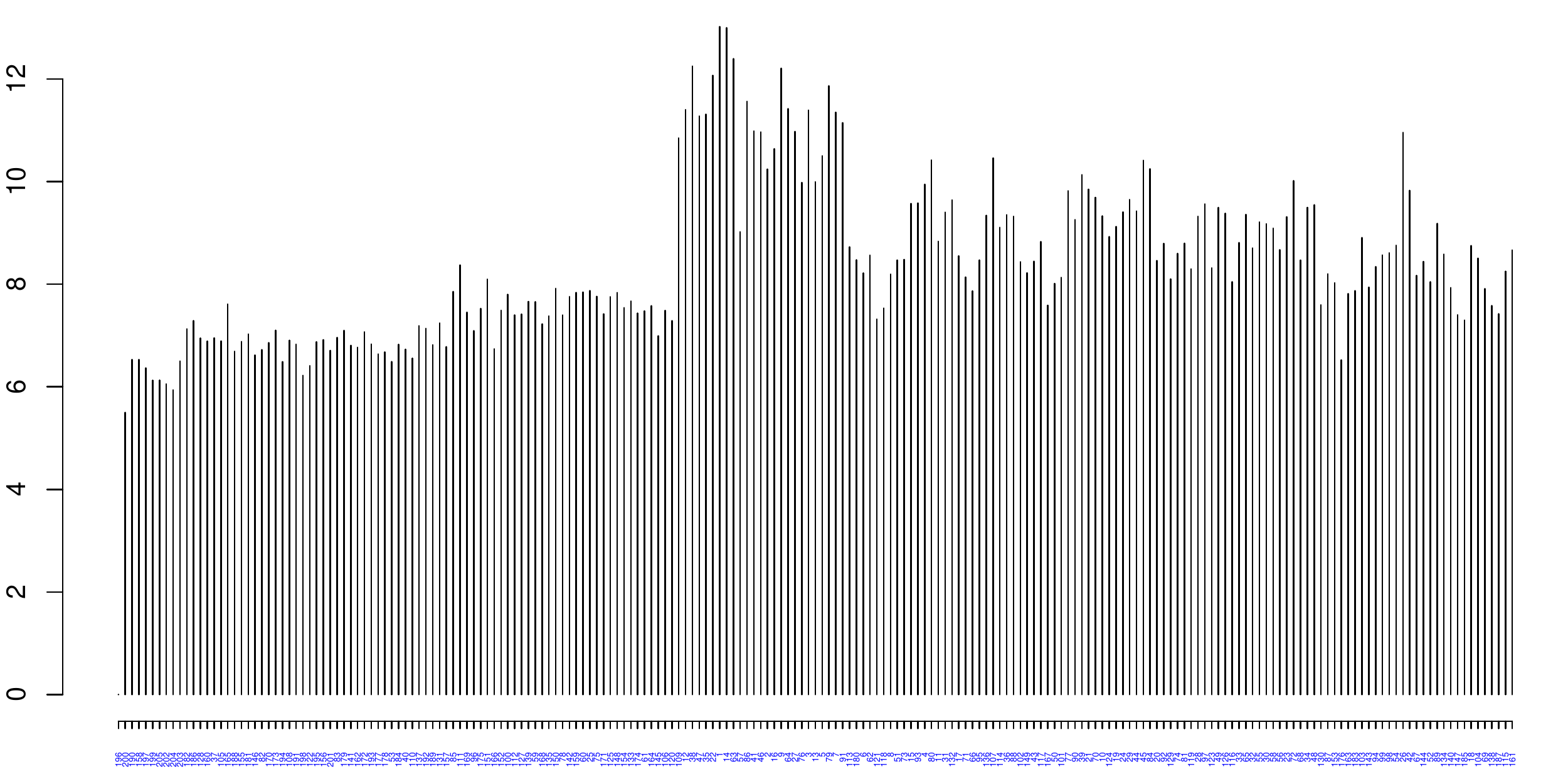}} 
	\caption{Comparing topics learned via cLDA-AGS, LDA-CGS, and 
		HDP-CRF algorithms. Left-column shows \texttt{hclust} 
		dendrograms built 
		from topic-to-topic similarity matrices calculated based on 
		the 
		\emph{manhattan} distance for all three algorithms. For 
		each 
		method, right-column shows barplots of topic-sizes (i.e. 
		the 
		total number of 
		words assigned to a topic). Topics in the barplot $x$-axis  
		were ordered based on the order of topics in the 
		corresponding dendrogram leaf-nodes to ease comparison. 
		HDP found redundant set of topics and many topics with 
		too low topic-size---e.g. the children of the first/left 
		child in the dendrogram in Plot (e). See discussion in 
		Section \ref{sec:eval-performance}.        
	}
	\label{fig:topics-hclust-dendrogram-size}
\end{figure}

\begin{figure}[htbp!]
	\centering
	\includegraphics[width=.7\textwidth]
	{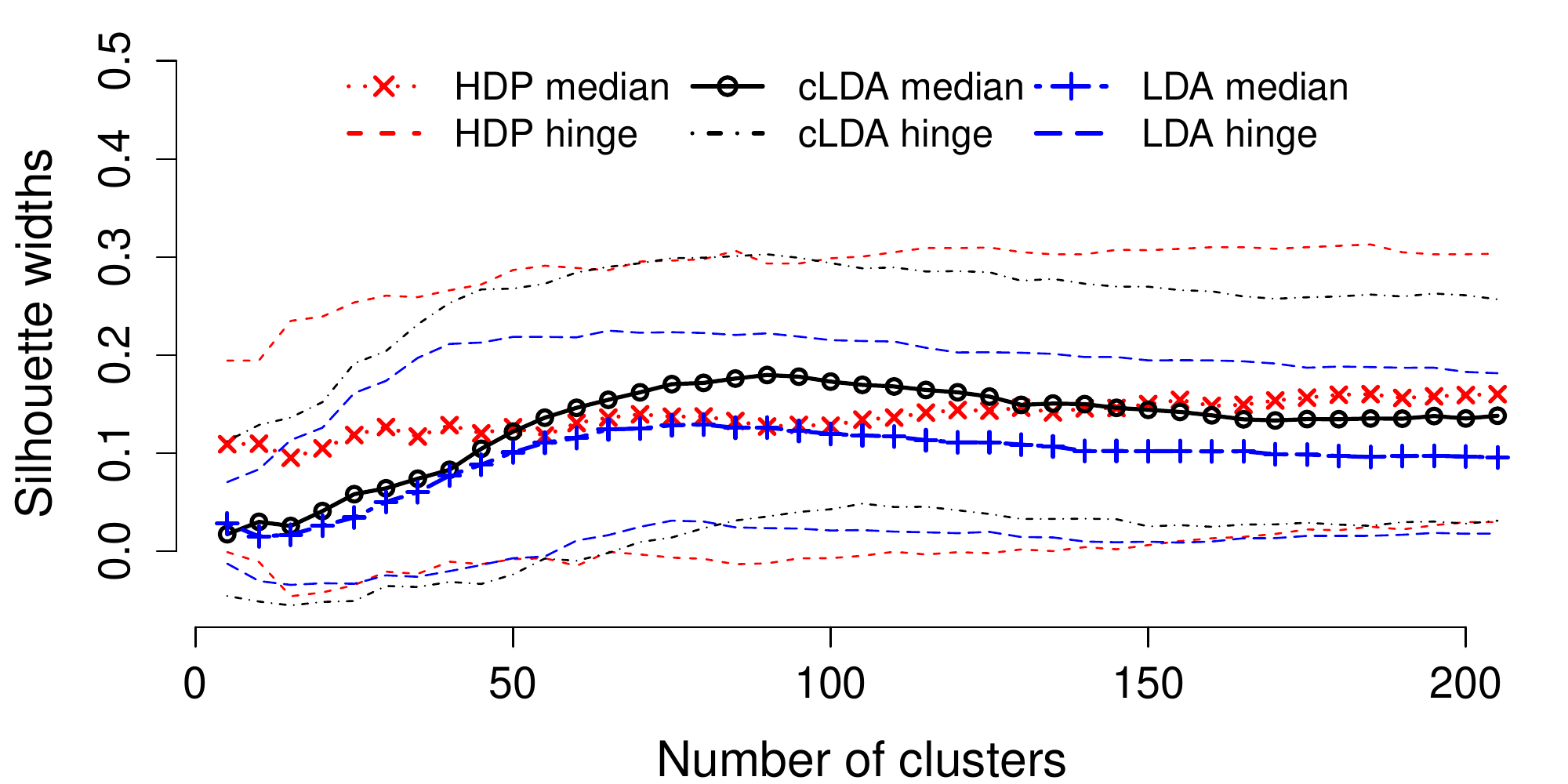}
	\caption{Boxplot statistics (median, lower hinge, and upper 
		hinge) of silhouette widths computed on \emph{documents}' 
		\texttt{hclust} clusters with various values of the number 
		of clusters, for corpus NIPS $00$-$18$. Algorithm 
		\texttt{hclust} was applied on documents' topic 
		distributions estimated via cLDA, LDA, and HDP sampling 
		algorithms.   
		Silhouette widths of clusters based on HDP are relatively 
		constant with different values of the number of clusters. 
		HDP estimated large set of minute topics ($K = 204$), which 
		may have helped clustering documents. cLDA performs better 
		than LDA, and is comparable with HDP or better than HDP 
		with the right number of \texttt{hclust} clusters (e.g. 
		from $50$ to $150$). See discussion in Section 
		\ref{sec:eval-performance}.  
	}
	\label{fig:clda-lda-hdp-doc-hclust-silhouette}
\end{figure}

\begin{figure}[htbp!]
	\centering
	\includegraphics[width=.7\textwidth]
		{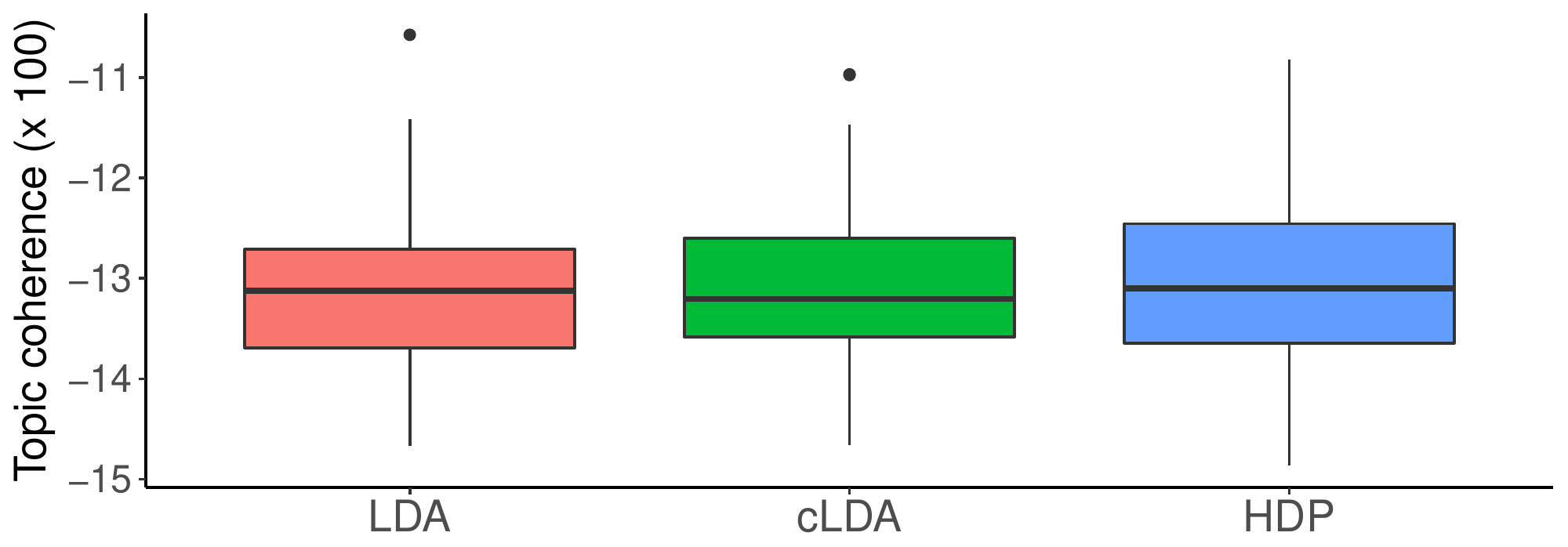}
	\caption{Estimated topic \emph{coherences} for models 
		LDA, cLDA, and HDP  for corpus NIPS $00$-$18$. See 
		discussion in Section \ref{sec:eval-performance}.   }
	\label{fig:lda-clda-hdp-topic-coherences} 
\end{figure}

\begin{figure}[t!] 
	\centering
	\includegraphics[width=.98\linewidth]
	{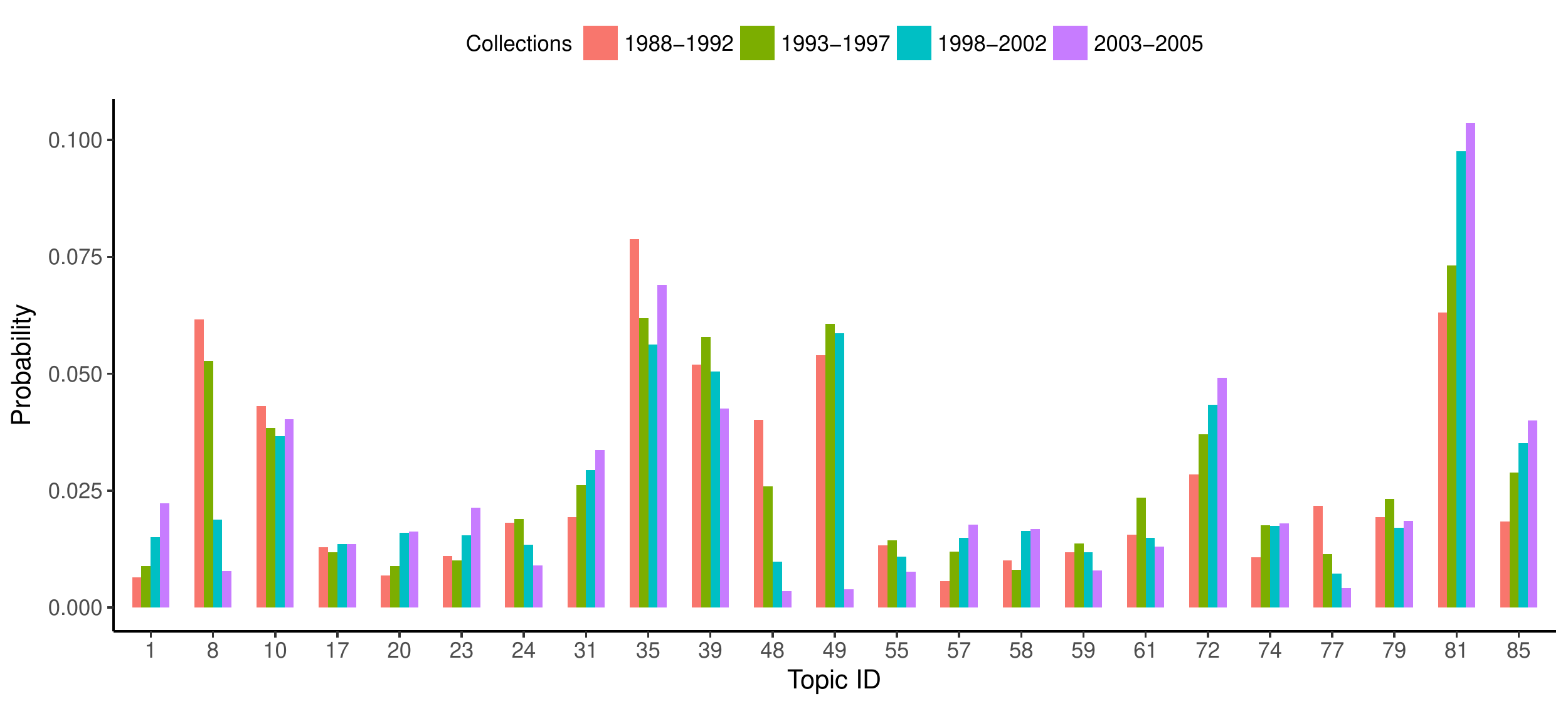}
	\caption{Estimates of topic distributions for four collections 
		(defined on timespans $1988$-$1992$, $1993$-$1997$, $1998$-$2002$, 
		and $2003$-$2005$) of the NIPS $00$-$18$ corpus via the cLDA AGS 
		algorithm. 
	}
	\label{fig:nips-clda-pi-samples}
\end{figure}

\begin{figure*}[ht!] 
	\centering 
	\subfloat[topic $3$ (Science)]{
		\includegraphics[width=.3\linewidth]
		{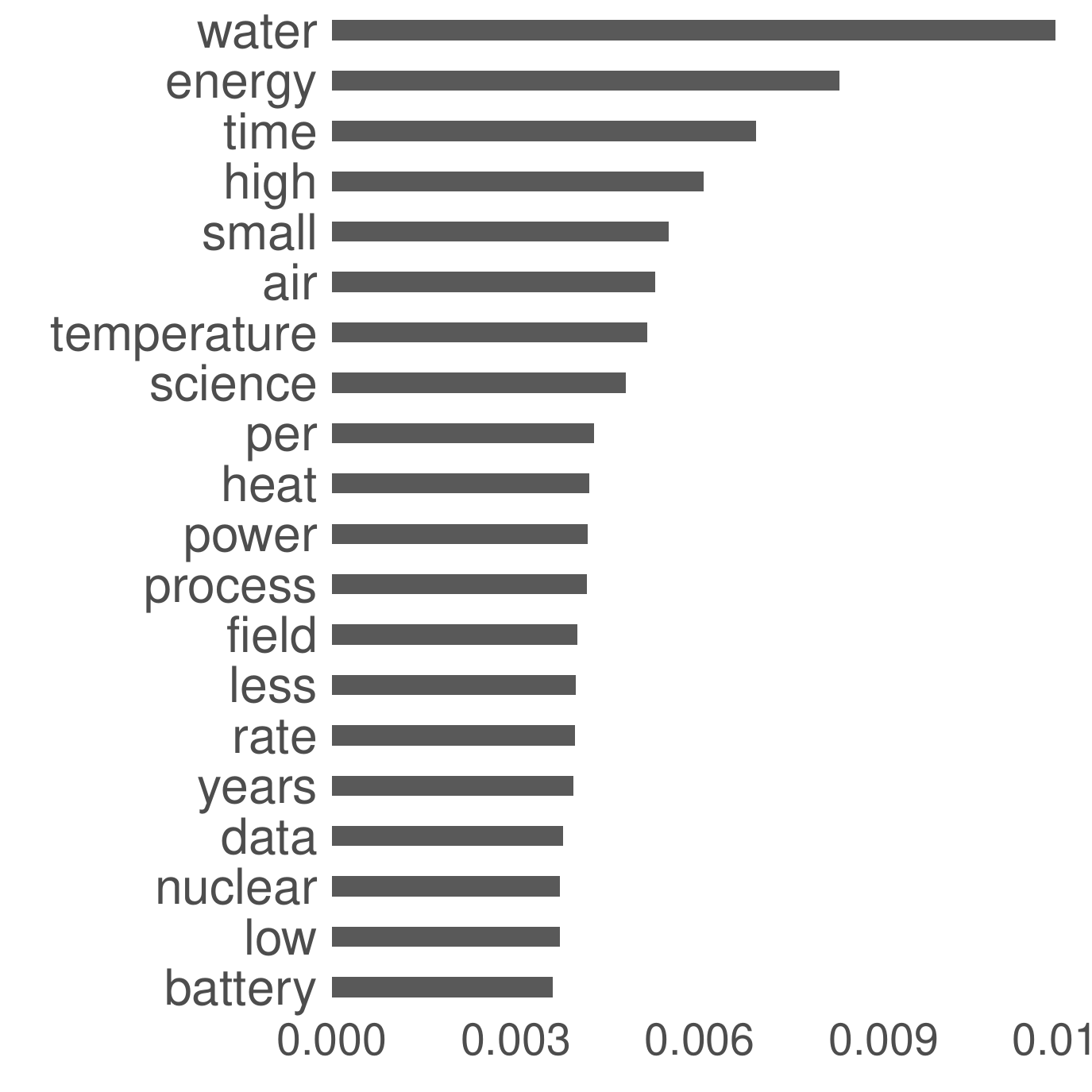}}\hfill%
	\subfloat[topic $10$ (Uninformative)]{
		\includegraphics[width=.3\linewidth]
		{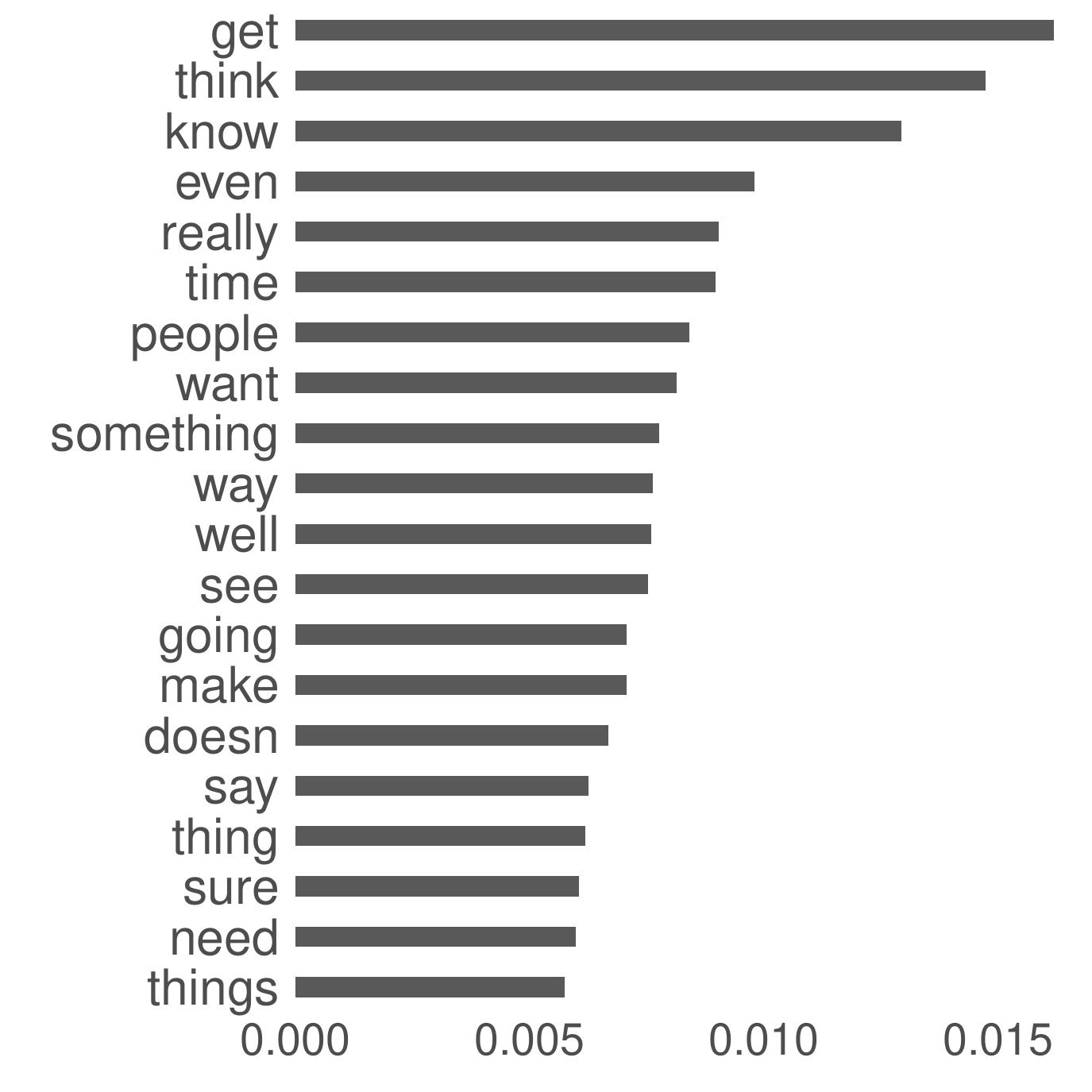}}\hfill%
	\subfloat[topic $21$ (Recreation)]{
		\includegraphics[width=.3\linewidth]
		{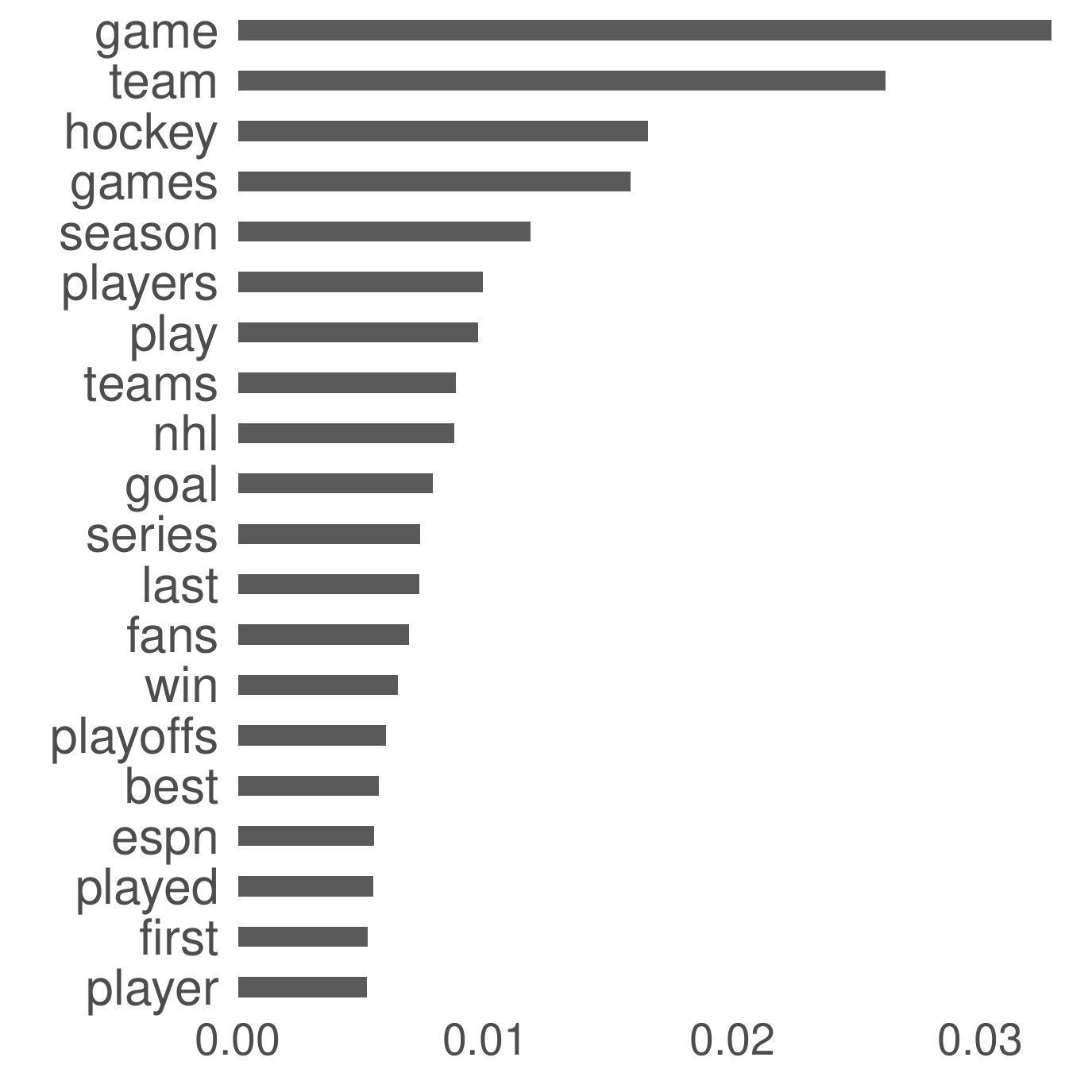}}\\
	\subfloat[topic $13$ (Computers)]{
		\includegraphics[width=.3\linewidth]
		{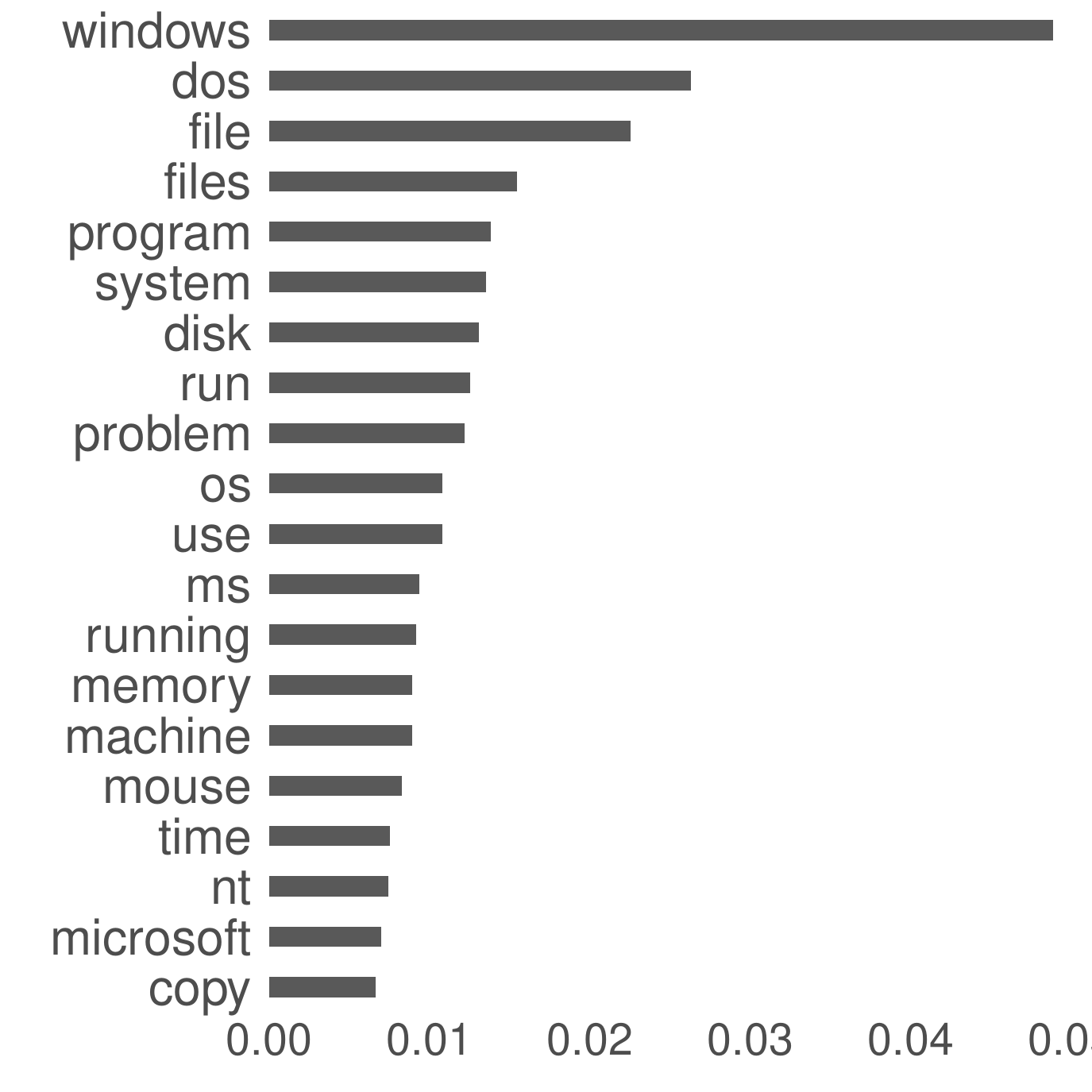}}\hfill%
	\subfloat[topic $28$ (Politics, gun trade)]{
		\includegraphics[width=.3\linewidth]
		{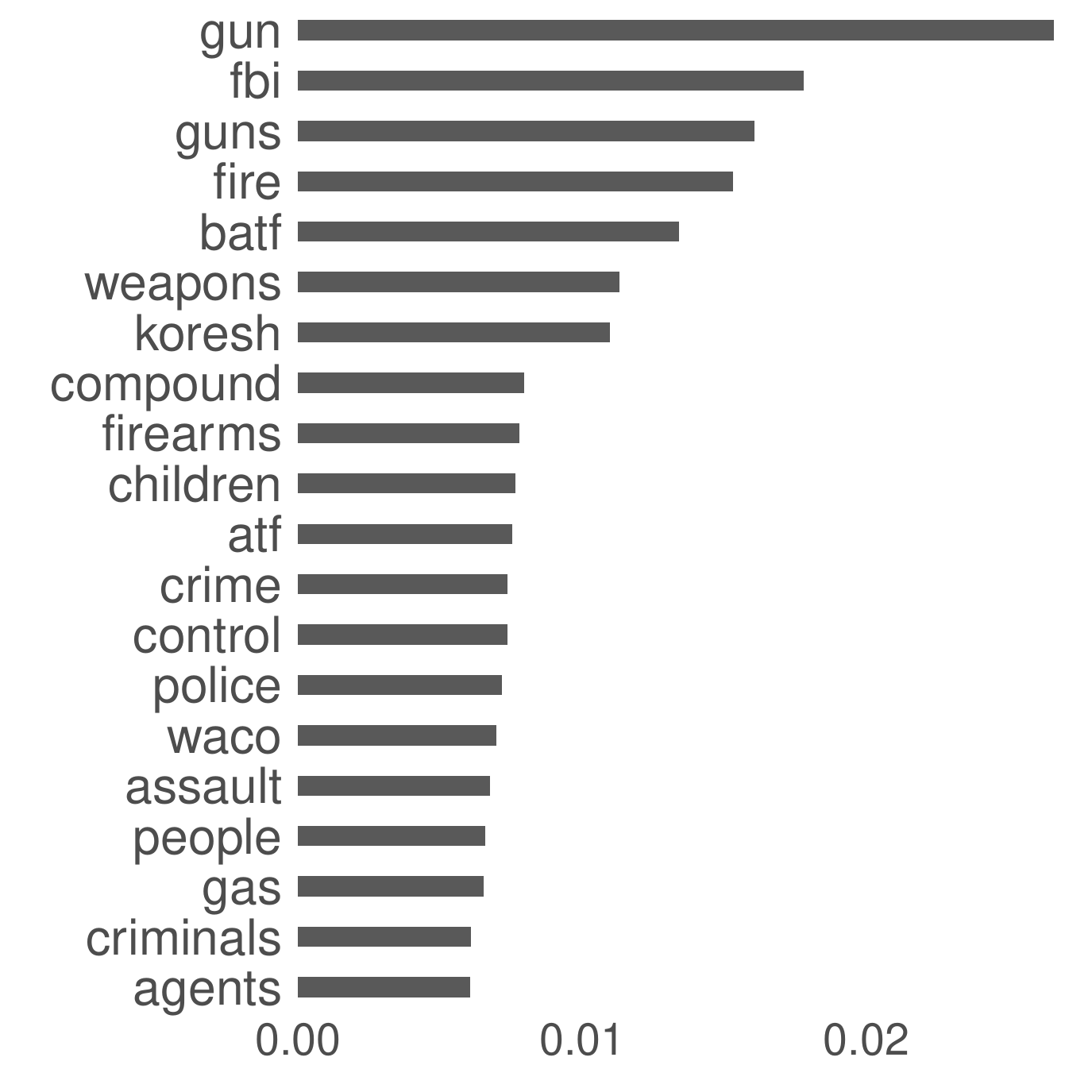}}\hfill%
	\subfloat[topic $29$ (Computers)]{
		\includegraphics[width=.3\linewidth]
		{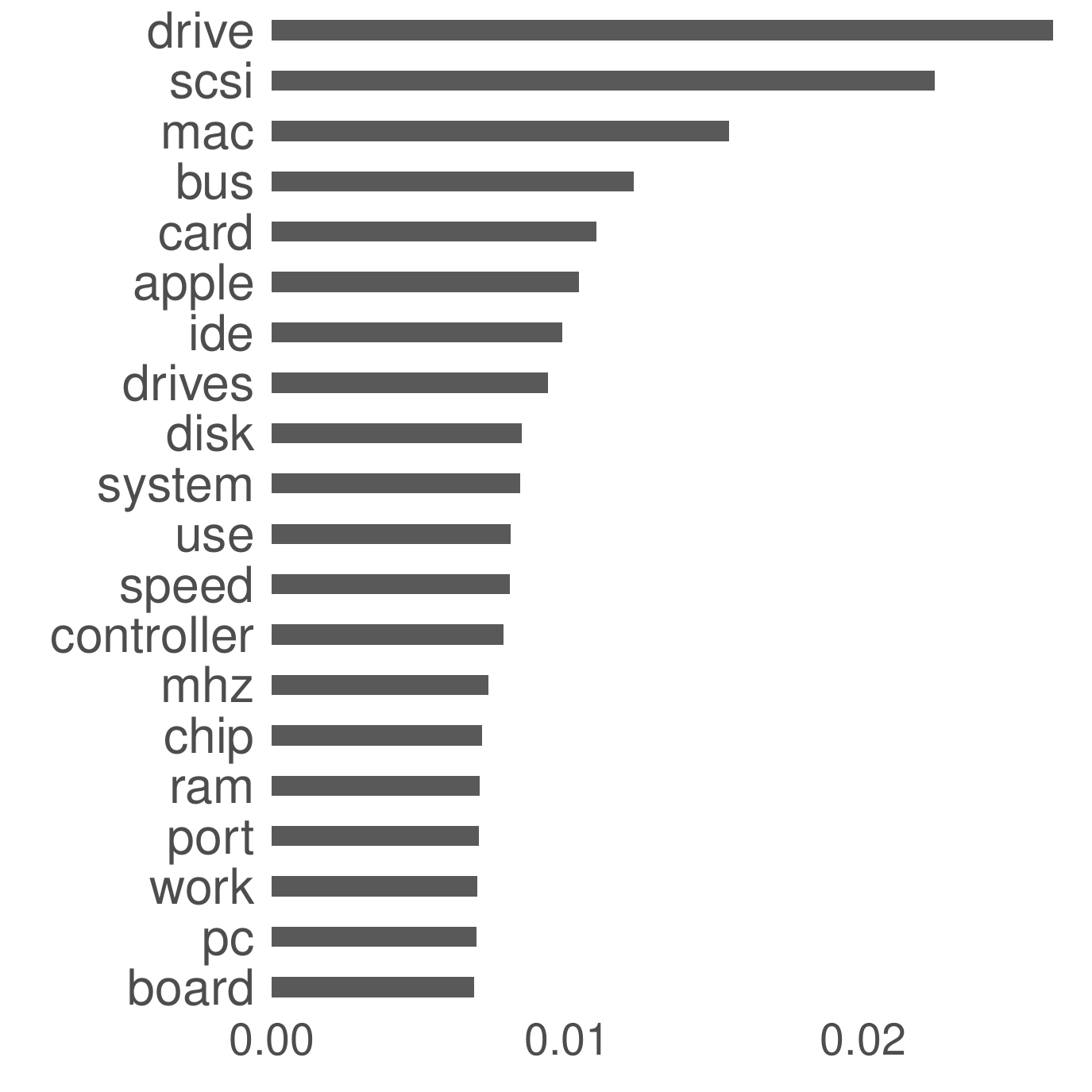}}
	\caption{$20$ most probable words for three selected topics from a 
		$30$-topic cLDA model trained on corpus $16$newsgroups.	The 
		$x$-axis gives the corresponding (estimated) probabilities of words 
		given a topic.}
	\label{fig:16news-topics-new}
\end{figure*}

%
%

\begin{figure}[t!] 
	\centering
	\includegraphics[width=.8\linewidth]
	{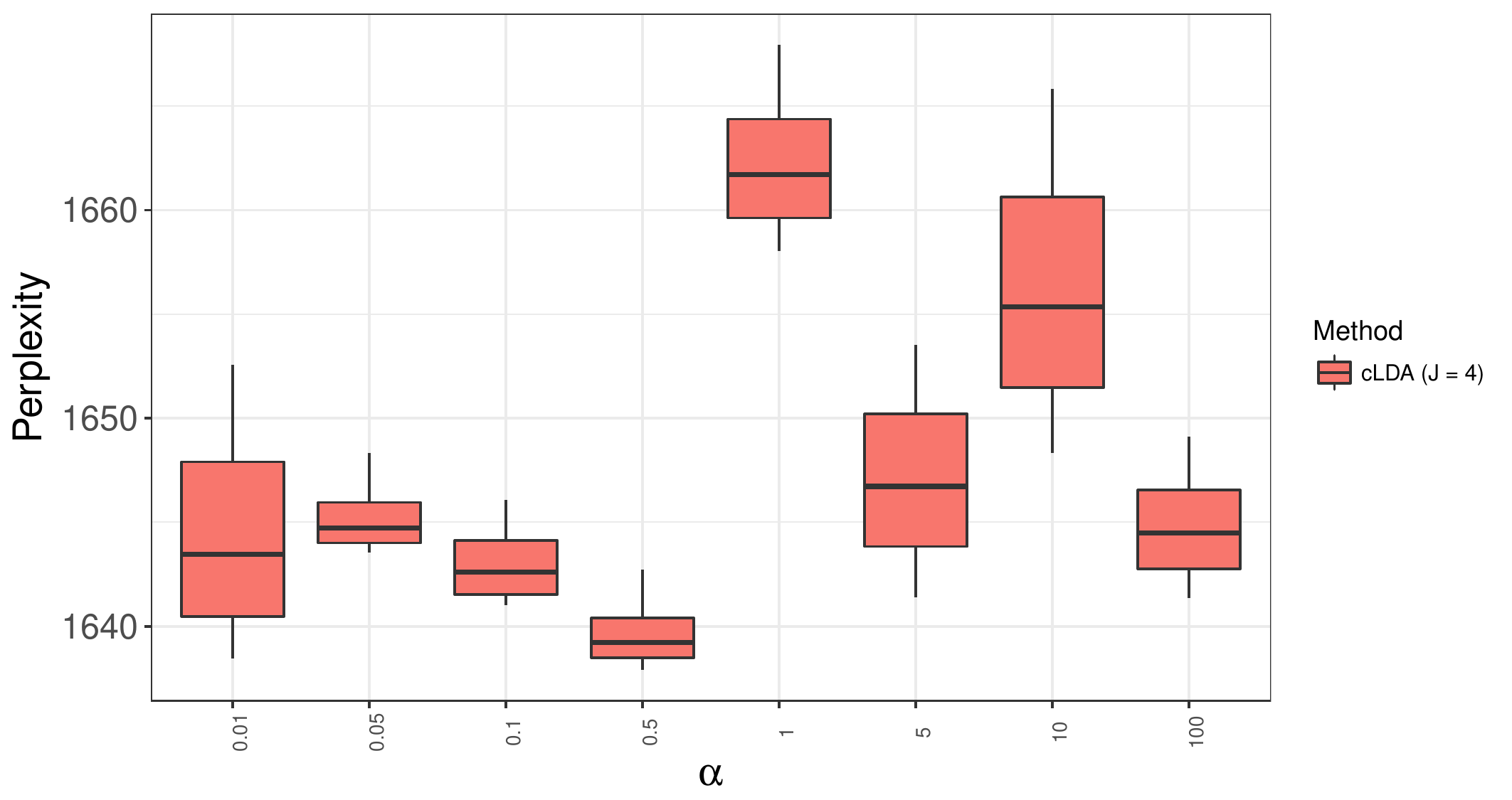}
	\caption{Boxplots of perplexity scores of cLDA models with 
	various values of $\alpha$ keeping $\eta$ and $\gamma$ fixed, 
	for corpus NIPS $00$-$18$.}
	\label{fig:nips-clda-alpha-sensitivity}
\end{figure}